# The Gaia-ESO Public Spectroscopic Survey: Implementation, data products, open cluster survey, science, and legacy.,⋆

S. Randich[1], G. Gilmore[2], L. Magrini[1], G. G. Sacco[1], R. J. Jackson[3], R. D. Jeffries[3], C. C. Worley[2], A. Hourihane[2], A. Gonneau[2], C. Viscasillas Vázquez[4], E. Franciosini[1], J. R. Lewis[†2], E. J. Alfaro[5], C. Allende Prieto[6,7], T. Bensby[8], R. Blomme[9], A. Bragaglia[10], E. Flaccomio[11], P. François[12], M. J. Irwin[2], S. E. Koposov[13,2], A. J. Korn[14], A. C. Lanzafame[15], E. Pancino[1,16], A. Recio-Blanco[17], R. Smiljanic[18], S. Van Eck[19], T. Zwitter[20], M. Asplund[21], P. Bonifacio[23], S. Feltzing[8], J. Binney[22], J. Drew[24], A. M. N. Ferguson[25], G. Micela[11], I. Negueruela[26], T. Prusti[27], H.-W. Rix[28], A. Vallenari[29], A. Bayo[30,31], M. Bergemann[28,32], K. Biazzo[33], G. Carraro[34], A. R. Casey[35], F. Damiani[11], A. Frasca[36], U. Heiter[14], V. Hill[17], P. Jofré[37], P. de Laverny[17], K. Lind[38], G. Marconi[39], C. Martayan[39], T. Masseron[6,7], L. Monaco[40], L. Morbidelli[1], L. Prisinzano[11], L. Sbordone[39], S. G. Sousa[41], S. Zaggia[29], V. Adibekyan[41], R. Bonito[11], E. Caffau[23], S. Daflon[42], D. K. Feuillet[8,28], M. Gebran[43], J. I. González Hernández[6], G. Guiglion[44], A. Herrero[6,7], A. Lobel[9], J. Maíz Apellániz[45], T. Merle[19], Š. Mikolaitis[4], D. Montes[46], T. Morel[47], C. Soubiran[48], L. Spina[29], H. M. Tabernero[49], G. Tautvaišienė[4], G. Traven[20], M. Valentini[44], M. Van der Swaelmen[1], S. Villanova[50], N. J. Wright[3], U. Abbas[51], V. Aguirre Børsen-Koch[52], J. Alves[53], L. Balaguer-Núnez[54], P. S. Barklem[55], D. Barrado[56], S. R. Berlanas[26], A. S. Binks[57,3], A. Bressan[58], R. Capuzzo–Dolcetta[59], L. Casagrande[60], L. Casamiquela[48], R. S. Collins[13], V. D'Orazi[29], M. L. L. Dantas[18], V. P. Debattista[61], E. Delgado-Mena[41], P. Di Marcantonio[62], A. Drazdauskas[4], N. W. Evans[2], B. Famaey[63], M. Franchini[62], Y. Frémat[9], E. D. Friel[64], X. Fu[65], D. Geisler[50,66,67], O. Gerhard[68], E. A. González Solares[2], E. K. Grebel[69], M. L. Gutiérrez Albarrán[46], D. Hatzidimitriou[70,71], E. V. Held[29], F. Jiménez-Esteban[45], H. Jönsson[72], C. Jordi[54], T. Khachaturyants[61], G. Kordopatis[17], J. Kos[20], N. Lagarde[48,73], L. Mahy[9], M. Mapelli[29], E. Marfil[45], S. L. Martell[74], S. Messina[36], A. Miglio[75,10], I. Minchev[44], A. Moitinho[76], J. Montalban[75], M. J. P. F. G. Monteiro[41,77], C. Morossi[62], N. Mowlavi[78], A. Mucciarelli[75,10], D. N. A. Murphy[2], N. Nardetto[17], S. Ortolani[34], F. Paletou[79], J. Palouš[80], E. Paunzen[81], J. C. Pickering[82], A. Quirrenbach[83], P. Re Fiorentin[51], J. I. Read[84], D. Romano[10], N. Ryde[8], N. Sanna[1], W. Santos[42], G. M. Seabroke[85], A. Spagna[51], M. Steinmetz[44], E. Stonkuté[86], E. Sutorius[13], F. Thévenin[17], M. Tosi[10], M. Tsantaki[1], J. S. Vink[87], N. Wright[3], R. F. G. Wyse[88], M. Zoccali[89], J. Zorec[90], D. B. Zucker[91], and N. A. Walton[2]

*(Affiliations can be found after the references)*

## ABSTRACT

*Context.* In the last 15 years different ground-based spectroscopic surveys have been started (and completed) with the general aim of delivering stellar parameters and elemental abundances for large samples of Galactic stars, complementing *Gaia* astrometry. Among those surveys, the Gaia-ESO Public Spectroscopic Survey (GES) was designed to target 100,000 stars using FLAMES on the ESO VLT (both Giraffe and UVES spectrographs), covering all the Milky Way populations, with a special focus on open star clusters.

*Aims.* This article provides an overview of the survey implementation (observations, data quality, analysis and its success, data products, and releases), of the open cluster survey, of the science results and potential, and of the survey legacy. A companion article reviews the overall survey motivation, strategy, Giraffe pipeline data reduction, organisation, and workflow.

*Methods.* We made use of the information recorded and archived in the observing blocks; during the observing runs; in a number of relevant documents; in the spectra and master catalogue of spectra; in the parameters delivered by the analysis nodes and the working groups; in the final catalogue; and in the science papers. Based on these sources, we critically analyse and discuss the output and products of the Survey, including science highlights.

*Results.* The GES has determined homogeneous good-quality radial velocities and stellar parameters for a large fraction of its more than 110,000 unique target stars. Elemental abundances were derived for up to 31 elements for targets observed with UVES. The analysis and homogenisation strategies have proven to be successful; several science topics have been addressed by the GES consortium and the community, with many highlight results achieved.

*Conclusions.* The final catalogue will be released through the ESO archive in the first half of 2022, including the complete set of advanced data products. In addition to these results, the GES will leave a very important legacy, for several aspects and for many years to come.

## 1. Introduction

In the last 15 years or so several ground-based stellar spectroscopic surveys have been undertaken; the common broad goal was the detailed investigation of the structure, formation, and evolution of the Milky Way (MW) Galaxy and its component populations, complementing the exquisite astrometry and photometry of the *Gaia* space mission (Gaia Collaboration et al. 2016, 2018, 2021). These observational programmes include RAVE (Steinmetz et al. 2020b), APOGEE and APOGEE II (Majewski et al. 2016, 2017), LAMOST (Zhao et al. 2012), GALAH (De Silva et al. 2015), and the Gaia-ESO Survey (GES; Gilmore et al. 2012; Randich et al. 2013). These surveys have different properties, and are characterised by a variety of spectral resolutions, spectral ranges, limiting magnitudes, sampled populations, and selection functions. Whilst we refer to the above papers for a detailed description of the characteristics of the various surveys, we focus here on GES.

The GES is a large public spectroscopic survey that was devised in the context of the call for public spectroscopic surveys issued by the European Southern Observatory (ESO) in 2011, following the recommendations of a number of European strategic documents and the outcome of the 2009 ESO Workshop on wide-field spectroscopic surveys (Melnick et al. 2009). GES was designed to exploit the capabilities of the FLAMES instrument (Pasquini et al. 2002) on the ESO Very Large Telescope, using both Giraffe and UVES spectrographs; GES is unique with respect to the other spectroscopic surveys in several ways. Specifically, it is the only stellar spectroscopic survey that **i.** has been performed on an 8m class telescope, meaning that it has reached much fainter targets (larger volumes and/or, at a given distance, intrinsically less luminous and lower mass stars); **ii.** has systematically covered all populations and all types of stars in the MW, from the halo (although with relatively few stars), the thin and thick discs, and the Galactic Bulge, to young star clusters and star forming regions in the solar vicinity; from pre-main sequence (PMS) stars to old turn-off (TO) stars and evolved giants; from very cool stars to hot massive stars; **iii.** has used different settings (spectral coverage) and instruments optimised for the different types of stars and science drivers; **iv.** has used multiple pipelines to analyse the same sets of spectra and has combined and homogenised the results using internal calibrators; **v.** has put a particular focus on open star clusters (OCs), observing large unbiased samples comprising hundreds of stars in more than 60 clusters, well sampling the age-distance-metallicity parameter space; **vi.** has analysed ESO archive samples, homogeneously with the survey targets; **vii.** has paid specific attention to deriving high-precision radial velocities (RVs), with the initial goal of reaching ∼ 300 m/s in members of nearby clusters; and **viii.** in addition to stellar parameters (effective temperature -$T_{\rm eff}$- and surface gravity -logg), metallicity ([Fe/H]), and elemental abundances, has delivered key products such as a spectroscopic gravity index, chromospheric activity tracers, and mass accretion rate diagnostics (see Damiani et al. 2014; Lanzafame et al. 2015).

All these peculiar properties, as well as the strategy and analysis approach, are the basis of the success of GES and contribute to enhancing its legacy value; at the same time, however, they imply a complex data flow and have created a number of challenges that the Survey consortium has dealt with during the ten years of the project. The overall survey motivation, strategy, organisation, and workflow are described in the companion paper Gilmore et al. (submitted; hereafter GRH22). This paper is complementary to GRH22 and provides a broad overview of the aspects that were not included there, focusing mostly on the survey's implementation and outputs, including the science and science potential, and describing in detail the OC survey.

This paper is meant to be a reference for all the science papers coming from the GES consortium and, at the same time, to be a primer for people interested in understanding the overall structure of GES and in using data from the ESO archive. Several papers more focused on technical issues have been written or are in preparation, and address specific aspects of GES, from target cluster selection to data reduction, calibration strategy, spectrum analysis, and homogenisation. A complete list of references is provided in GRH22.

The paper is structured as follows: In Sect. 2 we present an overview of the Gaia-ESO observations and the quality of the acquired spectra; in Sect. 3 we summarise information on the analysis cycles and data releases, and on the delivered products; in Sect. 4 we discuss the success of the chosen strategy and data flow. Section 5 focuses on the open cluster survey, and Sect. 6 presents a few science highlights. A discussion of the science potential, legacy of GES, and conclusions are given in Sects. 7 and 8.

## 2. Observations

### 2.1. Observing runs

Originally, 300 observing nights were allocated to the project; after the fourth year review that occurred in September 2015, an additional 40 nights were granted to compensate for the time lost due to technical problems and bad weather (see below), for a total of 340 observing nights. GES observations were carried out in Visitor mode by a dedicated team who alternated at the ESO Paranal site (see Table 10 in GRH22).

The observations were distributed across 12 ESO periods (from ESO P88 to P100, P99 being skipped), with typically six observing runs per period and five to seven nights per run. GES observations started on December 31, 2011, and were completed in January 2018, in ESO P100, after 64 observing runs. Details of the observations, including observing blocks (OBs) completed during each run and night, exposure times, information on airmass, and seeing, can be found by querying the ESO observation schedule (https://www.eso.org/sci/observing/telesalloc.html) and specifying the following run IDs: 188.B-3002, 193.B-0936, 197.B-1074.

About 19 % of the allocated time was completely lost due to bad weather (13.6 %), technical problems (3.2 %), or target of opportunity (ToOs) observations (2.3 %); the number of effective nights is hence about 275. Weather conditions when the observations could be carried out were reasonably good; 35% and 42% of the time was characterised by photometric and clear conditions, while only about 23% of the nights were affected by thin clouds (19 %) and thick clouds (4 %). The median seeing was around 0.9 arcsec, very much in line with typical conditions at Paranal. A summary of the weather conditions and seeing statistics is given in Fig. 1.

### 2.2. Observed sample

As described in detail in GRH22, the observed targets include MW fields, science OCs, and a variety of calibrators, such as RV standards, *Gaia* benchmark stars, open and globular clusters, and COROT and Kepler 2 red giants (see also Pancino et al. 2017a). Before each run started, OBs for the three different target categories were prepared by the working group in charge (WG0; see GRH22) and sent to the observer. In typical OBs 80–100 and 20 Giraffe fibres were allocated to science targets and the sky, respectively, while for UVES the fibres allocated to targets and the sky were 7 and 1. Decisions on the fields and OBs to be observed during each night were made based on the priorities indicated by the PIs, target coordinates, and weather conditions.

⋆ Based on data products from observations made with ESO Telescopes at the La Silla Paranal Observatory under programmes 188.B-3002, 193.B-0936, and 197.B-1074.

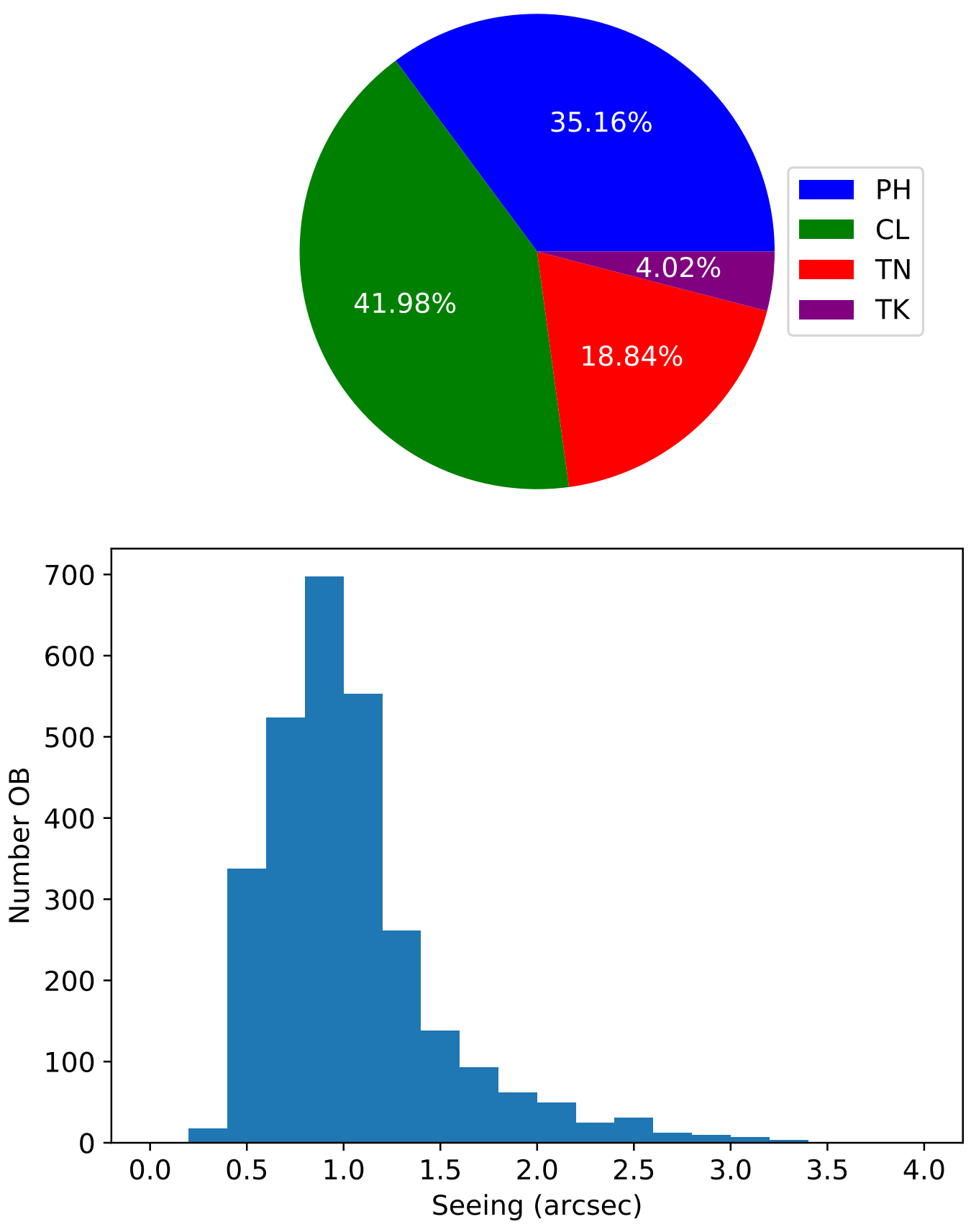

**Fig. 1.** Summary of the observing conditions. The top panel shows the percentages of the observing time during which the following weather conditions were experienced: PH = photometric, CL = clear, TN = thin cirrus cloud, TK = thick cirrus cloud. The bottom panel shows the distribution of DIMM Seeing measured during each OB.

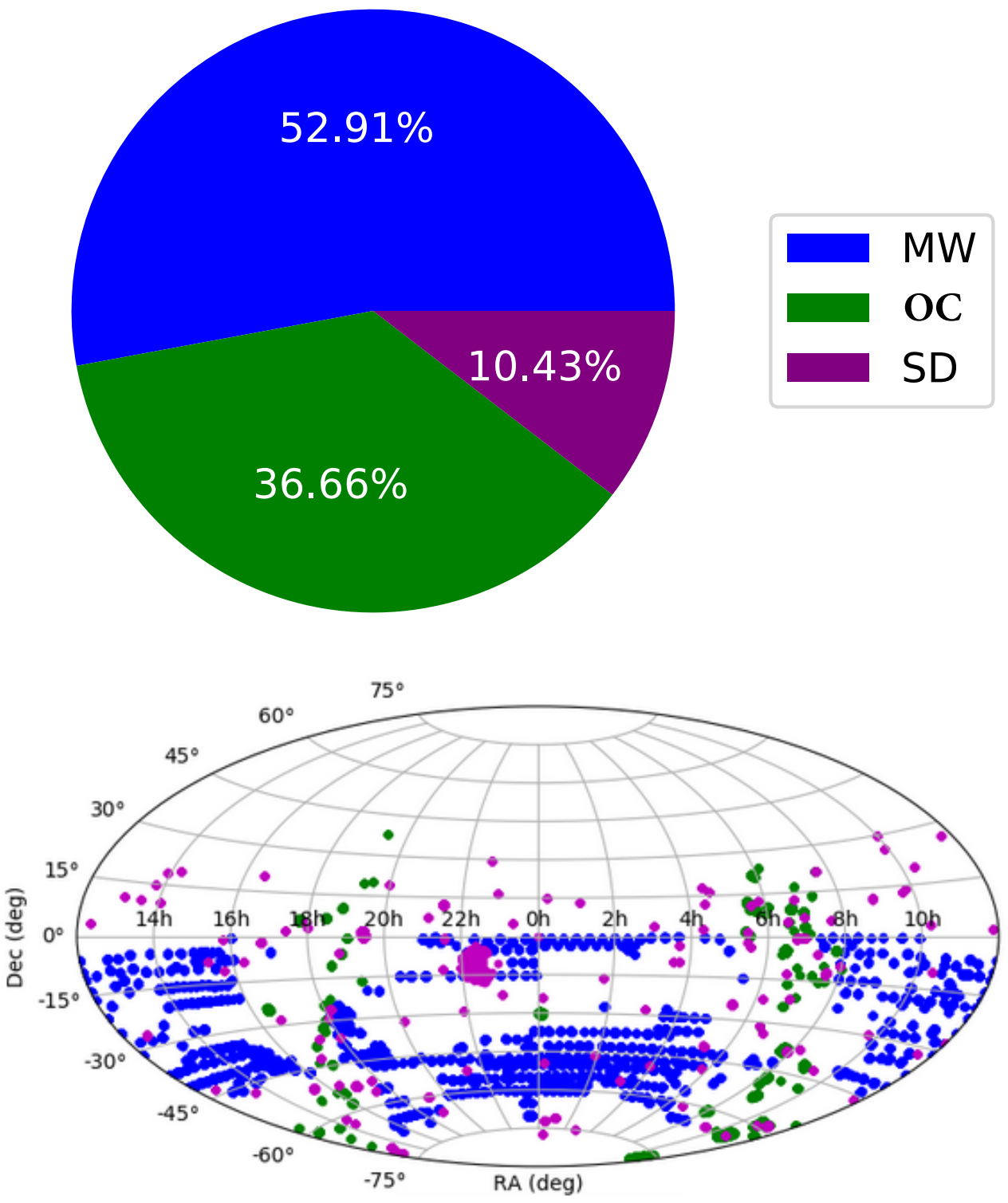

**Fig. 2.** Overview of the observed fields and targets. The top panel shows the fraction of time dedicated to the different types of field: Milky Way, science open clusters, standards and calibrations (SD). The bottom panel shows the sky distribution for all the observations broken down by field type, using the same colour scheme. This figure is also shown in GRH22.

The final fraction of the time dedicated to the different fields and samples and their distribution on the sky is shown in Fig. 2 (see also GHRH22 for the distribution in Galactic coordinates). Figure 3 shows the fraction of spectra obtained with each of the Giraffe and UVES gratings. Figure 2 shows that about 53 % and 37 % of the time was devoted respectively to MW and science cluster observations (corresponding to ~ 145 and 100 of the 275 effective nights), while 10 % (~ 30 nights) was used for calibrations; figure 3 instead indicates that most of the targets were observed with HR10, HR21, and HR15N.

In total, GES observed slightly less than 2000 OBs, many of which (in particular the cluster ones) were repeated several times; this resulted in 185940 co-added spectra[1] for a final sample of 110463 unique stars. Table 1 summarises the number of co-added spectra for the different settings and fields, along with information on the resolution, wavelength coverage, and median signal-to-noise ratio (SNR). Most of the MW stars were observed with both HR10 and HR21; in addition, a fraction of the targets (20–25 %), the calibrators and OC fields in particular, were observed with more than one instrument or set-up (e.g. UVES and Giraffe, or HR10/21 and HR15N).

The magnitude distribution of the targets is shown in Fig. 4. The distribution of Milky Way stars peaks at fainter magnitudes and is narrower than the distribution of the cluster sample. This difference is the result of the target selection strategies used for the two samples, which are discussed in detail in GHR22 and Bragaglia et al. (2022) for the MW and the OCs, respectively. The broader magnitude distribution that characterises the cluster sample is due to its heterogeneity (by design), and in particular to the large range of cluster distances and to the variety of spectral types and evolutionary phases that were covered: from O to M type; from PMS phases to evolved giants.

### 2.3. Data quality

After pipeline data reduction, detailed quality control (QC) of the reduced spectra was performed prior to releasing them for the spectrum analysis. A small number of UVES spectra still had defects after this step (e.g. order merging issues) and were removed from the sample, along with spectra (both Giraffe and UVES) with SNR < 2. These add up to slightly more than 4000

[1] We consider the set of two spectra for UVES lower and upper CCDs as one individual spectrum.

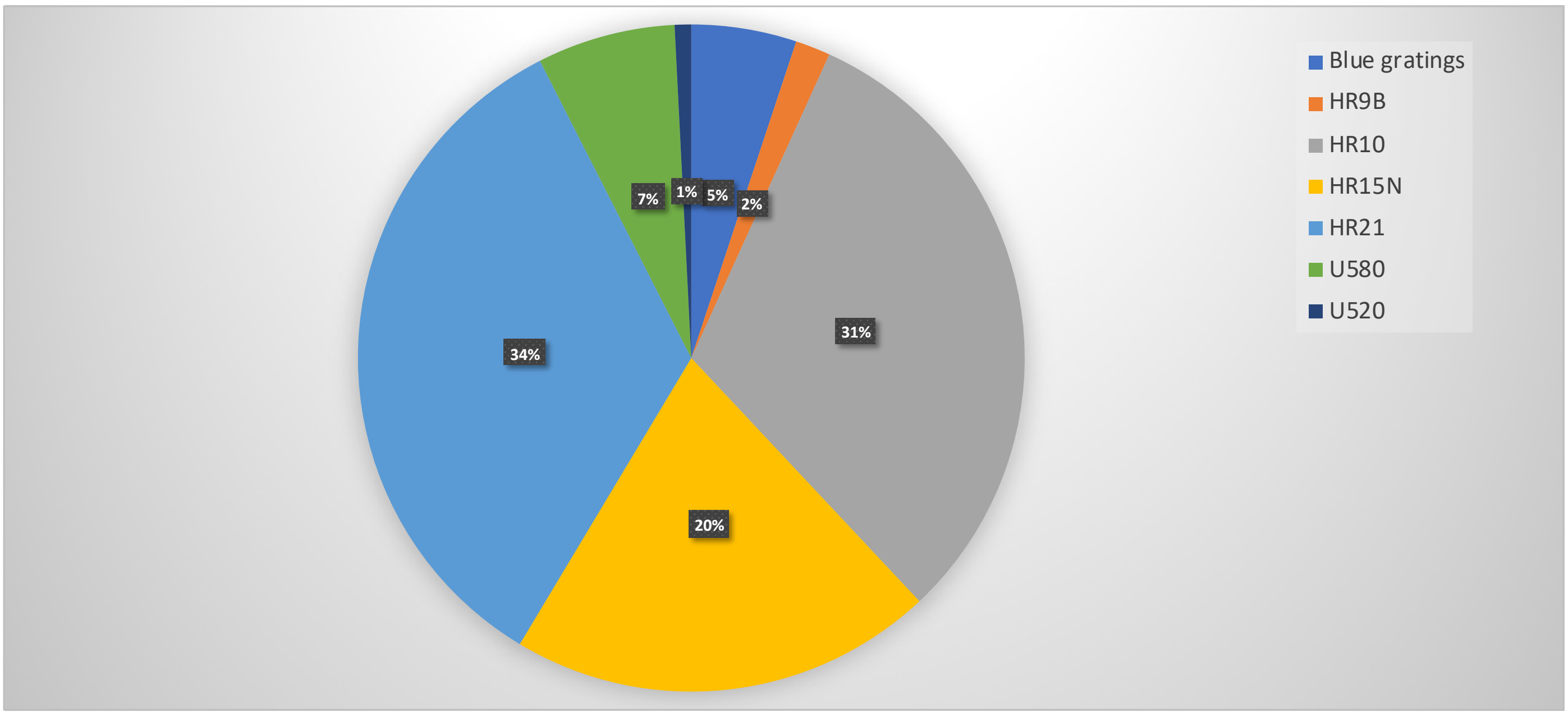

**Fig. 3.** Fraction of stars observed with each of the Giraffe and UVES gratings. The blue gratings refer to HR3, HR4, HR5A, HR6, and HR14A.

| Target type | Instrument | Grating | Spectral Range (nm) | Resolving Power (R) | N Spectra | Median SNR | SNR 1*st* quartile |
|---|---|---|---|---|---|---|---|
| **MW** | Giraffe | HR10 | 534-562 | 19800-21500 | 53798 | 12 | 6 |
| | | HR21 | 848-900 | 16200-18000 | 53446 | 29 | 16 |
| | UVES | 580 | 477.1-678.5 | 47000 | 3332 | 45 | 31 |
| **MW Bulge** | Giraffe | HR10 | – | – | 114 | 47 | 42 |
| | Giraffe | HR21 | – | – | 5707 | 84 | 69 |
| | UVES | 580 | – | – | 318 | 92 | 70 |
| **OCs** | Giraffe | HR15N | 647-679 | 17000-19200 | 35840 | 38 | 20 |
| | | HR3 | 403-420 | 24800-31400 | 2160 | 42 | 19 |
| | | HR4 | 419-439 | 20350-24000 | 1188 | 63 | 42 |
| | | HR5A | 434-459 | 18470-20250 | 2055 | 51 | 30 |
| | | HR6 | 454-476 | 20350-24300 | 2054 | 52 | 27 |
| | | HR9B | 514-536 | 25900-31750 | 2630 | 32 | 18 |
| | | HR14A | 631-670 | 17740-18000 | 2036 | 82 | 48 |
| | UVES | 520 | 414.0-621.0 | 47000 | 323 | 145 | 83 |
| | | 580 | 477.1-678.5 | 47000 | 1626 | 76 | 48 |
| **Calibration targets** | Giraffe | HR3 | – | – | 104 | 113 | 135 |
| | | HR5A | – | – | 89 | 194 | 102 |
| | | HR6 | – | – | 85 | 185 | 117 |
| | | HR9B | – | – | 617 | 209 | 112 |
| | | HR10 | – | – | 6438 | 47 | 30 |
| | | HR14A | – | – | 103 | 285 | 135 |
| | | HR15N | – | – | 3962 | 74 | 50 |
| | | HR21 | – | – | 5862 | 92 | 63 |
| | UVES | 520 | – | – | 486 | 90 | 50 |
| | | 580 | – | – | 1157 | 93 | 55 |

**Table 1.** Number of co-added spectra, divided by field, instrument, and set-up.

spectra; the total number 185490 mentioned above is after the low-quality spectra were discarded.

The distribution of SNR for the final co-added spectra obtained with Giraffe and UVES is shown in Figs. 5 and 6, respectively. The median and first quartile values are listed in Table 1. The figures and table show that the SNR is good for most of the observed samples and, with the exception of the MW HR10 observations, the median values are always above $\sim$ 30 for Giraffe; the median SNR and even the first quartile values for the UVES spectra are above about 40 for virtually all the different target categories. Most importantly, in most cases the achieved SNR values meet the initial goal indicated in the ESO proposal, confirming the success of the observing strategy.

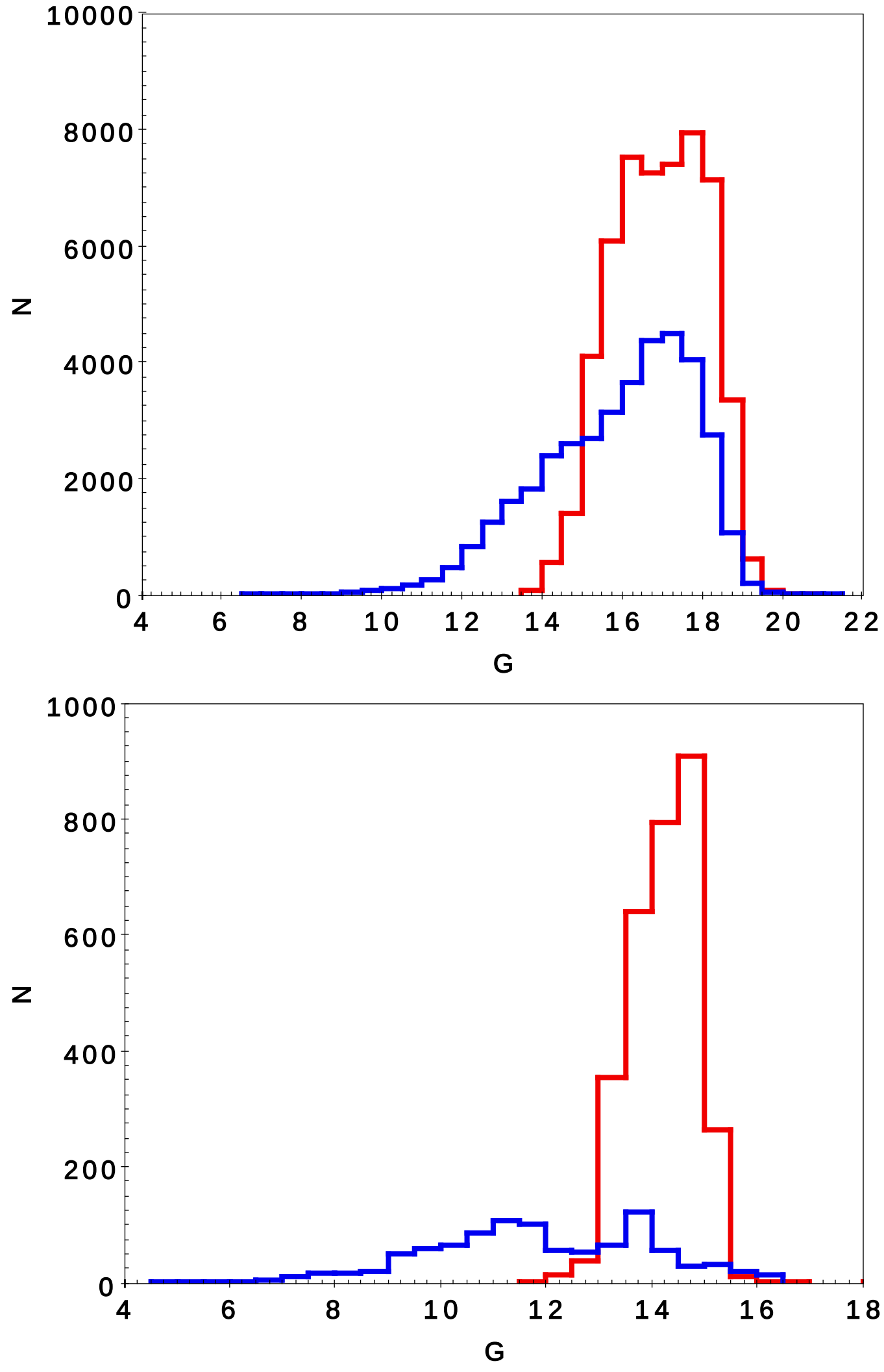

**Fig. 4.** Distribution of G magnitudes for the stars observed with Giraffe (top panel) and UVES (bottom panel). OCs and MW samples are shown in blue and red, respectively.

### 2.4. Archive spectra

Gaia-ESO has invested considerable efforts in the analysis of relevant data retrieved from the ESO FLAMES archive. We mention in passing that spectra of benchmark stars obtained with different (also non-ESO) instruments and telescopes (e.g. UVES in single object mode; HARPS; NARVAL at Pic du Midi; Espadons at CFHT) were also retrieved and processed. The total number of archive spectra is 9051, corresponding to 5954 individual stars.

The archive spectra were processed and analysed with the GES pipelines to ensure maximum consistency across all datasets. Most of the archive data are cluster observations, aimed both to complement the science OC sample (in terms of target stars and clusters) and to benefit calibrations; the samples extracted from the archive also include some Bulge observations, as well as the Giraffe solar atlas (i.e. several solar spectra obtained with the different settings). We finally note that spectra obtained with HR14B and HR5B gratings were retrieved from the archive; whilst these are not GES set-ups, the spectra were nevertheless useful to complete the NGC 3293 archive dataset. A summary of the FLAMES archive spectra is provided in Table 2.4.

## 3. Analysis cycles and internal releases

As explained in GRH22, the GES data flow is organised in working groups (WGs) and analysis cycles, followed by internal releases (iDR) of the data products to the consortium and subsequent Phase 3 releases to ESO. Each cycle includes all the steps indicated in GRH22 (their Figs. 5 and 6), from pipeline data reduction and radial velocity determination, down to final product homogenisation. In the following we summarise the analysis cycles, releases, and content since the survey started. In each cycle the UVES data reduction and RV determination were performed by the INAF–Arcetri team, following Sacco et al. (2014), and updated for the last cycle, as described in Appendix A. Giraffe data processing was carried out by the Cambridge/CASU team and is presented in detail in GRH22. As detailed in that paper, the spectrum analysis was performed by five distinct working groups, each of which included a number of nodes that employed different analysis techniques. The lists of nodes and methods are shown in Tables 10 and 11 of GHR22. Reduced spectra and quality information, a version controlled line list (including atomic and molecular data), and a grid of synthetic spectra were made available to the nodes prior to the analysis. Information on the line list can be found in Heiter et al. (2021); synthetic spectra were computed (or interpolated) from MARCS models, using the Turbospectrum V14.1 code for spectral synthesis and using the relevant version of the line list (see e.g. de Laverny et al. 2012).

### 3.1. Gaia-ESO analysis and releases

Six analysis cycles and internal releases were carried out. We provide below summary information on the first five releases (iDR1 to iDR5), while the last cycle (iDR6) is described in Sect. 3.2.

**iDR1**: Analysis of the first six months of observations (up to June 2012) for a total of about 11000 stars; the analysis was performed using version 3.0 of the line list and version 0.0 of the grid. The analysis was completed in March 2013; only WG recommended parameters were derived, while no WG15 homogenisation was performed and almost no RVs were delivered. Products were released to the consortium in August 2013.

**iDR2**: Analysis of the first 18 months of observations (up to June 2013); versions 4.0 and 3.0 of the line list and grid, respectively, were used. iDR2 was completed in April 2014. Homogenisation of stellar parameters and RVs was achieved for the first time, resulting in a final set of recommended results per star rather than per spectrum. In addition, for the first time a dictionary was introduced by Working Group 14, namely a classification scheme for the different types of outliers and peculiarities (see GRH22 for more details). Products for almost 18,000 stars were released to the consortium in July 2014.

**iDR3**: Incremental analysis of selected spectra obtained between July 2013 and December 2013. No updates were made to methods, tools, and pipelines. iDR3 was completed in September 2014. Homogenisation of stellar parameters and RVs was achieved. Products were released to the consortium in January 2015.

**iDR4**: Analysis of the first 31 months of observations (up to July 2014). Updated versions of the line list and grid of synthetic spectra were made available (versions 5.0 and 4.0, respectively). Homogenisation of abundances (in addition to parameters) was achieved for the first time. Improvements with respect to the previous cycles include the determination of recommended parameters, homogenisation, and error estimates. iDR4 delivered advanced products for approximately 55,000 stars; those products were made available to the consortium through the Survey archive (WFAU database; see GRH22) in February 2016.

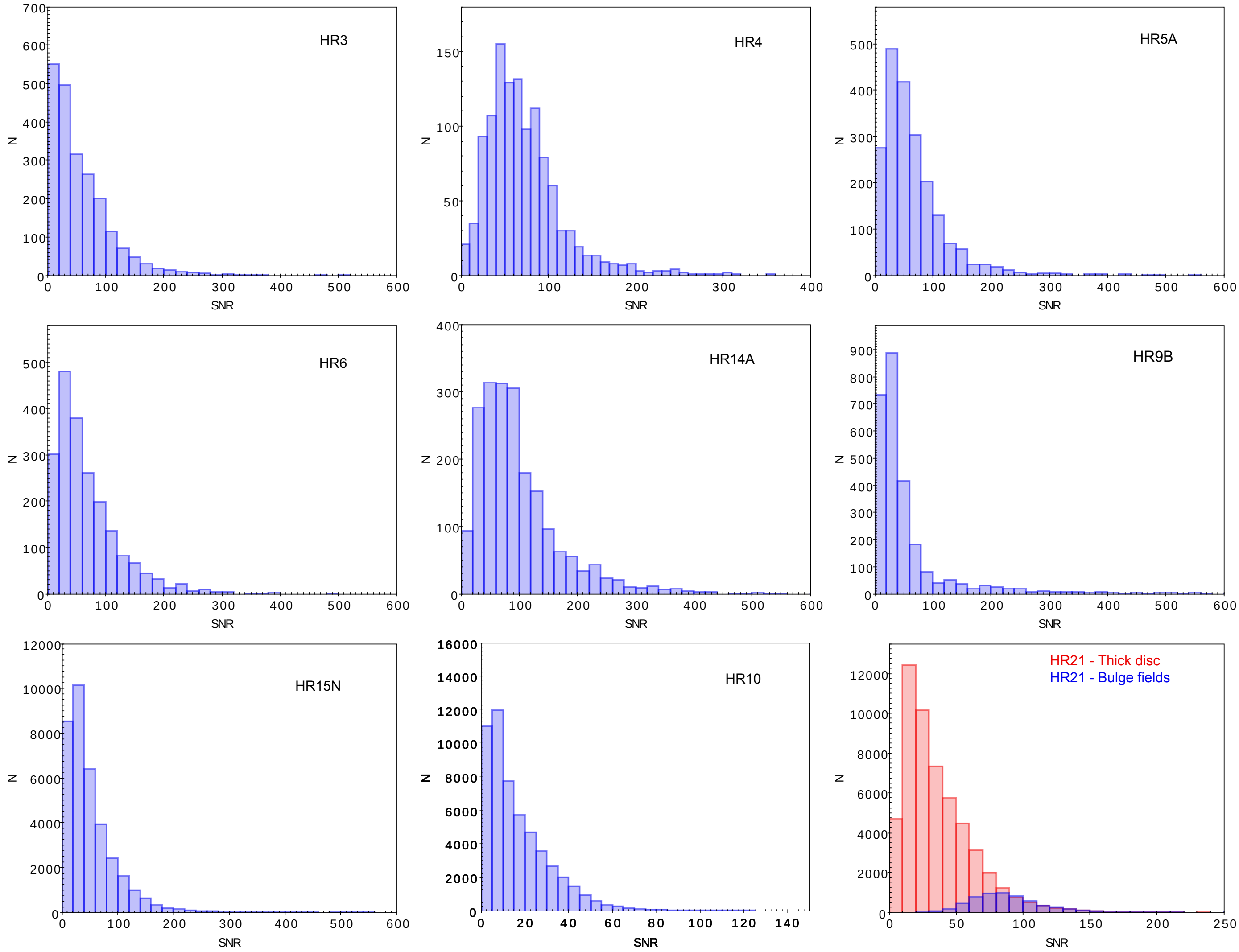

**Fig. 5.** Distributions of SNR of the co-added spectra observed with Giraffe. From top left to bottom right: HR3, HR4, HR5A, HR6, HR14A, HR9B, HR15N, HR10, HR21. For HR21 the thick disc–halo and Bulge–inner fields are shown in red and blue, respectively. See Table 1 for the information on the targets observed with the different settings.

**iDR5**: Analysis of the first 48 months of observations (up to December 2015, plus selected additional key samples observed after this time, including K2 fields and fields in the newly added HR4 set-up; see Blomme et al. 2022). For the determination of the stellar parameters line list version 5.0 was used. For the abundance determinations the final version (6.0) of the line list was employed, which introduced minor changes to atomic lines other than Fe. This version of the line list is described in detail in Heiter et al. (2021). No changes were introduced to the grid of synthetic spectra with respect to iDR4. The final file, which included products for slightly more than 80,000 objects, was made available to the consortium in November 2017.

### *3.2. The last cycle: iDR6*

The last scheduled observing run of the Survey was completed at the end of January 2018 and iDR6 started a few months later, in coincidence with *Gaia* Data Release 2. Immediately after the release, a set of *Gaia* priors (i.e. Bayesian inferences on the stellar parameters of the iDR6 targets determined based on the *Gaia* parallaxes and photometric colours) was produced. The parameter priors were made available to the analysis nodes that had the capability to accept priors as input in their codes, with the aim of improving the quality of the results by providing a more accurate starting point for the parametrisation. Line list version 6.0 was employed in iDR6, as well as the same grid of synthetic spectra used for iDR4 and iDR5.

Updated homogeneous lithium curves of growth (COGs) were developed for iDR6 covering the entire parameter space of Gaia-ESO lithium observations (see Franciosini et al. 2022). Specifically, the Li COGs were computed for the following parameter ranges: $3000 \leq T_{\rm eff} \leq 8000$ K; $0.5 \leq \log g \leq 5$; $-2.5 \leq$ [Fe/H] $\leq +0.5$; Li abundances from A(Li)= $-1.0$ to A(Li)= $+4.0$ in steps of 0.2 dex, except for [Fe/H] $< -1.00$, where abundances were limited to A(Li) $\leq +3.0$. A set of corrections for the Fe I 6707.4 line, which is blended with the Li line in Giraffe spec-

**Table 2.** FLAMES spectra retrieved from the ESO archive, divided by type, instrument, and set-up.

| Type | Instrument/Grating | N Spectra | Median SNR | SNR 1*st* quartile |
|---|---|---|---|---|
| **Science open clusters** | Giraffe/HR14B | 106 | 49 | 37 |
| | Giraffe/HR15N | 684 | 80 | 49 |
| | Giraffe/HR3 | 106 | 51 | 37 |
| | Giraffe/HR4 | 106 | 74 | 57 |
| | Giraffe/HR5B | 106 | 63 | 45 |
| | Giraffe/HR6 | 106 | 98 | 74 |
| | Giraffe/HR9B | 997 | 18 | 12 |
| | UVES/520 | 11 | 228 | 159 |
| | UVES/580 | 236 | 73 | 38 |
| **Bulge** | Giraffe/HR21 | 228 | 126 | 110 |
| **Calibration open clusters** | Giraffe/HR14A | 199 | 46 | 35 |
| | Giraffe/HR15N | 292 | 88 | 58 |
| | Giraffe/HR9B | 186 | 70 | 45 |
| | UVES/580 | 124 | 49 | 26 |
| **Calibration globular clusters** | Giraffe/HR10 | 100 | 48 | 43 |
| | Giraffe/HR15N | 852 | 115 | 73 |
| | Giraffe/HR21 | 1738 | 49 | 36 |
| | UVES/520 | 113 | 17 | 9 |
| | UVES/580 | 184 | 74 | 47 |

tra or in UVES spectra of rapid rotators (vsin $i$ larger than about 15 km/s), was also derived.

iDR6 took considerably longer than the previous analysis cycles because the dataset was larger and a more detailed quality control was carried out. In particular, the final products of this last release were checked by a group of independent reviewers consisting of a few Co-Investigators of the survey who were not involved in the spectrum analysis. This additional quality control, which was initially carried out on a preliminary version of the parameters, led to the discovery of unpredicted systematic errors in the metallicities measured from the spectra of both Giraffe and UVES. The systematic errors that were affecting both the low- and high-metallicity ends of the distribution were corrected by a re-processing of stellar parameters using a new procedure for the homogenisation. This procedure is described in detail in Worley et al. (in preparation).

The final catalogue delivers products for 114,917 targets (including those retrieved from the ESO archive) and is structured in 422 (first extension) plus 13 (second extension) columns that provide information on the targets (including the *Gaia* eDR3 source ID and distance from the *Gaia* EDR3 source); the recommended set-ups and working group, which is relevant for the stars observed with more than one setting and/or analysed by more than one WG; radial and rotational velocities with errors; stellar parameters, $\gamma$ spectroscopic gravity index (see Damiani et al. 2014), and metallicity with uncertainties; abundances for up to 31 elements (He, Li, C, N, O, Na, Mg, Al, Si, S, Ca, Sc, Ti, V, Cr, Mn, Co, Ni, Cu, Zn, Sr, Y, Zr, Mo, Ba, La, Ce, Pr, Nd, Sm, Eu) with their uncertainties; equivalent width of the lithium absorption 6707.8 Å line; $H_\alpha$ and $H_\beta$ emission equivalent widths and chromospheric flux, $H_\alpha$ 10% width and mass accretion rate. Finally, two columns with two sets of flags are included in the catalogue, both technical flags (e.g. SNR, reduction and analysis issues) and more scientific flags for phenomenological classification (e.g. binarity, variability, emission lines, asymmetric line profiles, peculiar and/or enhanced abundances, lambda Bootis-type stars; blue stragglers are instead not classified with flags). These flags are detailed in GHR22 and Sect. 4.5.

The spectral type distribution as a function of surface gravity and metallicity of stars included in the final catalogue is shown in Fig. 7. In Fig. 8 we instead plot, for the different instruments and gratings, the fraction of stars for which radial and rotational velocities, stellar parameters, and other properties were derived and are included in the final catalogue. Figure 9 is similar, but the percentage of stars with individual element abundance measurements for UVES580, HR10+HR21, and HR15N are shown.

Figure 8 indicates that RVs were derived for the vast majority of the samples, with the exception of the targets observed with HR9B, the blue gratings (HR3, HR4, HR5A, HR6, and HR14A, see Blomme et al. 2022), and U520, for which the percentage falls below (or far below) 80%. This is mainly due to the warmer temperatures (and fewer lines) of the targets observed with these gratings which make RV determination more challenging. Stellar parameters and [Fe/H] were derived for at least 70 % of the stars observed with each grating and/or setting, with the percentage increasing to more than 90% for U580. Considering all gratings and both UVES and Giraffe, the catalogue includes effective temperatures for about 85 % of the targets, and gravity and metallicity for about 80 % of them. Abundances have been measured for about 90 % of the stars observed with U580 for most elements, with the exception of a few difficult ones for which the percentage of targets with measured abundances remains well below 50 %. As expected, the lower resolution, shorter spectral range, and typically lower SNR values of the Giraffe spectra allowed the measurement of abundances of fewer elements and in fewer stars. We highlight, however, that GES was very successful in measuring lithium from Giraffe HR15N spectra, which was indeed one of the initial goals of the survey. We finally note that, although for a much smaller fraction of stars, abundances were also measured from U520 (noticeably helium) and the blue gratings.

### 3.3. Phase 3 data delivery to ESO

At the time of writing five phase 3 releases to the ESO archive were completed; a sixth and final one is in progress and is planned in Spring 2022. Detailed information, including the release content and a release description document, can be found on the ESO webpage at the link:
*http : //eso.org/rm/publicAccess#/dataReleases*.
We summarise here the main features.
**ESO-DR2** was published in July 2015. This release covers observations obtained in the period 31.12.2011–31.12.2013; it includes 27359 spectra corresponding to 14947 unique targets. For a fraction of the stars for which spectra were delivered, advanced products were also released. When a star was observed with more than one setting and/or with multiple exposures, more than one spectrum is delivered per star (i.e. HR10 and HR21, or HR15N and UVES580). As for the internal releases, in such cases only one recommended set of parameters (one row of data) is written to the catalogue.
**ESO-DR3 and DR3.1** were published in December 2016 and May 2017. DR3 covers observations obtained by the Survey in the period 31.12.2011–19.07.2014, plus some archival data; 44210 spectra for 25533 unique targets were submitted, including 2342 ESO archive spectra. As for DR2, for a fraction of the stars for which spectra were delivered, advanced products were also released, including RVs (for 96% of stars), stellar parameters ($T_{\rm eff}$ for 76% of the stars; logg for 47% of the stars), metallicity ([Fe/H] for 57% of the stars); lithium equivalent width, $H_\alpha$ emission information, gravity index, and individual abundances for a number of elements (with abundances delivered for 1% of the stars for N, and 45% for Li). DR3.1 includes only spectra.
**ESO-DR4** was published in December 2020. This release only includes spectra, covering the complete set of Gaia-ESO observations made in the period 31.12.2011–26.01.2018. The total number of submitted data files is 190200, comprising spectra of 114500 unique targets. Included in this sample are 7143 ESO archive spectra. The SNR and other selection criteria used for previous releases were relaxed. Released DR4 spectra included are subject only to a SNR selection threshold of 2.

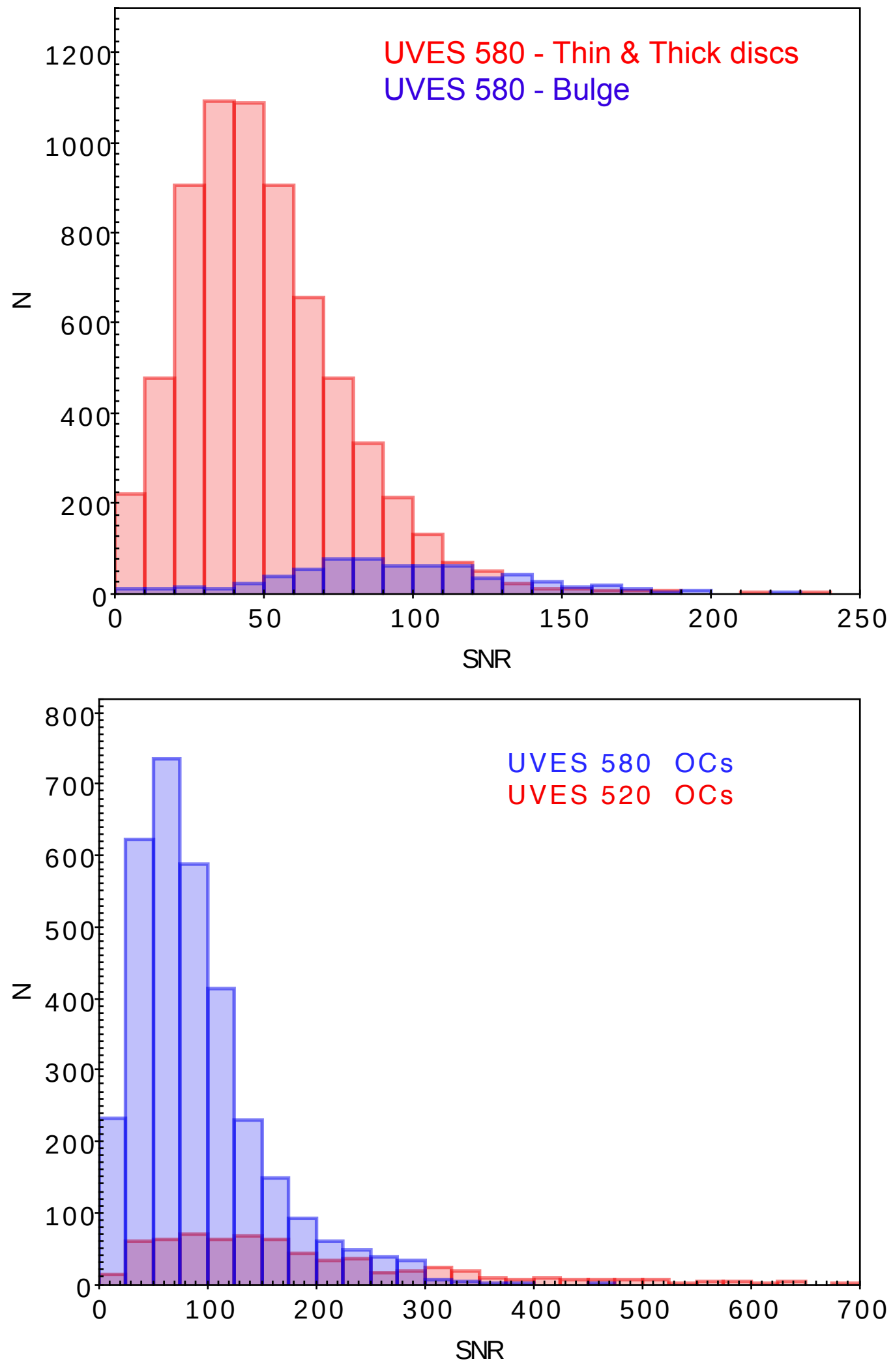

**Fig. 6.** Distributions of SNR of the co-added spectra for the MW samples observed with UVES 580 (top panel: red for thin and thick discs; blue for Bulge) and science cluster targets observed with UVES 520 and 580 (bottom panel: blue for U580; red for U520).

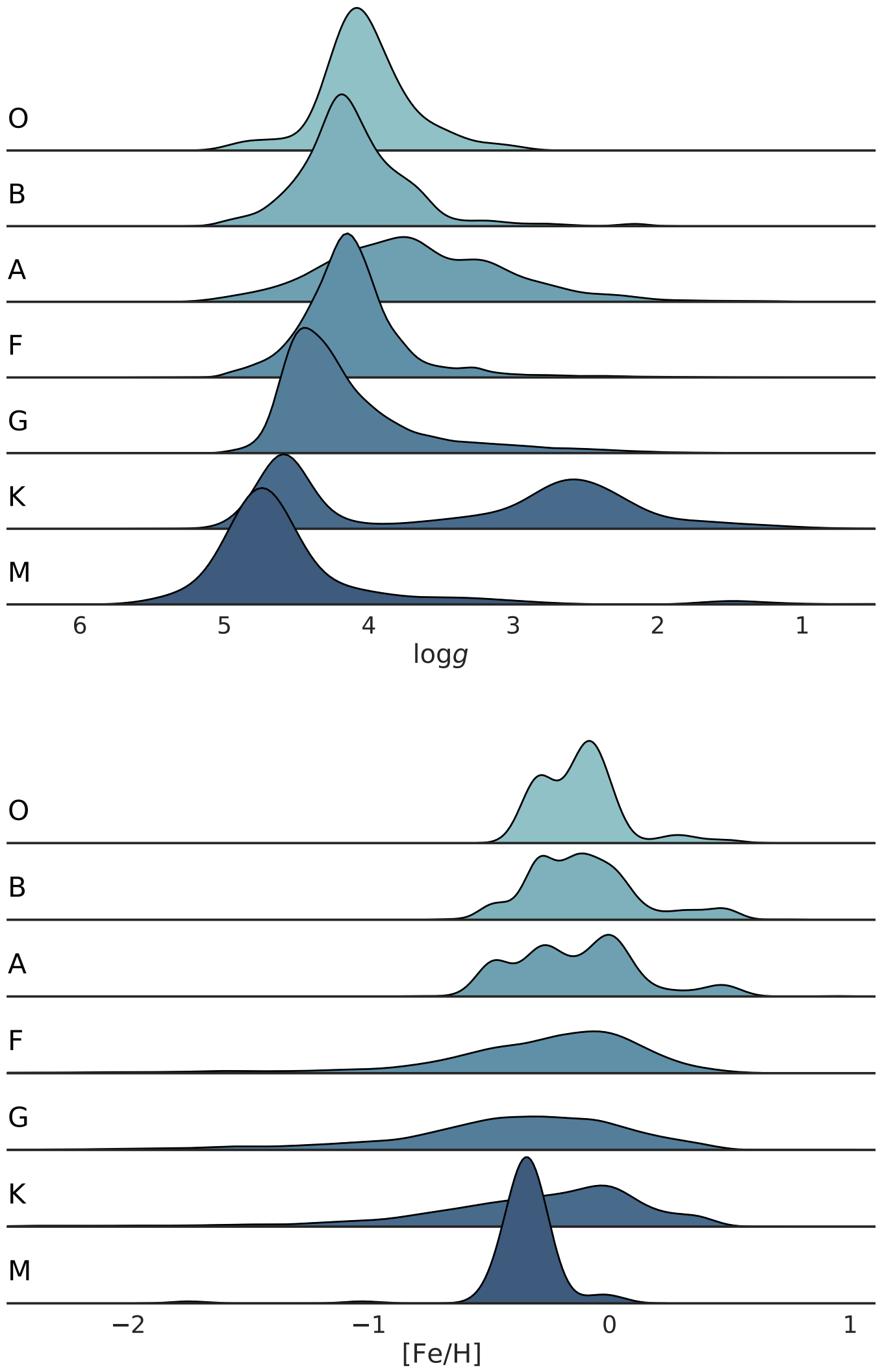

**Fig. 7.** Spectral type distribution of the stars analysed in iDR6 as a function of logg (upper panel) and of [Fe/H] (bottom panel).

With **DR4.1**, the iDR6 catalogue of RVs was published in October 2021, including more than 110,000 stars, covering 97% of all the objects whose spectra were published in DR4.

## 4. Data products: Success of the multi-pipeline approach

A complete discussion of the quality and the validation of the data products, uncertainties, accuracy and precision will be included in a forthcoming paper focusing on the homogenisation;

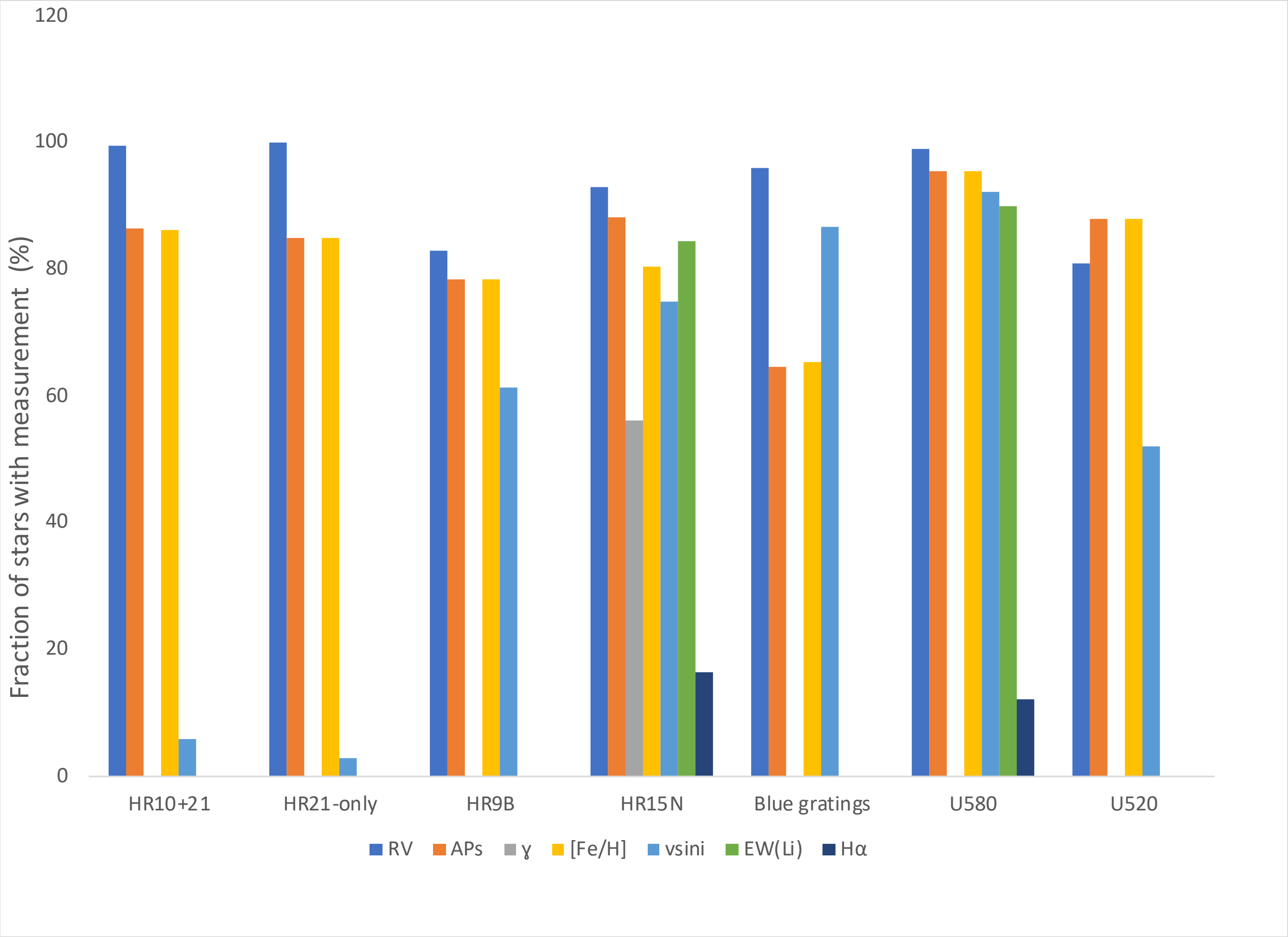

**Fig. 8.** Percentages of stars for which the different main products were derived, divided by set-up. The 'blue gratings' are HR3, HR4, HR5A, HR6, and HR14A together, the set-ups used for hot stars. The archive data are considered in the determination of the percentages.

in Appendixes A and B we provide an update of Sacco et al. (2014) and Jackson et al. (2015) on the UVES data reduction, RV determination and precision, and on the Giraffe RV precision.
In the following sections we instead show via a few examples how the GES analysis approach (multiple pipelines, nodes, methods for the same star or spectrum, see GRH22, followed by two homogenisation steps) has in general allowed us to get better results than would have been obtained employing one single pipeline.

### 4.1. Stellar parameters

In this section we focus on effective temperatures and surface gravities. Figures 10 and 11 display the Kiel diagrams (logg vs $T_{\rm eff}$) of a sample of stars observed in the MW fields, with metallicity in the range -0.2<[Fe/H]<+0.2. More specifically, we compare the final recommended parameters and the results of the individual nodes with a sample of suitable isochrones. The results obtained from the UVES (WG11, Fig. 10) and Giraffe (WG10, Fig. 11) spectra are shown. The figures clearly indicate that the GES approach, based on a dataset of calibrators and reference stars to combine the node results, has generated a final set of results better than each of the individual sets. Each pipeline was more heavily weighted in the region of the parameter space where it produces the best results. The extent of the improvement is even more evident in Figure 11 showing the results of the two WG10 nodes that analysed the dedicated set-ups, HR10 and HR21, for MW field stars. In Figure 12, we present the Kiel diagrams of six example open clusters available in the latest GES internal data release. Both the final recommended stellar parameters and the node parameters for high-probability member stars are shown (P>0.9 from Jackson et al. 2022). The dispersion of the stellar parameters around the isochrone decreases considerably with the final recommended parameters, leading, in many cases, to an excellent agreement along the whole evolutionary sequence.

### 4.2. Metallicity

In Figure 13 we compare the recommended and node metallicities with the literature for a sample of calibrating OCs selected for the calibration strategy, as described in Pancino et al. (2017a). More specifically, we compare our results with the values reported in Pancino et al. (2017a). For the five clusters we adopt the membership analysis of Jackson et al. (2022) and select stars with membership probability >0.9. The agreement between the recommended GES values and the literature reference values is extremely good. The most discrepant results of some nodes do not affect the final values, demonstrating again the effectiveness of the multi-pipeline approach.

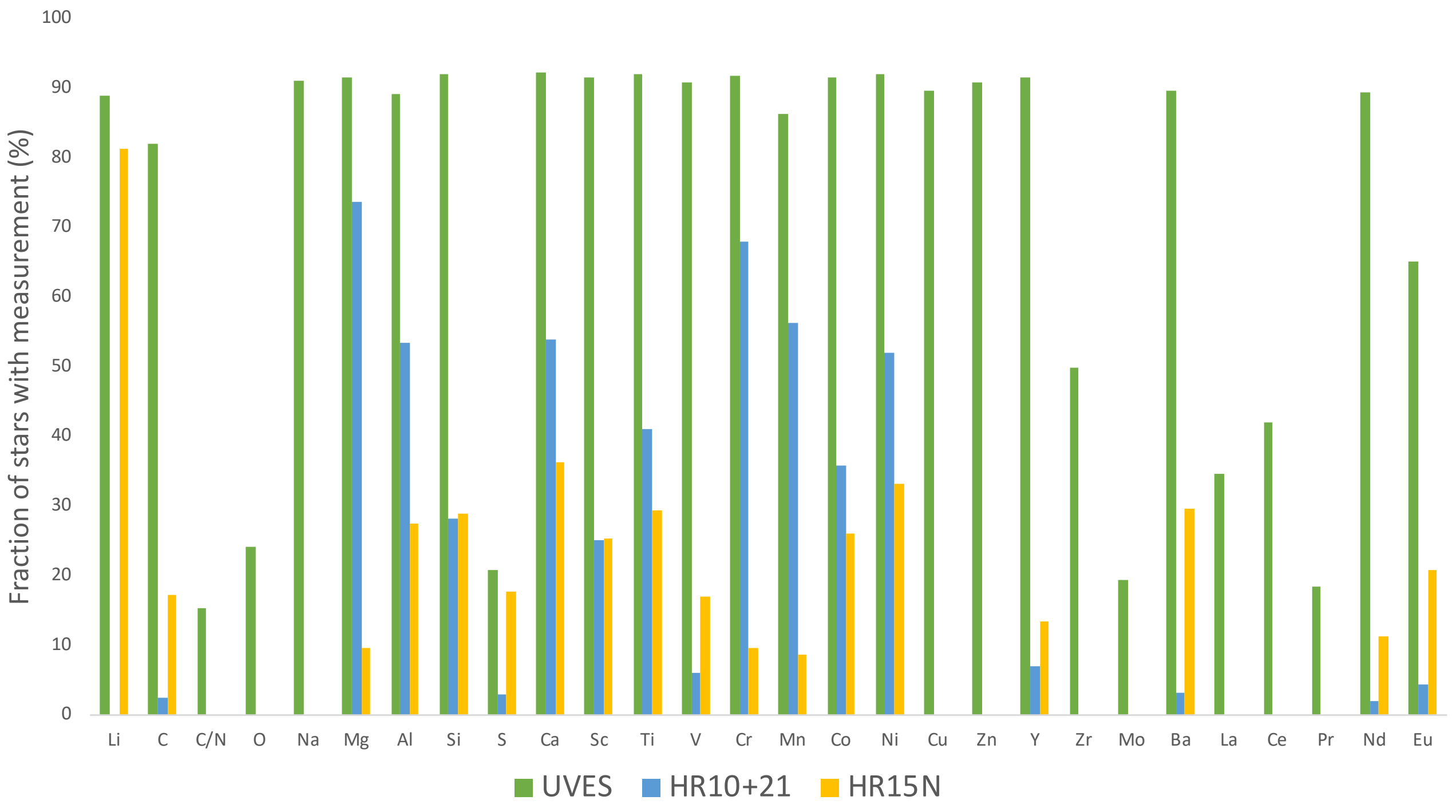

**Fig. 9.** Percentage of stars for which the different elements were determined from UVES spectra, combined Giraffe HR10/HR21, and HR15N. Elements measured in a very tiny fraction of stars (He, N, Sr, Sm) are not included.

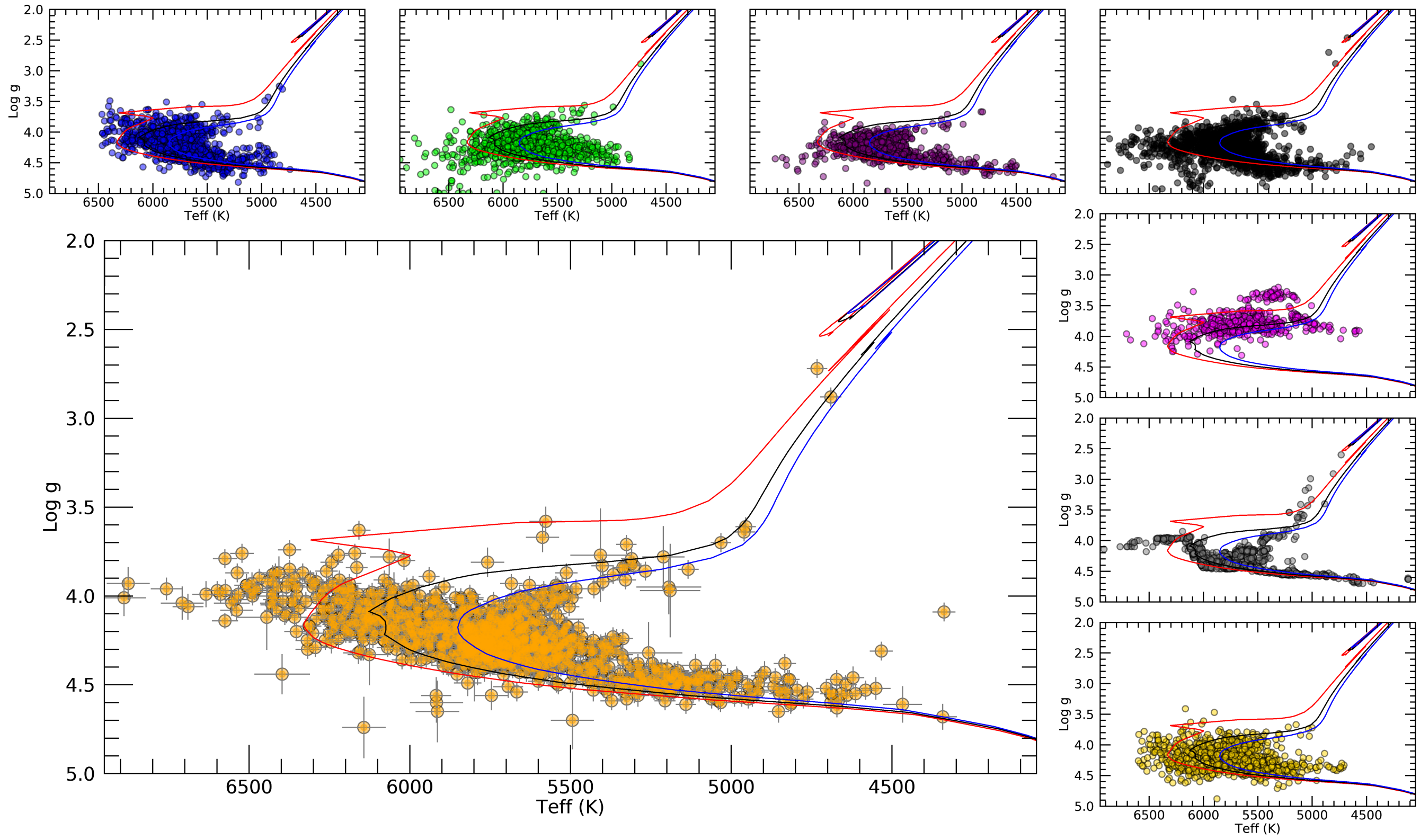

**Fig. 10.** logg vs $T_{eff}$ diagrams of Milky Way field stars in the metallicity range -0.2<[Fe/H]<+0.2, available in iDR6. In the large panel we show the recommended stellar parameters from WG11 (in orange), while in the small panels we plot the results of the seven individual nodes contributing to the final parameters. Parsec isochrones (Bressan et al. 2012) at solar metallicity, and at three different ages are shown: 2 Gyr (in red), 5 Gyr (in black), 7 Gyr (in blue).

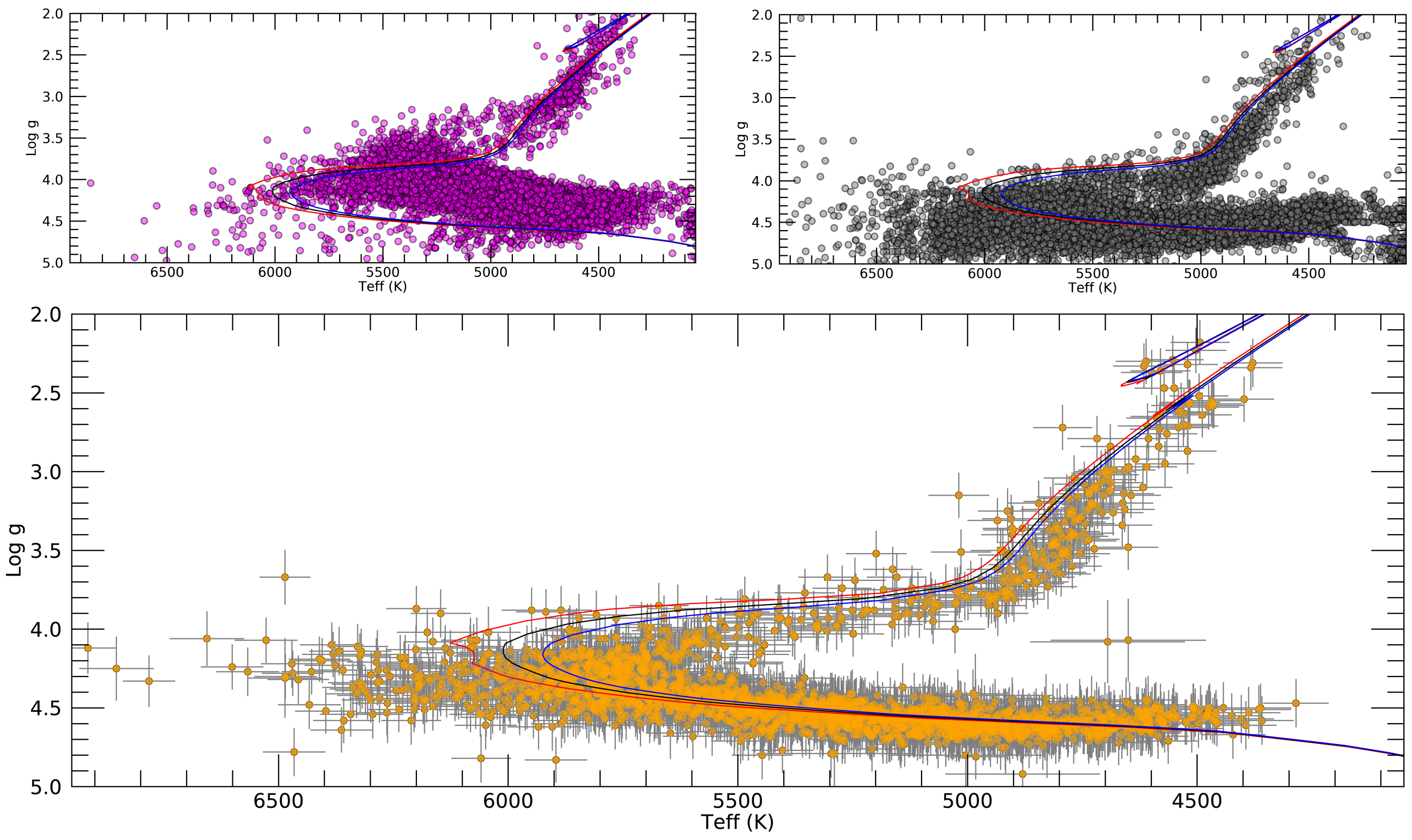

**Fig. 11.** logg vs $T_{eff}$ diagrams of Milky Way field stars in the metallicity range -0.2<[Fe/H]<+0.2, available in iDR6. In the large panel we show the recommended stellar parameters from WG10 (in orange), while in the two upper panels we show the results of two individual nodes contributing to the final parameters for the MW field stars (different colours and symbols). Parsec isochrones (Bressan et al. 2012) at solar metallicity, and at three different ages are shown: 2 Gyr (in red), 5 Gyr (in black), 7 Gyr (in blue).

### *4.3. Elemental abundances*

In Figs. 14, 15, and 16, we show the abundances of three elements, Si, Ni, and Ba, belonging, respectively, to the $\alpha$, iron-peak, and neutron-capture groups. We present their abundance ratios [El/Fe] as a function of [Fe/H] for high-probability members (P>0.9, as above) of eight intermediate-age and old open clusters. We display, as in the previous figures, both the recommended abundances, and the abundances from the nodes that contributed to them. Note that the node abundances shown in the figures give an indication of what would be obtained by using a single pipeline. They are calculated directly by each node, combining the abundances obtained from various spectral lines. The homogenisation process instead considers the node abundances line by line and combines them to produce the final recommended value. Therefore, lines giving discrepant results are discarded in the determination of the final abundances.

For each cluster, we indicate in the figure the mean value, and 1, 2, and 3 times the standard deviation. With these plots we can estimate the improvement on the precision achieved using the recommended abundances: for most clusters the recommended abundance ratios of cluster members are within 1-2$\sigma$ of the average, even for difficult elements, such as barium. The improvement over the use of a single pipeline is certainly evident (see e.g. the [Ba/Fe] in Be 32). More difficult to estimate is the accuracy, since many of our clusters have no reference values for their abundances.

In Figs. 17, 18, and 19, [El/Fe] as a function of [Fe/H] is shown for the MW targets and for the same elements as in Figures 14, 15, and 16. We plot both the recommended abundances, and the abundances derived by the node pipelines. We compare them with the sample of abundances derived for about 700 stars in the solar neighbourhood in Bensby et al. (2014). We selected abundances from spectra with SNR≥100, applying a cut on the uncertainties of the stellar parameters ($\delta(T_{eff})$<100 K, $\delta$(logg)<0.1 dex, $\delta$([Fe/H])<0.1 dex), resulting in a sample of about 1400 stars. The figures show that the multi-pipeline approach allowed us to identify nodes that do not produce quality results for some elements and, by combining the results of the remaining nodes, to obtain products of equivalent quality to or higher than those of the individual nodes. Some nodes deliver, in general, good-quality abundances, while others have poor results (see e.g. [Ni/Fe] in the third panel of Fig. 18). However, thanks to the homogenisation process those poor results did not affect the final quality of the recommended results. The higher quality of the recommended abundances is also noticeable for elements with few or weak absorption lines, such as Ba, for which the quality of recommended abundances is in better agreement with literature abundances than those of individual nodes (see Figure 19). In Fig. 17, where [Si/Fe] versus [Fe/H] is shown, we see the separation between the thin and thick disc, while in Figure 18 the expected almost-flat behaviour of [Ni/Fe] versus [Fe/H] is present, with a slight increase in [Ni/Fe] at high metallicity, as already appreciable in the sample of Bensby et al. (2014). At low metallicity there is an increase in the scatter of [Ni/Fe] due to the presence of different Galactic populations.

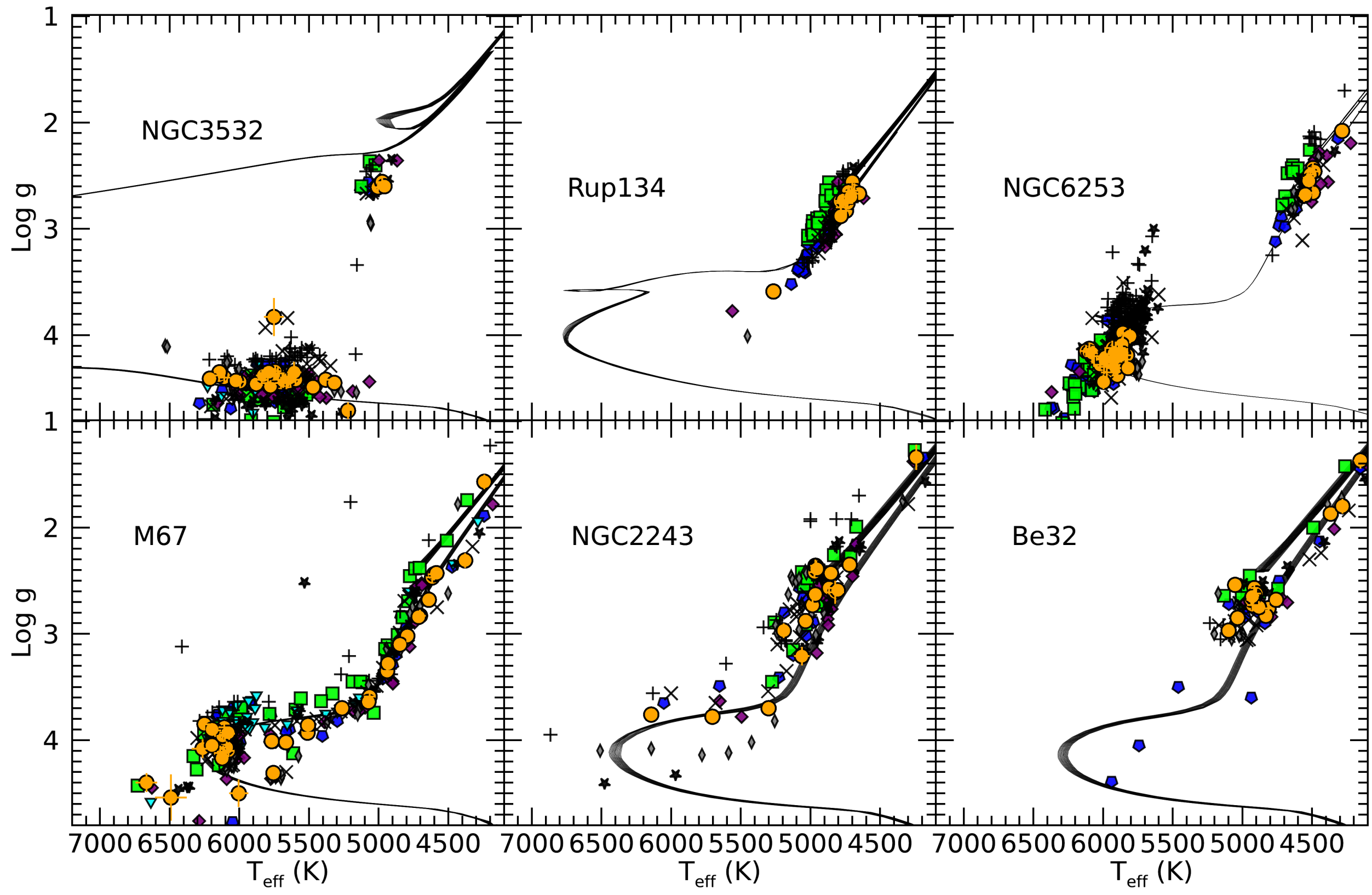

**Fig. 12.** Kiel diagrams of six representative open clusters, ordered by age, available in iDR6. Shown are their recommended stellar parameters from WG 11 (in orange), and the results of individual nodes (in different colours). Parsec isochrones (Bressan et al. 2012) corresponding to the cluster ages and encompassing the range of the cluster mean metallicity plus and minus 1$\sigma$ are shown as black continuous curves.

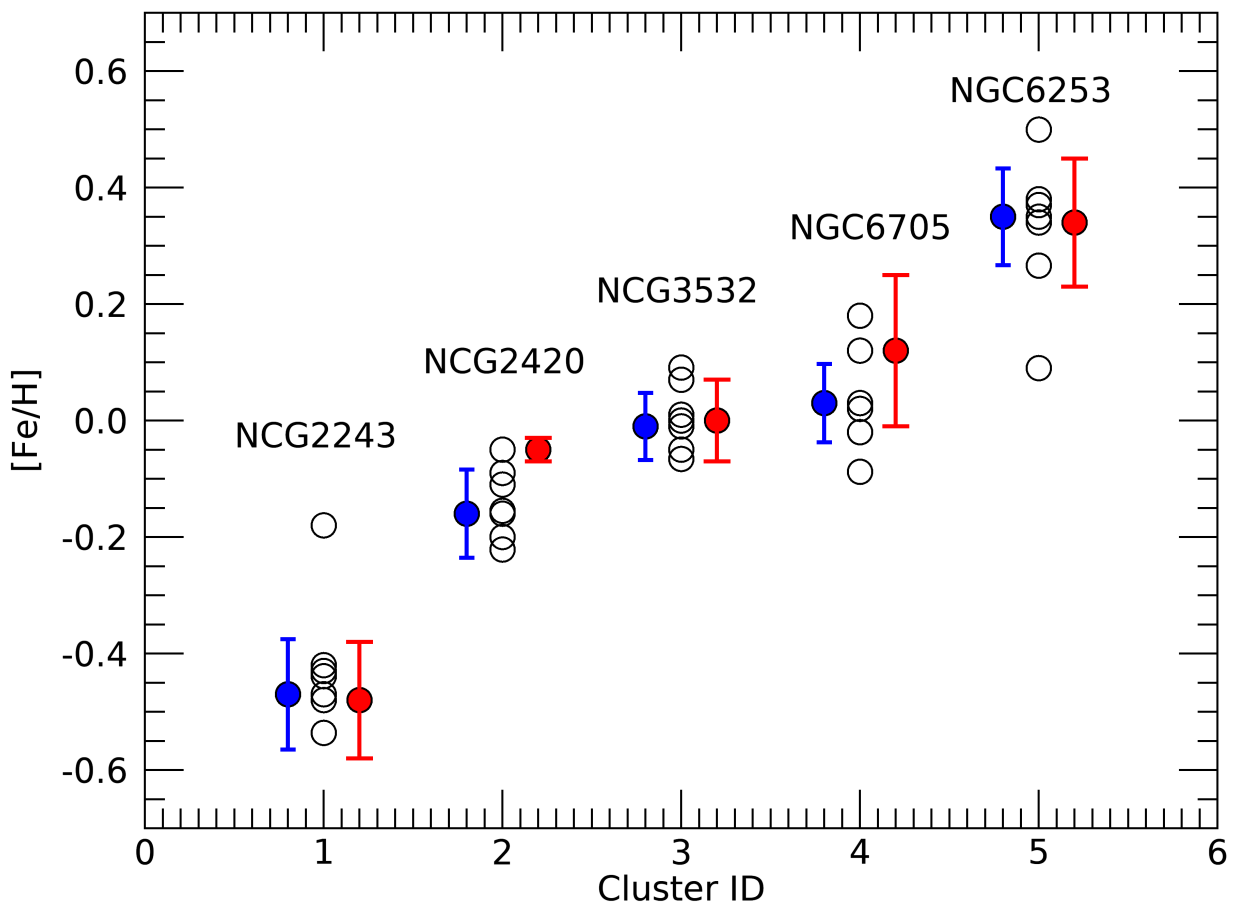

**Fig. 13.** [Fe/H] for confirmed members of five calibration open clusters in Pancino et al. (2017a). The node results are shown as empty circles, the reference values from Pancino et al. (2017a) as red circles, and the final GES values (from the UVES results) as blue circles. The points are arbitrarily shifted for better visualisation.

Finally, in Fig. 20 we show the recommended results for oxygen abundance in the [O/Fe]-[Fe/H] plane. Contrary to the other elements, for oxygen, and for C (from molecular bands), N, and Li we did not use the multi-pipeline approach, but the determination of abundances was perfomed by a single specialised node: Vilnius for the elements CNO and Arcetri for Li. The agreement with the literature values for the oxygen results is remarkable.

### 4.4. Solar abundance scale

Whilst stars belonging to clusters provide an excellent tool for measuring the precision of GES abundances, the Sun remains one of the main references for measuring their accuracy. Along with the Sun, we can also use the abundances obtained for the solar-type stars in the M 67 cluster, which has a similar chemical composition to the Sun (see e.g. Randich et al. 2006; Liu et al. 2016; Nissen & Gustafsson 2018). In Fig. 21 we hence plot elemental abundances as a function of atomic number for the solar composition from Grevesse et al. (2007) and GES recommended solar and average values for M67 dwarf members. With the exception of very few elements (e.g. S and Cu), the figure shows the excellent agreement (better than 0.1 dex) between the literature solar scale of Grevesse et al. (2007), which is the one used to compute the MARCS model atmospheres used in the GES analysis, and Sun and M67 abundances.

### 4.5. Flags

A sophisticated system of flags (hereafter detailed flags; see GRH22) was designed within the Gaia-ESO survey, and applied from the very first data releases, to report and keep track of issues occurring during the analysis (TECH flags) and also to indicate physical peculiarities on a given target (PECULI flags). The TECH flags covered a broad range of topics (SNR, data reduction, determination and quality of stellar parameters and chemical abundances).The syntax of the flags allowed us to quickly identify the issue (prefix), to trace the emitting working group (WG ID) and node (node ID), and in some cases to have extra information (suffix). However, this system is too sophisticated for the end-users wanting to quickly use the Gaia-ESO data. A system of 12 simplified flags has thus been designed for the DR6 release of the Gaia-ESO survey.

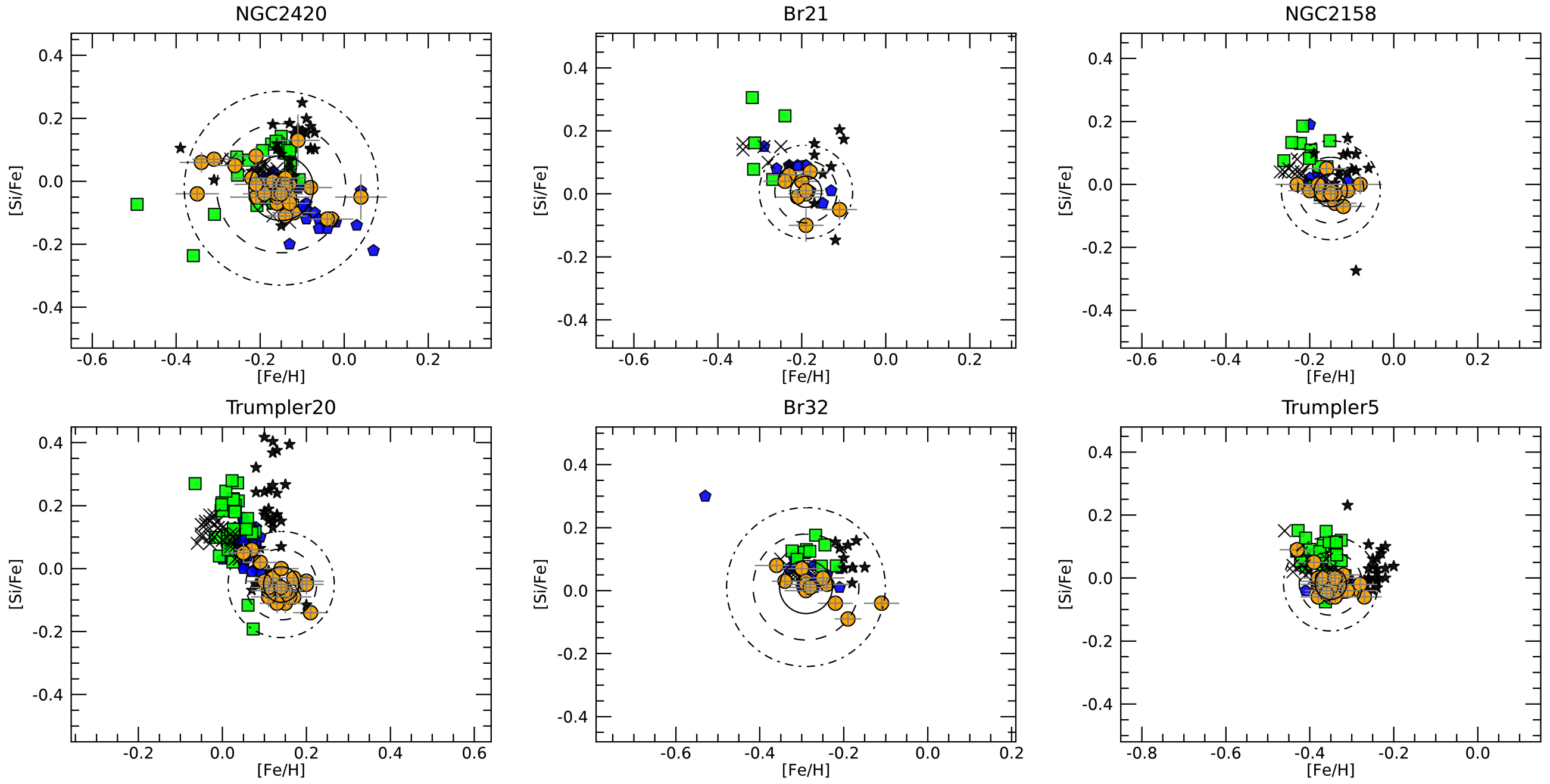

**Fig. 14.** [Si/Fe] vs [Fe/H] for members of six intermediate-age and old open clusters available in iDR6. The recommended abundances are shown in orange, while the results of individual nodes are shown in different colours. The circles give the mean abundance of each clusters and their radii denote $1\sigma$, $2\sigma$, and $3\sigma$.

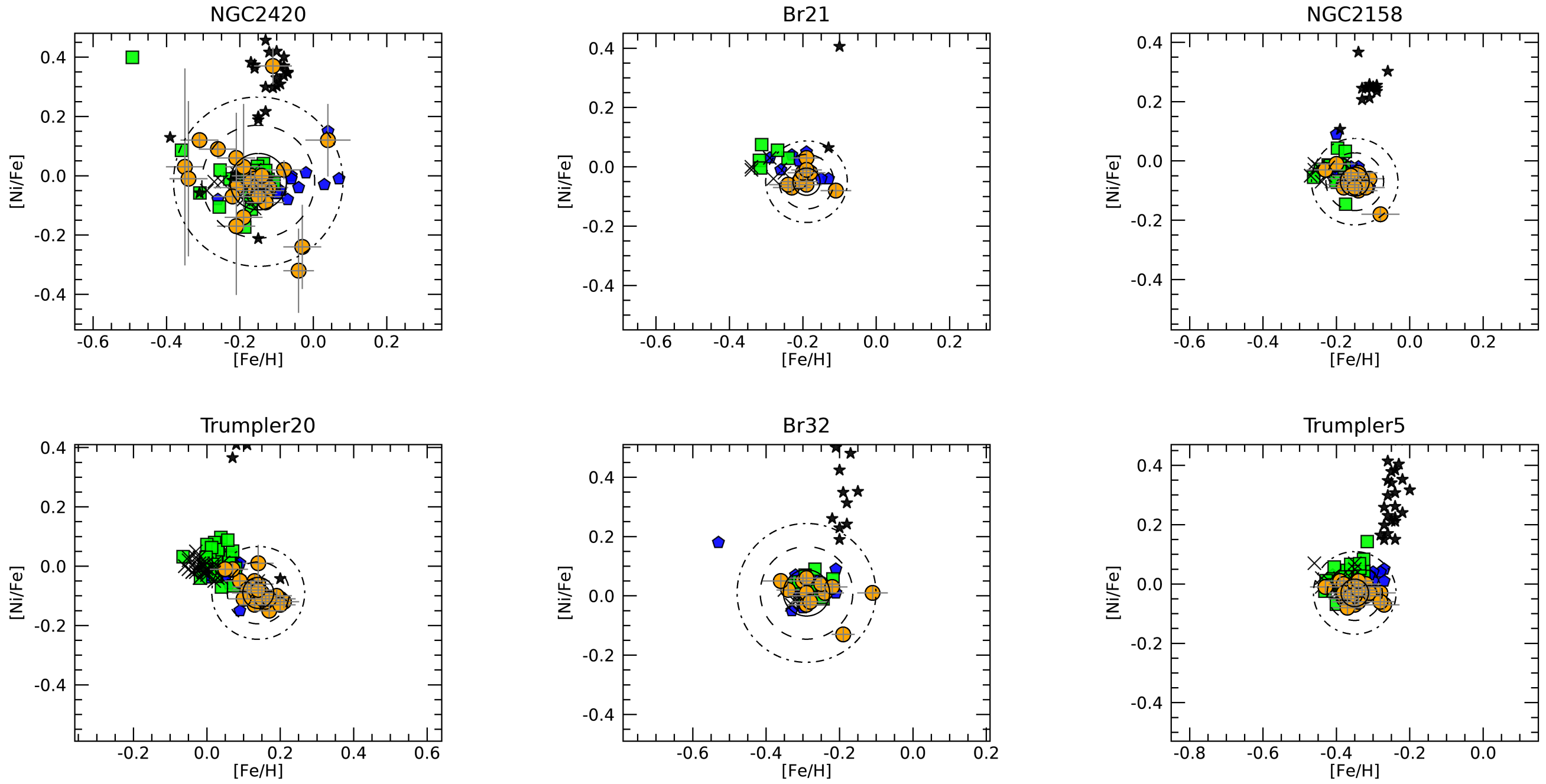

**Fig. 15.** Same as Fig. 14, but for [Ni/Fe] vs [Fe/H].

Basically, all TECH flags have been translated into simplified flags. These simplified flags are meant to allow the end-users to quickly filter the data. Therefore, they should allow the rapid rejection of objects with non-physical or highly suspicious results, completing the information already carried by the error bars associated with the measured products. It should be kept in mind that simplifying implies losing valuable information, and it is thus mandatory that the detailed flags are kept and made avail-able in the final releases such that the end-users have the ability to check them if needed.

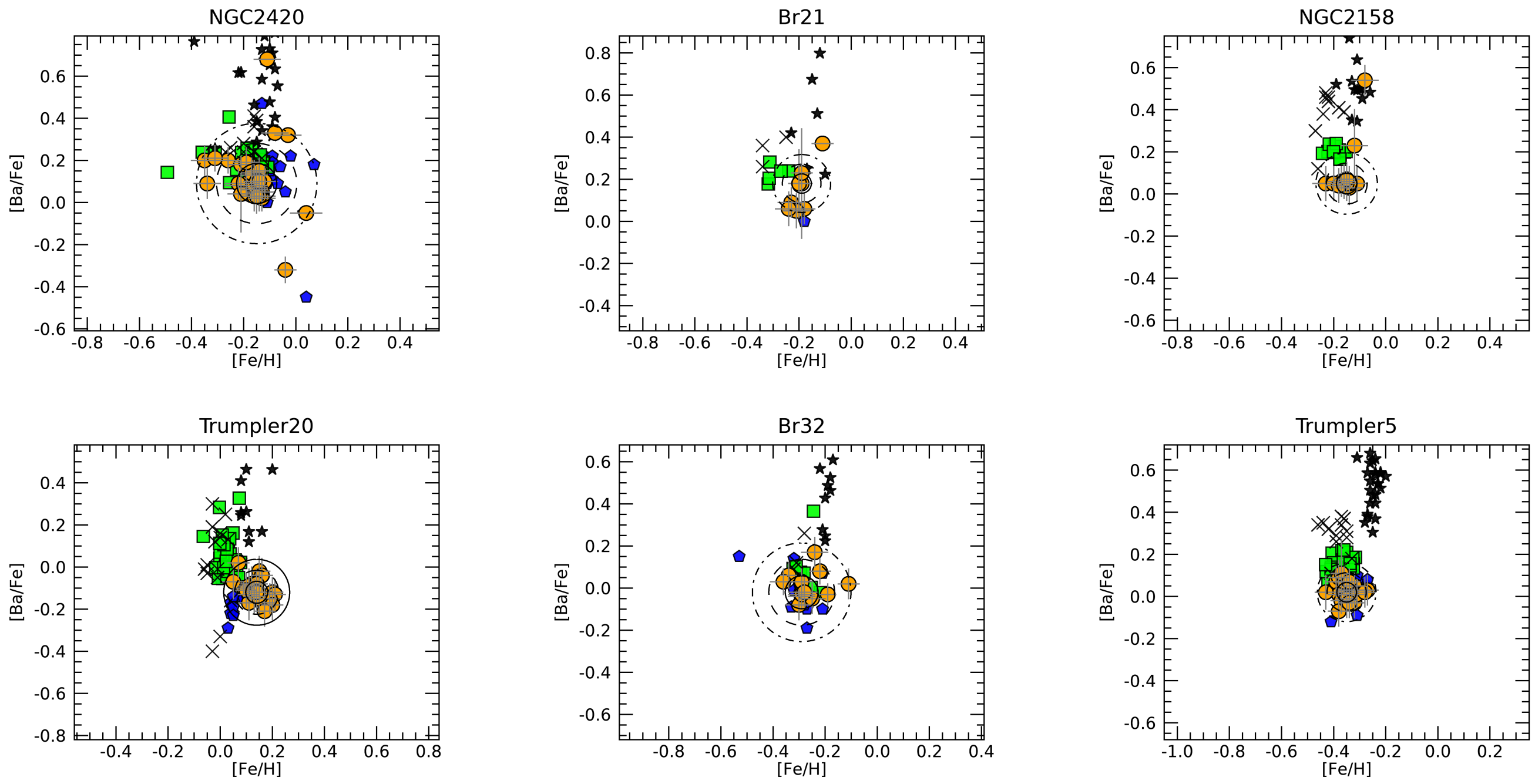

**Fig. 16.** Same as Fig. 14, but for [Ba/Fe] vs [Fe/H].

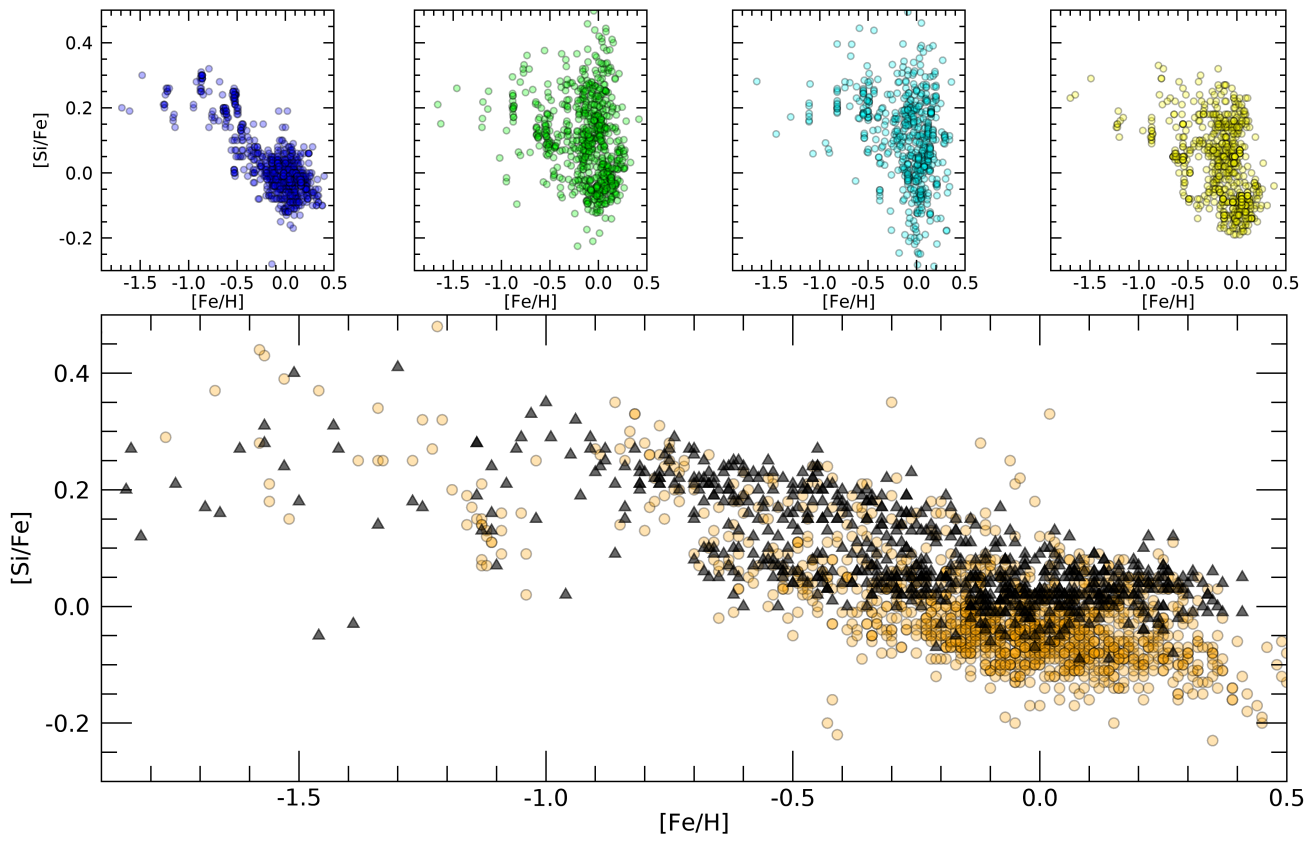

**Fig. 17.** [Si/Fe] vs [Fe/H] for iDR6. The recommended abundances from WG11 are shown in orange, while the results of the individual nodes are shown in different colours. The black triangles are the abundances from Bensby et al. (2014).

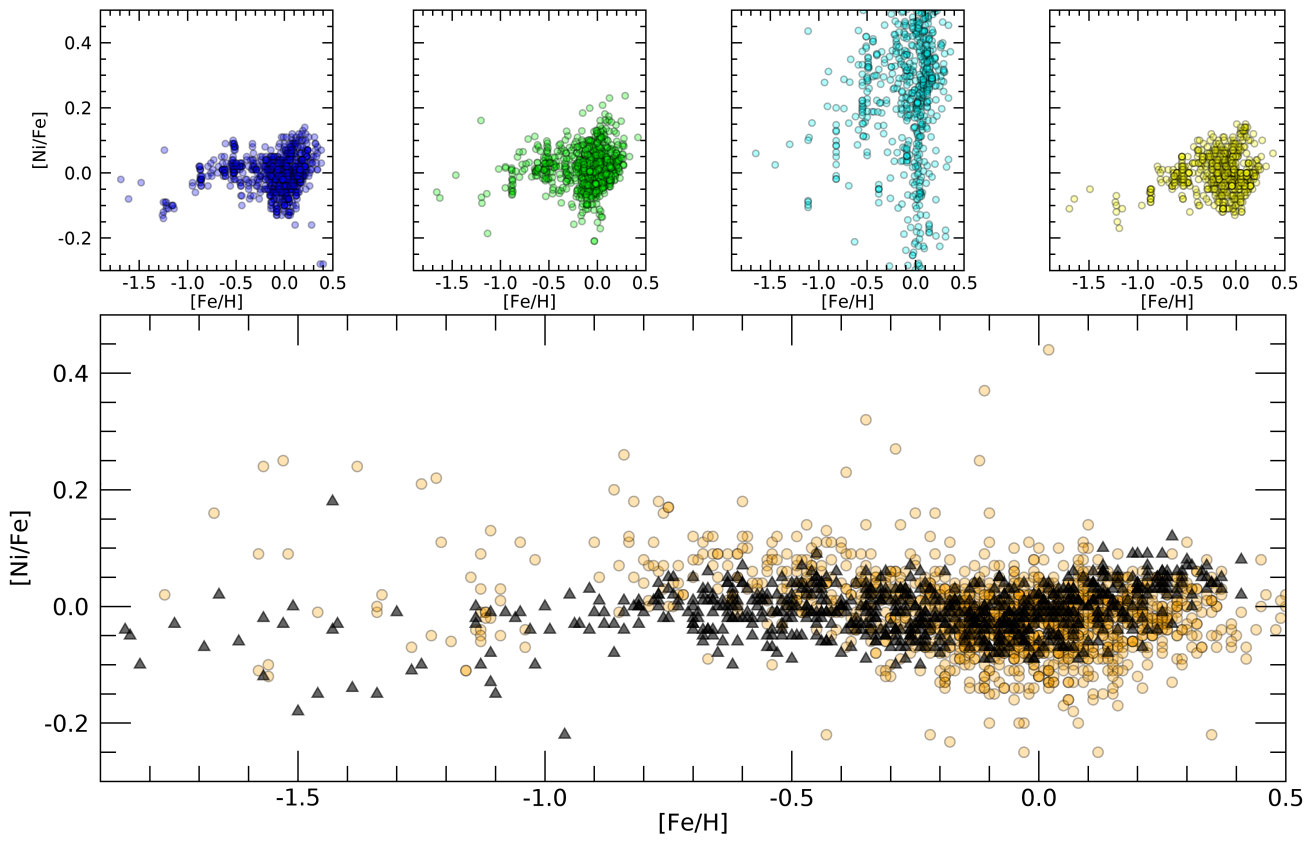

**Fig. 18.** Same as Fig. 17, but for [Ni/Fe] vs [Fe/H].

During the process of reducing the detailed flags to the simplified flags, a conservative approach was adopted, meaning that the problems might be less severe than indicated by the simplified flags. For example, the SSP (some suspicious parameters) or IPA (incomplete parameter) flags are sometimes raised when some, though not all, analysis nodes provided unreliable parameters or abundances, even though other nodes might well have provided reliable results.

The simplified flags associated stellar parameters only deal with the effective temperature, the surface gravity, the metallicity and microturbulence. The simplified flags indicating at least one stellar parameter (resp. one abundance) should help identifying suspiscious parameters (resp., abundances). It is not possible to have a limited set of simplified flags and at the same time have a detailed assessment of each stellar parameter (resp. abundance). It means that the end-users need to make some further checks (e.g. based on the detailed flags) to decide which abundances can be kept when an object has the flag 'some suspicious abundances' raised.

There is a dedicated simplified flag for the RV, on the one hand, and the rotational velocity, on the other hand. The detailed flags can tell the end-user if the object is suspected of being a SB1 or a SB2, and what the specific emission lines are, if any.

The simplified flags consist of a three-letter acronym whose meaning is easily recoverable or can be easily guessed without looking at the documentation. They are coded with Booleans (FALSE/TRUE), each in an individual column, allowing the end-users to easily sort from them.

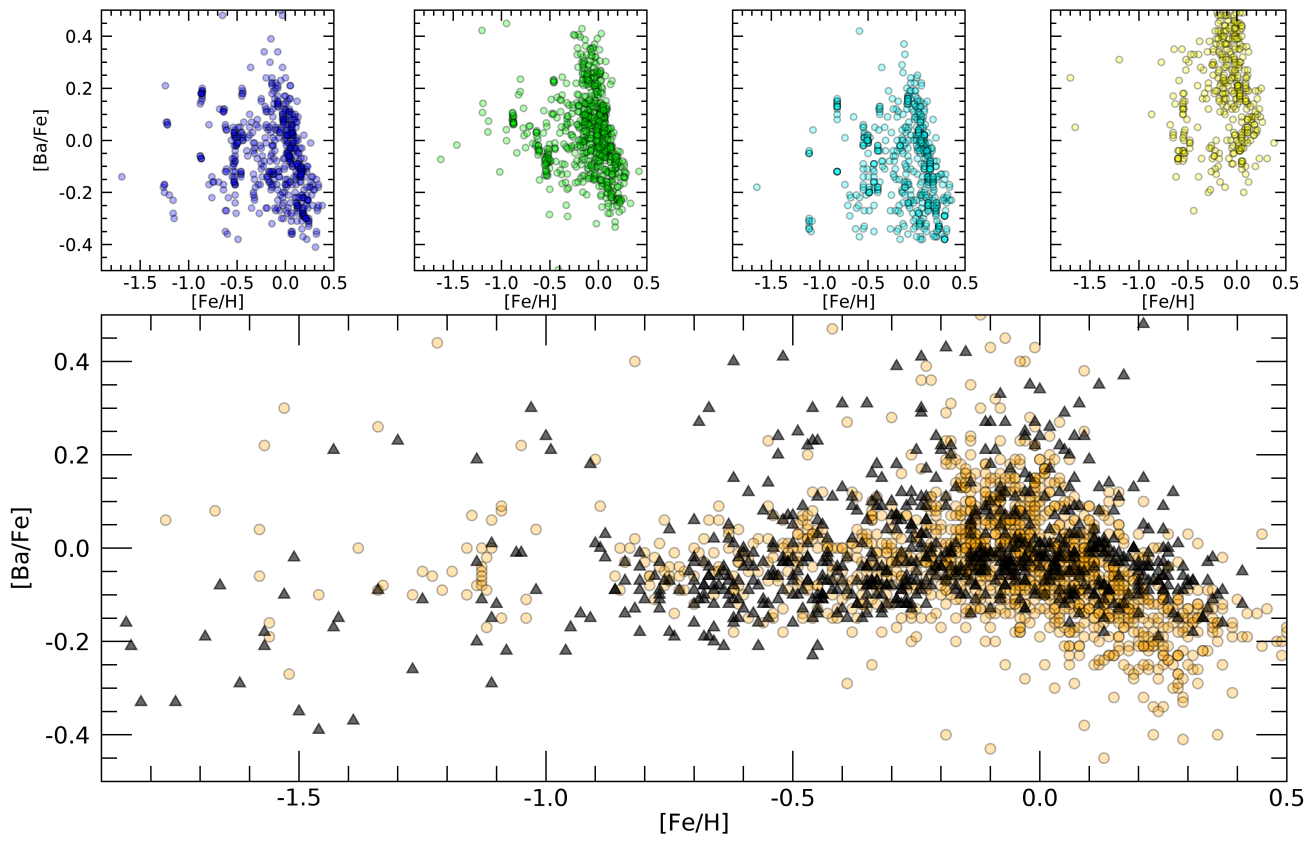

**Fig. 19.** Same as Fig. 17, but for [Ba/Fe] vs [Fe/H].

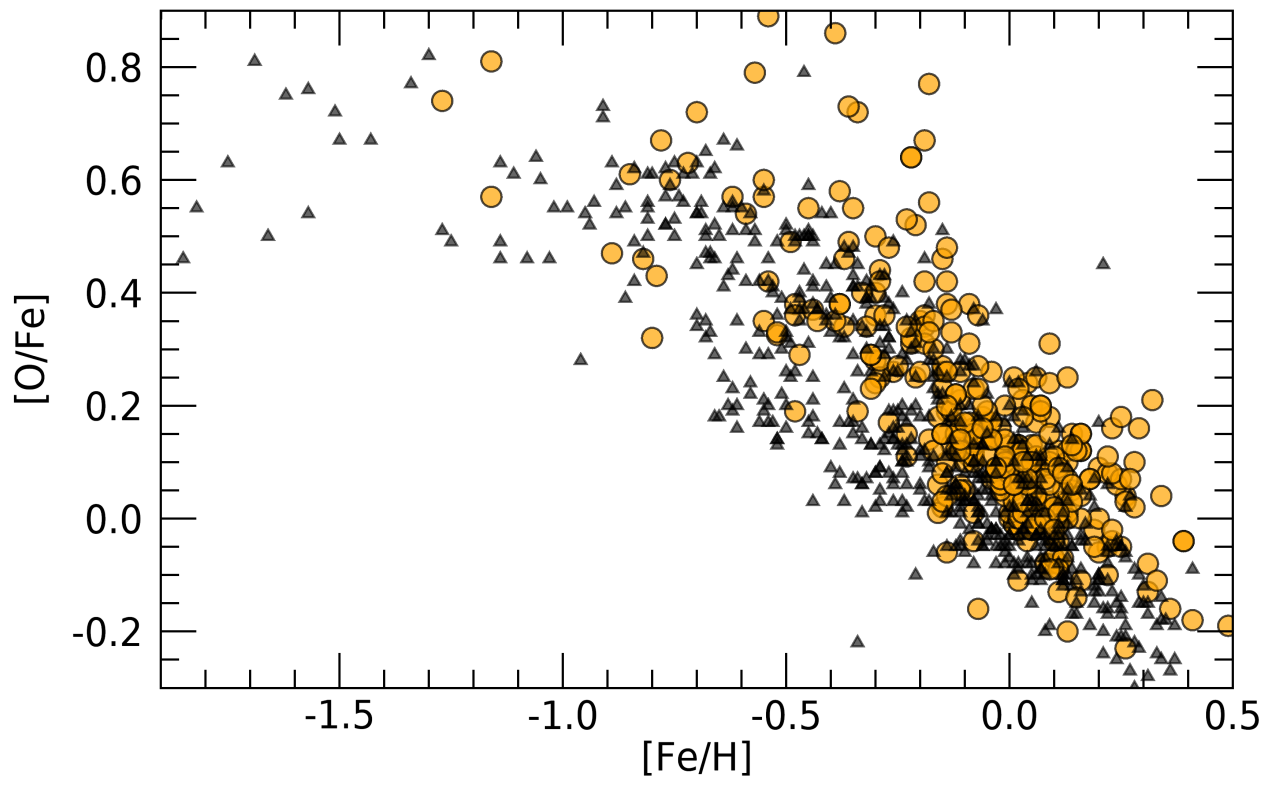

**Fig. 20.** Same as Fig. 17, but for [O/Fe] vs [Fe/H].

## 5. The open cluster survey

As anticipated in the Introduction, the GES put a particular emphasis on observations of open star clusters. Indeed it is the survey that targeted the largest sample of OCs and largest samples of targets within each cluster, the only one that covered all the different types of clusters (including the massive ones) and different evolutionary phases employing homogeneous and unbiased target selection criteria (see Bragaglia et al. 2022) and analysis. This, along with *Gaia* astrometry and photometry, has allowed us to enter a new domain of cluster research. In particular, the GES OC survey aimed to complement *Gaia* parallaxes and proper motions with crucial and extremely precise information on RVs, lithium, and chemistry in general, enabling full exploitation of the potential of *Gaia* and OCs for a variety of scientific issues, such as the formation and evolution of clusters themselves, stellar physics and evolution, the metallicity and abundance distribution in the thin disc at different ages, and the calibration of stellar ages (see GRH22 for a more detailed discussion of the science drivers). Investigation of the scientific issues listed in the previous section requires surveying statistically significant samples of members of OCs on different time- and spatial scales, from very young clusters to the oldest, from the smallest scale of the internal structure of nearby clusters to the larger MW disc scales, and from the hot stars in massive clusters to the coolest lowest mass members of the young nearby clusters. The last are crucial for the *Gaia* connection, to investigate internal kinematics and dynamics, and to put further constraints on stellar evolutionary models; the distant clusters are essential to trace the chemical structure and evolution of the MW thin disc.

The Gaia-ESO cluster selection was therefore devised to adequately cover the cluster parameter space, and to observe unbiased samples of stars in different evolutionary stages or with different masses in each cluster. We refer to Bragaglia et al. (2022), Blomme et al. (2022), and GRH22 for a detailed description of the target selection within each cluster and observational strategy, while we summarise in this section the main global properties of the OC sample.

Cluster selection was optimised to fine-sample the age-[Fe/H]-radial distance-mass parameter space. OCs in all phases of evolution (except embedded ones), with ages in the interval $\sim$ 1 Myr–8 Gyr were included, encompassing different environments and star formation conditions. The final sample includes 62 clusters for a total of 40304 individual stars; additional data were retrieved from the ESO archive, including both additional spectra of stars in OCs already covered in the GES sample and 18 complementary clusters, for a total of 1740 stars. The total science sample thus includes 80 clusters and slightly more than 42,000 targets. The archive sample mostly include giants observed with UVES, while the GES data, as noted, cover stars from the PMS phases to evolved giants. We also recall that a few OCs were both targeted by GES and retrieved from the archive for calibration purposes; they are not considered here, but they are presented in Pancino et al. (2017a). The sample of science and archive clusters are listed in Tables 3 and 4. The number of observed spectra and individual targets are included in the table. We note that, due to the inclusive and unbiased selection criteria, a large or very large fraction of non-members was eventually identified (see Jackson et al. 2020; Bragaglia et al. 2022; Jackson et al. 2022); this was expected and indeed the non-members can be (and have been) exploited to address a variety of science topics (see e.g. Casey et al. 2016; Magrini et al. 2021a; Romano et al. 2021). In particular, these non-members cover a critical age range (1–2 Gyr) that is not sampled by the MW sample, due to selection effects.

The average [Fe/H] listed in the last column of the two tables was derived from high-probability cluster members and considering [Fe/H] values from UVES spectra in most cases. In a few instances of young clusters (Trumpler 14, 25 Ori, Chamaeleon I, NGC 2244, NGC 6530, and $\rho$ Oph) we considered Giraffe metallicities of the relatively warm slowly rotating members ($T_{eff} \geq 4200$ K; vsin $i \leq 20$ km/s). The properties of the combined GES+ archive OC sample are shown in Figs. 22, 23, 24, and 25.

The figures show that our initial goal was succesfully achieved. The observed cluster sample covers a distance range from slightly more than 100 parsec to several kiloparsec from the Sun; the target clusters are distributed throughout the disc, from the innermost parts to its outskirts, and GES is the ground-based spectroscopic survey including the largest number of OCs in the inner Galaxy observed at high resolution. The archive sample nicely complements the GES one. The age range from about 1 Myr to almost 10 Gyr is fully sampled; young clusters (age $\leq$ 100 Myr) are preferentially located within 3 kpc from the Sun, while the older OCs extend to much larger distances. As discussed in Randich et al. (2013), Bragaglia et al. (2022), Blomme

**Table 3.** Open cluster sample. Information in Cols. 2–7 comes from Cantat-Gaudin et al. (2020) and the following: Bell et al. (2013) age of NGC 2244 and NGC 6530; Venuti et al. (2018) age of NGC 2264; Franciosini et al. (2022) age of $\gamma$ Velorum; Damiani et al. (2017b) age of Trumpler 14; Grasser et al. (2021) age of $\rho$ Oph; Galli et al. (2021) age of Cha I. Columns 8 and 9 list the number of spectra and targets, while the average [Fe/H] and standard deviation are given in Col. 10. In most cases [Fe/H] was derived in this paper as described in the text. Metallicity for NGC 3293, NGC 3766, and NGC 6649 was instead taken from Bragaglia et al. (2022). Clusters with an 'a' and/or 'c' superscript also have spectra retrieved from the archive or have been observed as calibration targets.

| Cluster | RA | DEC | log (age) (yr) | Distance (pc) | z (kpc) | $R_{GC}$ (kpc) | N spectra Giraffe/UVES | N stars | [Fe/H] |
|---|---|---|---|---|---|---|---|---|---|
| Blanco 1 | 0.853 | −29.958 | 8.02 | 240 | -236 | 8.3 | 431/37 | 463 | −0.03 ± 0.04 |
| 25 Ori | 81.198 | 1.655 | 7.13 | 344 | −108 | 8.6 | 307/29 | 294 | 0.0 ± 0.02 |
| Collinder 69, $\lambda$ Ori | 83.792 | 9.813 | 7.1 | 416 | −87 | 8.7 | 802/117 | 836 | −0.09 ± 0.06 |
| Berkeley 21 | 87.93 | 21.812 | 9.33 | 6417 | −278 | 14.7 | 737/13 | 744 | −0.21 ± 0.04 |
| Czernik 24 | 88.848 | 20.876 | 9.43 | 3981 | −154 | 12.3 | 340/6 | 346 | −0.11 ± 0.03 |
| Berkeley 22 | 89.618 | 7.763 | 9.39 | 6225 | −874 | 14.3 | 409/7 | 395 | −0.26 ± 0.06 |
| NGC 2141 | 90.734 | 10.451 | 9.27 | 5183 | −524 | 13.3 | 848/23 | 853 | −0.04 ± 0.04 |
| NGC 2158 | 91.862 | 24.099 | 9.19 | 4298 | 134 | 12.6 | 613/14 | 616 | −0.15 ± 0.04 |
| Berkeley 73 | 95.52 | −6.321 | 9.15 | 6158 | -1005 | 13.7 | 70/7 | 77 | −0.26 ± 0.03 |
| NGC 2232 | 96.888 | −4.749 | 7.25 | 315 | −40 | 8.6 | 2022/69 | 1866 | 0.02 ± 0.05 |
| NGC 2243[a,c] | 97.395 | −31.282 | 9.64 | 3719 | −1150 | 10.6 | 908/34 | 710 | −0.45 ± 0.05 |
| NGC 2244 | 98.045 | 4.914 | 6.6 | 1478 | −52 | 9.7 | 418/14 | 432 | −0.04 ± 0.05 |
| Trumpler 5 | 99.126 | 9.465 | 9.63 | 3047 | 54 | 11.2 | 1132/27 | 1138 | −0.35 ± 0.04 |
| NGC 2264[a] | 100.217 | 9.877 | 6.5 | 707 | 26 | 9.0 | 1759/118 | 1877 | −0.10 ± 0.03 |
| Berkeley 25[a] | 100.317 | −16.487 | 9.39 | 6780 | −1134 | 13.8 | 87/7 | 83 | −0.25 ± 0.06 |
| Berkeley 75 | 102.252 | −23.999 | 9.23 | 8304 | -1611 | 14.7 | 69/6 | 75 | −0.34 ± 0.05 |
| Berkeley 31 | 104.406 | 8.285 | 9.45 | 7177 | 642 | 15.1 | 616/14 | 616 | −0.29 ± 0.03 |
| Berkeley 30 | 104.438 | 3.229 | 8.47 | 5383 | 270 | 13.2 | 369/14 | 332 | −0.13 ± 0.01 |
| Berkeley 32[c] | 104.53 | 6.433 | 9.69 | 3072 | 236 | 11.1 | 588/46 | 438 | −0.31 ± 0.06 |
| Berkeley 36 | 109.105 | −13.196 | 9.83 | 4360 | −42 | 11.7 | 751/14 | 739 | −0.15 ± 0.02 |
| NGC 2355 | 109.247 | 13.772 | 9.0 | 1941 | 397 | 10.1 | 199/11 | 208 | −0.13 ± 0.03 |
| Haffner 10 | 112.156 | −15.364 | 9.58 | 3409 | 60 | 10.8 | 557/13 | 562 | −0.1 ± 0.03 |
| Czernik 30 | 112.796 | −9.945 | 9.46 | 6647 | 482 | 13.8 | 219/7 | 226 | −0.31 ± 0.01 |
| NGC 2425 | 114.577 | −14.885 | 9.38 | 3576 | 205 | 10.9 | 522/17 | 528 | −0.13 ± 0.03 |
| NGC 2420[c] | 114.602 | 21.575 | 9.24 | 2587 | 869 | 10.7 | 768/38 | 562 | −0.15 ± 0.02 |
| NGC 2451A | 115.736 | −38.264 | 7.55 | 195 | −24 | 8.4 | 1606/90 | 1656 | −0.08 ± 0.06 |
| NGC 2451B | 116.128 | −37.954 | 7.61 | 361 | −43 | 8.4 | 1606/90 | 1656 | −0.02 ± 0.06 |
| Berkeley 39[a] | 116.702 | −4.665 | 9.75 | 3968 | 694 | 11.5 | 896/28 | 899 | −0.14 ± 0.05 |
| NGC 2516 | 119.527 | −60.8 | 8.38 | 423 | −115 | 8.3 | 743/53 | 759 | −0.04 ± 0.04 |
| gamma Vel | 122.374 | −47.335 | 7.3 | 330 | −44 | 8.4 | 1321/80 | 1269 | −0.02 ± 0.05* |
| NGC 2547[a] | 122.525 | −49.198 | 7.51 | 396 | −59 | 8.4 | 450/67 | 477 | −0.03 ± 0.04 |
| IC 2391 | 130.292 | −52.991 | 7.46 | 148 | −17 | 8.3 | 398/52 | 434 | −0.06 ± 0.13 |
| Collinder 197 | 131.202 | −41.28 | 7.15 | 955 | 15 | 8.5 | 406/9 | 409 | +0.03 ± 0.015 |
| Alessi 43 | 132.631 | −41.738 | 7.06 | 917 | 24 | 8.5 | 1206/36 | 1225 | +0.02 ± 0.06 |
| Pismis 15 | 143.684 | −48.04 | 8.94 | 2559 | 127 | 8.6 | 332/11 | 333 | +0.02 ± 0.06 |
| ESO 92-5 | 150.801 | −64.755 | 9.65 | 12444 | −1625 | 12.8 | 205/7 | 212 | −0.29 ± 0.06 |
| NGC 3293[a] | 158.97 | −58.231 | 7.01 | 2710 | 3 | 8.0 | 3017/27 | 584 | 0.02 |
| IC 2602[a] | 160.613 | −64.426 | 7.56 | 149 | −12 | 8.3 | 1797/140 | 1841 | −0.06 ± 0.06 |
| Trumpler 14 | 160.986 | −59.553 | 6.3–6.6 | 2290 | −23 | 8.0 | 5443/45 | 1902 | −0.01 ± 0.06 |
| NGC 3532[c] | 166.417 | −58.707 | 8.6 | 498 | 12 | 8.2 | 1234/83 | 1145 | −0.03 ± 0.08 |
| NGC 3766 | 174.061 | −61.616 | 7.36 | 2123 | −1 | 7.7 | 1563/8 | 399 | −0.12 |
| Trumpler 20[a] | 189.882 | −60.637 | 9.27 | 3392 | 130 | 7.2 | 1452/41 | 1303 | 0.13 ± 0.05 |
| NGC 4815 | 194.499 | −64.96 | 8.57 | 3295 | −120 | 7.1 | 226/14 | 218 | +0.08 ± 0.14 |
| Pismis 18 | 204.227 | −62.091 | 8.76 | 2860 | 15 | 6.9 | 142/10 | 142 | +0.14 ± 0.03 |
| NGC 6005 | 238.955 | −57.439 | 9.1 | 2383 | −124 | 6.5 | 559/38 | 560 | 0.22 ± 0.03 |
| Trumpler 23 | 240.218 | −53.539 | 8.85 | 2590 | 21 | 6.3 | 167/19 | 165 | +0.20 ± 0.03 |
| NGC 6067 | 243.299 | −54.227 | 8.1 | 1881 | −72 | 6.8 | 812/32 | 780 | +0.03 ± 0.16 |
| rho Oph[a] | 246.0 | −23.8 | 5.5-6.8 | 139 | 42 | 8.0 | 313/23 | 311 | 0.03 ± 0.06 |
| NGC 6259 | 255.195 | −44.678 | 8.43 | 2314 | −61 | 6.2 | 606/21 | 494 | +0.18 ± 0.05 |
| NGC 6281 | 256.179 | −37.948 | 8.71 | 539 | 18 | 7.8 | 333/16 | 320 | −0.04 ± 0.03 |
| NGC 6405 | 265.069 | −32.242 | 7.54 | 459 | −6 | 7.9 | 696/25 | 560 | −0.02 ± 0.04 |
| IC 4665 | 266.554 | 5.615 | 7.52 | 354 | 104 | 8.0 | 545/34 | 567 | 0.01 ± 0.05 |
| Ruprecht 134 | 268.184 | −29.537 | 9.22 | 2252 | −64 | 6.1 | 661/38 | 680 | +0.27 ± 0.04 |
| NGC 6530 | 271.09 | −24.33 | 6.3 | 1325 | – | 6.8 | 1981/62 | 1984 | −0.02 ± 0.08 |
| NGC 6633[a] | 276.845 | 6.615 | 8.84 | 424 | 61 | 8.0 | 1687/81 | 1663 | −0.03 ± 0.04 |
| NGC 6649 | 278.359 | −10.399 | 7.85 | 2124 | −28.0 | 6.4 | 437/9 | 283 | −0.08 |
| NGC 6705 | 282.766 | −6.272 | 8.49 | 2203 | −106 | 6.5 | 2593/59 | 1066 | +0.03 ± 0.05 |
| NGC 6709 | 282.836 | 10.334 | 8.28 | 1041 | 85 | 7.6 | 795/73 | 730 | −0.02 ± 0.01 |
| Berkeley 81 | 285.419 | −0.454 | 9.06 | 3313 | -143 | 5.9 | 307/14 | 279 | +0.22 ± 0.06 |
| Berkeley 44 | 289.313 | 19.55 | 9.16 | 2863 | 167 | 7.0 | 86/7 | 93 | +0.22 ± 0.09 |
| NGC 6802 | 292.651 | 20.262 | 8.82 | 2573 | 43 | 7.1 | 198/13 | 197 | 0.14 ± 0.04 |
| Chamaeleon I | 297.2 | −15.4 | 6.2 | 189 | – | 8.0 | 674/49 | 709 | −0.03 ± 0.12 |

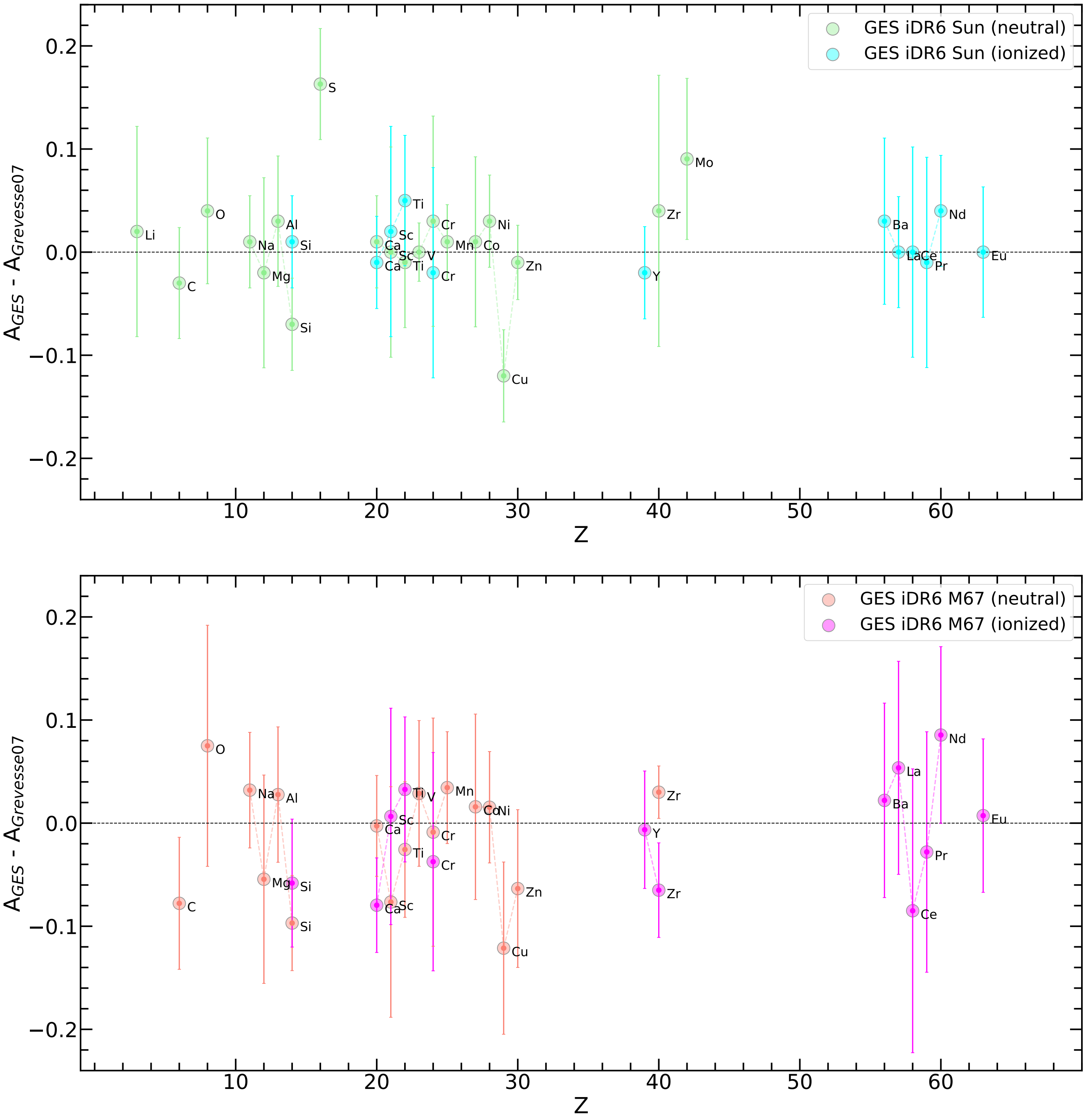

**Fig. 21.** Abundance difference vs atomic number between the solar composition in Grevesse et al. (2007) and in GES recommended values (neutral and ionised elements) in the Sun (upper panel) and in M 67 (bottom panel). For the latter the average abundances of dwarf cluster members are considered.

et al. (2022), and GRH22, for Giraffe target selection we adopted a magnitude-limited criterion, meaning that very cool low-mass stars were observed in the nearby clusters, while progressively warmer and higher-mass stars were targeted in the most distant clusters. The brightest cluster candidates were instead selected as UVES targets, these being luminous giants or hot main sequence stars in the distant clusters, or PMS-MS-TO stars in the nearby ones. Noticeably, the OC sample includes the Carina nebula, one of the most massive H ii regions known in the Galaxy that contains some of the most massive O stars known (see e.g. Damiani et al. 2017b, and references therein).

## 6. Science highlights

Thanks to the excellent data quality, as well as a bottom-up publication strategy within the consortium, since the first internal data release the GES has allowed a variety of significant results to be achieved. A summary of the topics that have been addressed by the Gaia-ESO consortium and references are provided below.

**Table 4.** Same as Table 3, but for clusters retrieved from the ESO archive. Only UVES spectra are considered.

| Cluster | RA | DEC | log age (yr) | Distance (pc) | z (pc) | $R_{GC}$ (kpc) | N stars | [Fe/H] |
|---|---|---|---|---|---|---|---|---|
| Berkeley 20 | 83.152 | 0.185 | 9.68 | 8728 | −2606 | 16.3 | 6 | −0.38 ± 0.04 |
| Collinder 110 | 99.677 | 2.069 | 9.26 | 2183 | −71 | 10.3 | 6 | −0.1 ± 0.04 |
| Ruprecht 4 | 102.248 | −10.524 | 8.93 | 4087 | −378 | 11.7 | 5 | −0.13 ± 0.01 |
| Berkeley 29 | 103.268 | 16.93 | 9.49 | 12604 | 1750 | 20.6 | 6 | −0.36 ± 0.07 |
| Ruprecht 7 | 104.456 | −13.227 | 8.37 | 5851 | −469 | 13.1 | 5 | −0.24 ± 0.03 |
| Tombaugh 2 | 105.773 | −20.82 | 9.21 | 9316 | −1115 | 15.8 | 12 | −0.24 ± 0.07 |
| NGC 2324 | 106.033 | 1.046 | 8.73 | 4214 | 242 | 12.1 | 8 | −0.18 ± 0.01 |
| NGC 2660 | 130.667 | −47.201 | 8.97 | 2788 | −146 | 8.9 | 5 | −0.05 ± 0.04 |
| M67 | 132.846 | 11.814 | 9.63 | 889 | 470 | 8.9 | 131 | 0.0 ± 0.02 |
| NGC 3960 | 177.644 | −55.679 | 8.9 | 2345 | 252 | 7.7 | 10 | 0.0 ± 0.01 |
| NGC 4337 | 186.022 | −58.125 | 9.16 | 2450 | 194 | 7.4 | 7 | 0.24 ± 0.03 |
| Collinder 261 | 189.519 | −68.377 | 9.8 | 2850 | −275 | 7.3 | 7 | −0.05 ± 0.07 |
| NGC 5822 | 226.051 | −54.366 | 8.96 | 1404 | 42 | 7.7 | 4 | 0.02 ± 0.02 |
| NGC 6192 | 250.077 | −43.355 | 8.38 | 1737 | 64 | 6.7 | 6 | −0.08 ± 0.07 |
| NGC 6404 | 264.916 | −33.224 | 8.0 | 2500 | −51 | 5.8 | 5 | 0.01 ± 0.06 |
| NGC 6583 | 273.962 | −22.143 | 9.08 | 2053 | −91 | 6.3 | 4 | 0.22 ± 0.01 |
| Ruprecht 147 | 289.087 | −16.333 | 9.48 | 323 | −71 | 8.05 | 6 | 0.12 ± 0.02 |
| NGC 6791 | 290.221 | 37.778. | 9.8 | 4231 | 800 | 7.9 | 8 | +0.23 ± 0.20 |

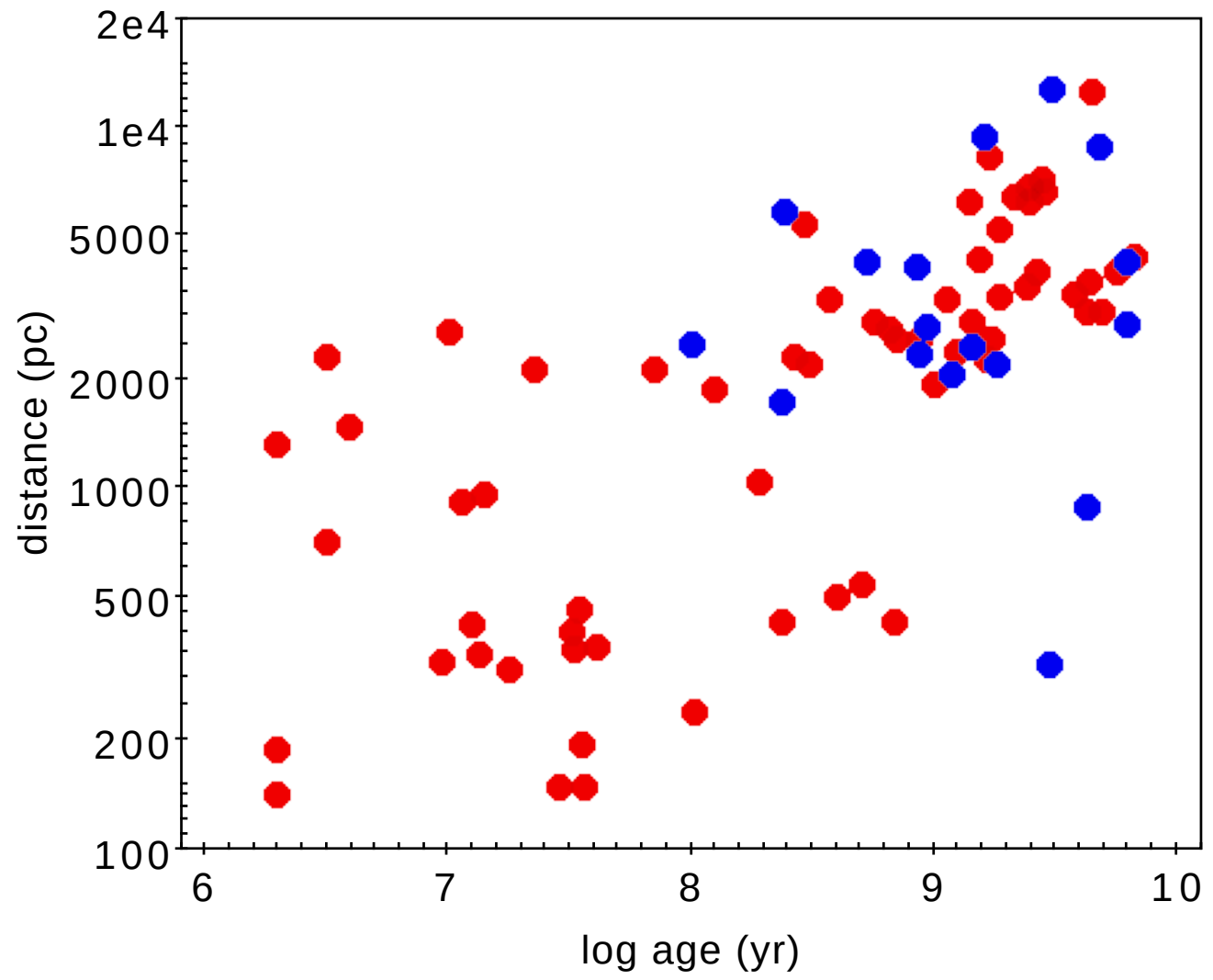

**Fig. 22.** Cluster distance as a function of age for GES and archive science open clusters (red and blue symbols, respectively).
.

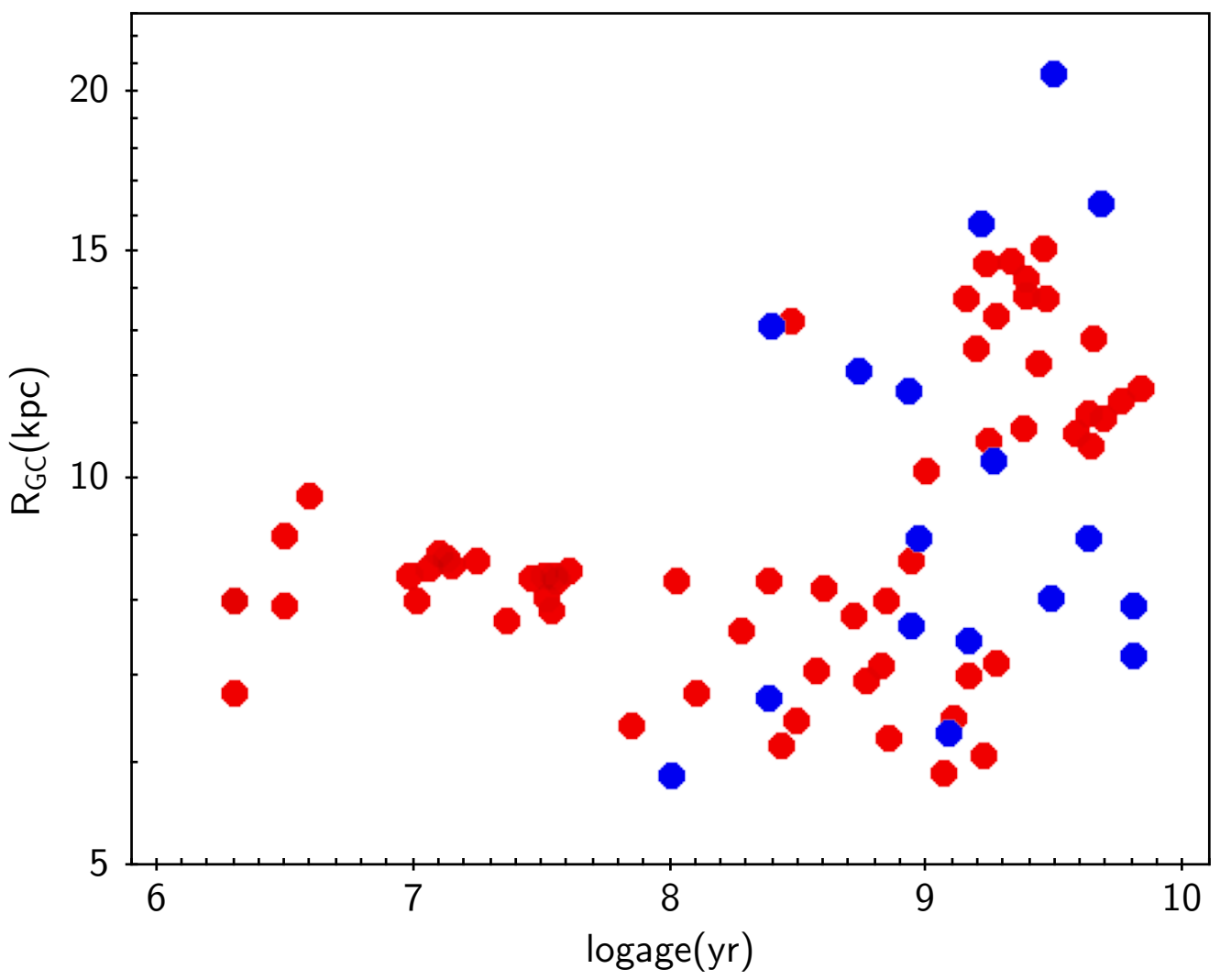

**Fig. 23.** Same as Fig. 22, but for Galactocentric distance.

### *6.1. The MW survey*

Milky Way science from Gaia-ESO has appeared in about 30 refereed papers to date, and in several works combining both cluster and MW data (see also Sect. 6.2). The main focus was on the properties of stellar populations of the Galactic discs, though studies of the Bulge, very metal-poor stars, and interstellar extinction were also performed. Highlights include the following:

- The early study by Bergemann et al. (2014) is of particular note as it has already become a very well cited article. This paper began the extension of analyses to include age estimates for field stars, opening direct studies of Galactic (local) chemical evolution.
- Howes et al. (2014) and Jackson-Jones et al. (2014) investigated the metal-poor stellar content of Gaia-ESO, at high (Jackson-Jones) and low (Howes) Galactic latitudes.
- Several Gaia-ESO studies have more generally analysed the chemical and kinematic properties of different MW subsamples, comparing the data to simulations and models in some cases. Some of these papers confirmed and extended the discreteness of the thin disc and thick disc in elemental abundance space, while others focused on the nucleosynthesis and main channel of production of chemical elements such as carbon (Recio-Blanco et al. 2014; Ruchti et al. 2015; Kordopatis et al. 2015; Guiglion et al. 2015; Rojas-Arriagada et al. 2016; Hayden et al. 2018; Fu et al. 2018; Thompson et al. 2018; Franchini et al. 2020, 2021).

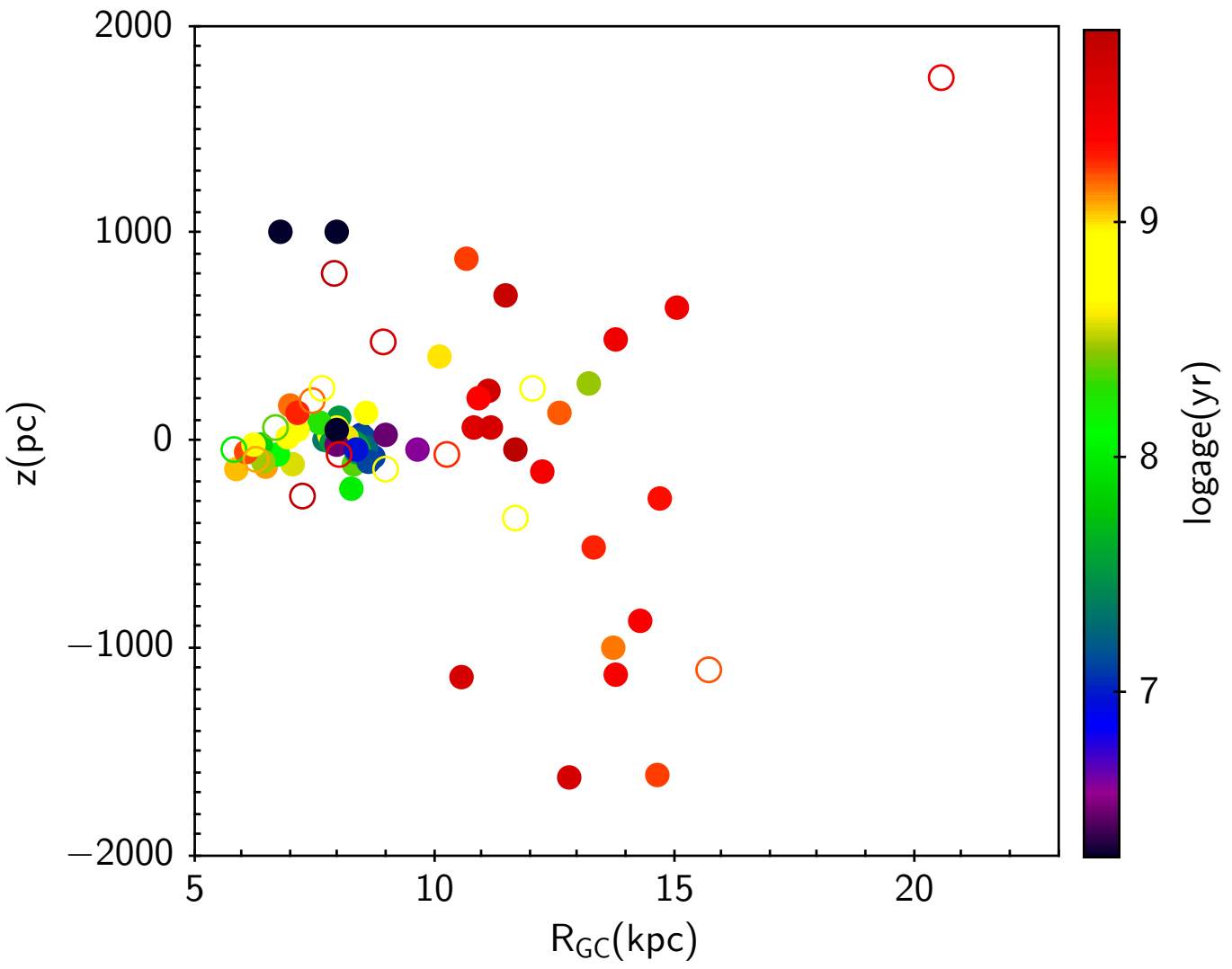

**Fig. 24.** Distance from the Galactic plane (z) as a function of Galactocentric distance for the open cluster sample. Filled and open circles denote GES and archive clusters, respectively. Symbols are colour-coded by age.
.

– Interesting results were obtained on Galactic extinction maps. Specifically, Schultheis et al. (2015) and Puspitarini et al. (2015) analysed the interstellar extinction distribution as a function of distance along observed lines of sight. Their agreement with other detailed studies was excellent, indicating the value of large surveys to map 3D dust distributions in the Galaxy.
– Beginning with early data from Gaia-ESO iDR1, Rojas-Arriagada et al. (2014) found evidence to support a boxy/peanut X-shaped component in the metal-rich population of the Galactic Bulge, with bar-like kinematics. Williams et al. (2016) investigated the metallicity and velocity distributions of a larger sample of metal-rich Bulge giants ([Fe/H] > 0) from the iDR4 sample and compared them with the expected properties of resonant orbits from simulations, while Recio-Blanco et al. (2017) found the existence of a bimodal distribution (at least) of [$\alpha$/Fe] in Bulge stars with Gaia-ESO iDR4. Rojas-Arriagada et al. (2017) probed the Bulge with a sample of 2500 red clump stars from GES iDR4 and confirmed the bimodality of the metallicity distribution function. They also found that the metal-rich sample was associated with a boxy/peanut Bulge formed via secular evolution of the thin disc. The origin of the metal-poor sample is less certain; the authors postulate an origin in an early prompt dissipative collapse dominated by massive stars but cannot rule out secular evolution of the thick disc.
– GES provided a unique opportunity to identify and characterise the distribution of spectroscopic multiple systems among different populations of the Galaxy, in clusters and the MW field. (Merle et al. 2017, 2019, 2020).
– Finally, calibration targets, in particular the globular clusters were also scientifically exploited (e.g. Lardo et al. 2015; Lind et al. 2015; San Roman et al. 2015; Pancino et al. 2017b; Sanna et al. 2020).

### 6.2. Open cluster science

Thanks to the unbiased target selection and observing strategy together with the superb data quality and homogeneous products, to the excellent RV precision, and to the unique lithium dataset (see below), GES is keeping its initial promises for open cluster science and is showing itself to be one of the most successful projects in this research area. Approximately 50 papers have been published so far; they address the original science drivers plus a number of serendipitous discoveries of significant impact. Highlights include the following:

– The detailed investigation of the kinematics and dynamics of young clusters and star forming regions, which has led to novel results, also anticipating later discoveries by *Gaia* (e.g. Jeffries et al. 2014; Sacco et al. 2015, 2017; Rigliaco et al. 2016; Bravi et al. 2018; Wright et al. 2019).
– The study of the structure, star formation histories, initial mass function, age spreads, and accretion properties in a number of very young clusters, a relevant topic with impact on our understanding of cluster and star formation (e.g. Frasca et al. 2015; Delgado et al. 2016; Prisinzano et al. 2016; Damiani et al. 2017a,b; Venuti et al. 2018; Prisinzano et al. 2019; Bonito et al. 2020).
– The determination of precise membership, detailed studies and characterisation of individual clusters, the derivation of Hertzprung–Russell diagrams for both young and old cluster members, and the determination of cluster ages also in combination with *Gaia* astrometry (e.g. Friel et al. 2014; Cantat-Gaudin et al. 2014; Donati et al. 2014; Tang et al. 2017; Overbeek et al. 2017; Randich et al. 2018; Hatzidimitriou et al. 2019).
– The use of elemental abundances (lithium in particular, but also other light elements) and their ratios to put constraints on stellar physics, atomic diffusion, and non-standard mixing processes at work in stellar interiors in different evolutionary phases, a modern topic to which GES made a very significant contribution (e.g. Tautvaišienė et al. 2015; Jackson et al. 2016; Smiljanic et al. 2016; Bouvier et al. 2016; Jeffries et al. 2017; Bertelli Motta et al. 2018; Lagarde et al. 2019; Semenova et al. 2020; Franciosini et al. 2020; Magrini et al. 2021a,b). We also mention the recent papers by Franciosini et al. (2022), Binks et al. (2021), and Binks et al. (2022), where the effect of starspots on PMS evolutionary models was further explored and the output from the stellar evolutionary code was compared with the observed colour-magnitude diagrams and Li depletion patterns of young clusters.
– The calibration and determination of stellar ages employing abundances and abundance ratios, another key topic on which GES has made and is making a great impact (e.g. Casali et al. 2019, 2020; Gutiérrez Albarrán et al. 2020; Binks et al. 2021; Randich & Magrini 2021, and references therein).

One of the primary goals of the GES cluster survey was their use as key tracers of the formation and evolution of the MW thin disc, which also allows constraints to be put on chemical evolution models and nucleosynthesis processes (see e.g. Magrini & Randich 2014; Randich 2020, and references therein). Since the first GES internal data release many articles investigating these issues using OCs have been proposed. A part of these papers addressed the traditional topic of the radial metallicity distribution and gradient (e.g. Magrini & Randich 2014; Jacobson et al. 2016); these studies also covered the determination for the first time of the present-day gradient based on [Fe/H] measurements in low-mass stars in young clusters (e.g. Spina et al. 2017, and references therein). A significant number of papers instead focused on the distribution and evolution of individual elements, often using cluster and MW samples, including lithium, CNO, and neutron capture elements, providing new relevant results (e.g. Magrini et al. 2014; Tautvaišienė et al. 2015; Magrini et al. 2015; Smiljanic et al. 2016; Magrini et al. 2017; Duffau et al. 2017; Magrini et al. 2018a,b; Randich et al. 2020; Romano et al. 2021).

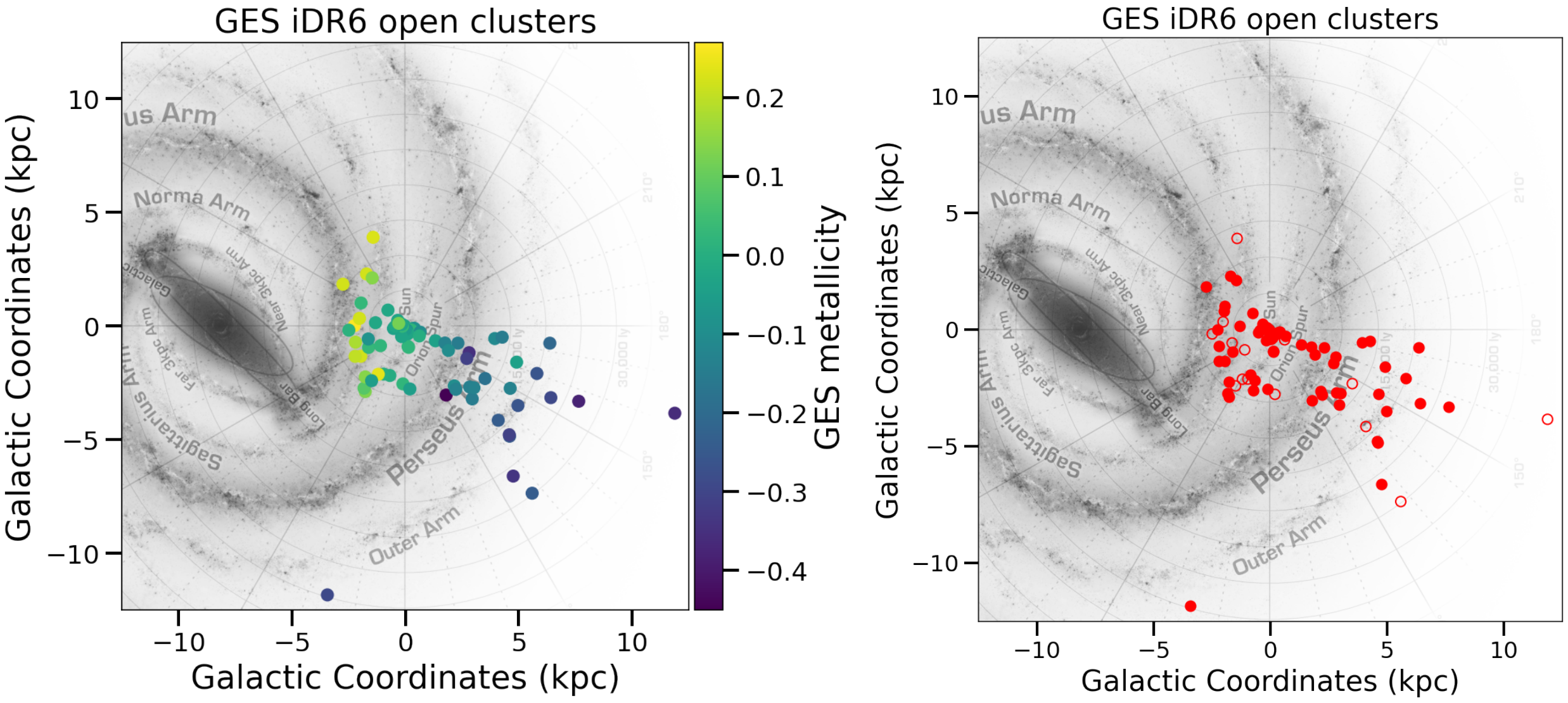

**Fig. 25.** Location in the Galactic disc of GES and archive science open clusters, colour-coded by their metallicity (left panel) and whether they belong to the GES sample (red filled circles) or the archive sample (red empty circles).

The last GES data release, which delivers homogenised products for 80 open clusters (GES plus archive), will represent an invaluable source for even more detailed and complete investigations of the metallicity and abundance distribution based on the complete cluster sample. In this context we also mention the work of Jackson et al. (2022) who combined GES and *Gaia* data to provide an astrometric determination of membership probabilities in most of these clusters. The membership probabilities are unbiased with respect to chemical abundances and photometric properties.

The data will be exploited in the coming months; we anticipate here a brief qualitative discussion on the radial metallicity distribution to show the potential of GES to address this topic. In Fig. 26 we plot metallicity ([Fe/H]) versus Galactocentric radius for the GES and ESO archive samples colour-coded by age. The plot shows the very well-known gradient, a decrease in the metallicity towards the outer regions of the disc. A few additional features should be noted: as reported in previous studies, the gradient seems to flatten out at $R_{GC} \geq 12$ kpc; the large age interval covered by the GES sample evidences that the gradient is prominent for the old (age $>$ 1 Gyr) and very old clusters, while it seems much shallower for the younger clusters, and the distribution becomes flat for the very youngest clusters ($\leq$ 100 Myr); the distribution in the inner parts of the disc appears bimodal, with all but one of the young clusters (younger than $\sim 200-300$ Myr) having a solar metallicity. The older clusters in the inner disc, again with one exception, all have supersolar metallicities; a scatter is present at Galactocentric distances between 11 and 13 kpc, with a few outer clusters showing metallicities well below the main trend. All these features certainly deserve further investigation; we thus refer to future papers for a thorough analysis, considering for example the cluster orbit and height above the plane, which may play a role.

### 6.3. Use of Gaia-ESO data from the community

Gaia-ESO data (spectra and/or the catalogue) published in the ESO archive have been extensively used by the broader community. A detailed description of the addressed science is not among the goals of this paper; we note, however, that a wide variety of topics have been covered, proving the value of Gaia-ESO as a public survey. An incomplete list of these topics includes the validation of analysis pipelines and/or the calibration of metallicity indicators (Boeche et al. 2018; Hanke et al. 2018; Usher et al. 2019; Steinmetz et al. 2020a); the determination of extinctions towards the Galactic thick disc and Bulge (Queiroz et al. 2020) and the study of the metal-poor population in the Bulge (Howes et al. 2014); the investigation of the properties, dark mass, metallicity variations, and extended turn-offs of globular clusters (Sollima et al. 2016; Baumgardt & Hilker 2018; Marino et al. 2018; Ferraro et al. 2018; Muñoz et al. 2021); the study of the evolution of lithium in stars, lithium rich stars, and the calibration of lithium abundances derived from lower resolution surveys (e.g. Zhou et al. 2019; Aoki et al. 2021; Gao et al. 2021); the investigation of the Galactic evolution of lithium (Grisoni et al. 2019); the measurement of the distribution of stellar spin axis orientations in the cluster NGC 2516 (Healy & McCullough 2020); open cluster membership and population studies (e.g. Fritzewski et al. 2019; Grasser et al. 2021).

## 7. The Gaia-ESO Survey science potential

As discussed in the previous section, the GES has already enabled novel and impact results to be obtained. However, its great science potential has not been fully exploited yet; the complete iDR6 dataset, along with *Gaia* eDR3 and future DR3 data and asteroseimology, will offer the community a great opportunity to address key topics in the stellar and Galactic archaeology science areas.

The GES iDR6 will enable further detailed insights in the general field of Galactic archaeology. Specifically, GES data will allow velocity-chemistry-position space to be probed, with a consistent well-defined selection function for the MW fields. This will enable analysis at many different levels by many communities. GES provides a clean homogeneous sample of Bulge, thin disc, thick disc, and halo turnoff stars, a unique product with targets probing larger distances than other surveys. Combining *Gaia* astrometry with the abundances and element ratios derived by GES will yield the age distribution function(s) for the Milky Way, delivering a robust contribution to the relative importance of assembly and accretion, and star formation histories. We highlight in particular the scientific potential of the the UVES parallel survey, which used UVES to target relatively bright FG-type stars during the Giraffe high-latitude survey. The colour–magnitude selection of this UVES parallel sample is defined to be unbiased against age and metallicity, and to provide a sample of all (accessible) stellar populations. We note that these stars are all observed and analysed consistently, delivering many chemical elements across a wide range of abundances.

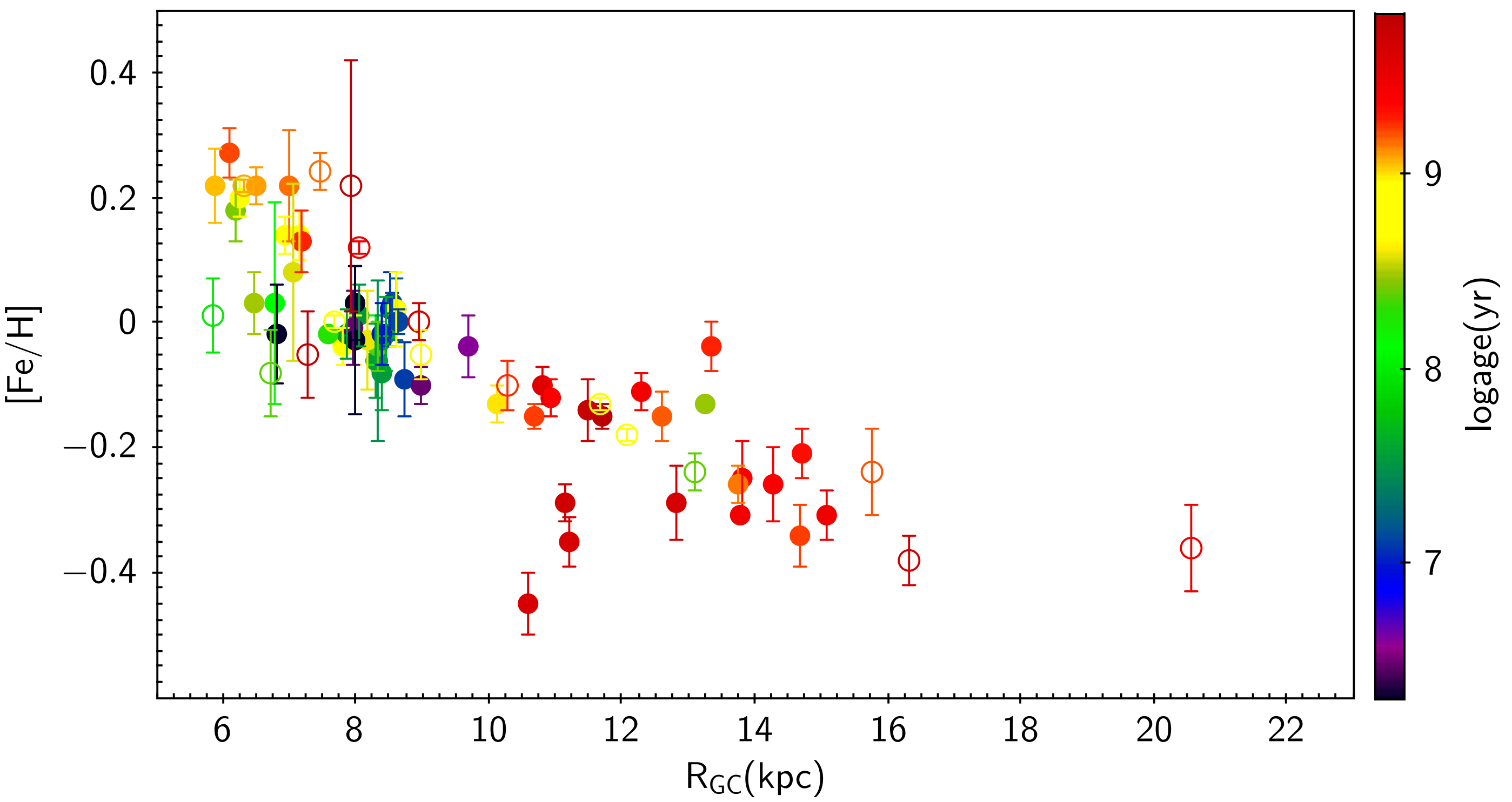

**Fig. 26.** [Fe/H] vs Galactocentric distance distribution of the GES (filled circles) and archive (open circles) OCs. The symbols are colour-coded by age.

Figure 27, which shows the iDR6 [Fe/H] distribution of our UVES parallel sample, confirms that we have achieved this goal. The sample contains stars covering the range $-3.5 <$[Fe/H]$<$ $+0.5$, covering all accessible populations. The super-solar tail (possible migration), the broad range of the thin disc with its rapid cut-off above solar, the thick disc extending down below $-1$ dex, and the halo extending down smoothly to very low abundances are all apparent. The sample is also in an excellent magnitude range for *Gaia*, so we have superb 6D phase-space data, complemented by the 6–8 extra dimensions of stellar parameters and abundances.

GES has also delivered a unique dataset of lithium abundances thanks to the Giraffe HR15N and UVES spectra (see Fig. 9). Lithium abundances or upper limits are available for more than 38,000 targets (about 1/3 of the full sample). Whilst this number is lower than that delivered by other spectroscopic surveys (e.g. GALAH), we note that targets with measured lithium cover all the evolutionary phases in the HR diagram well, noticeably including the PMS phases (see Fig. 28). This represents an invaluable source for lithium investigations themselves, but also for the determination of ages and population studies, and for models of stellar physics and evolution. As summarised in Sect. 6, a good number of GES publications focusing on lithium have already been published (see also Magrini et al. 2021c); however, the full lithium dataset available in iDR6 will allow many new studies to be performed (see e.g. the discussion in Randich & Magrini 2021).

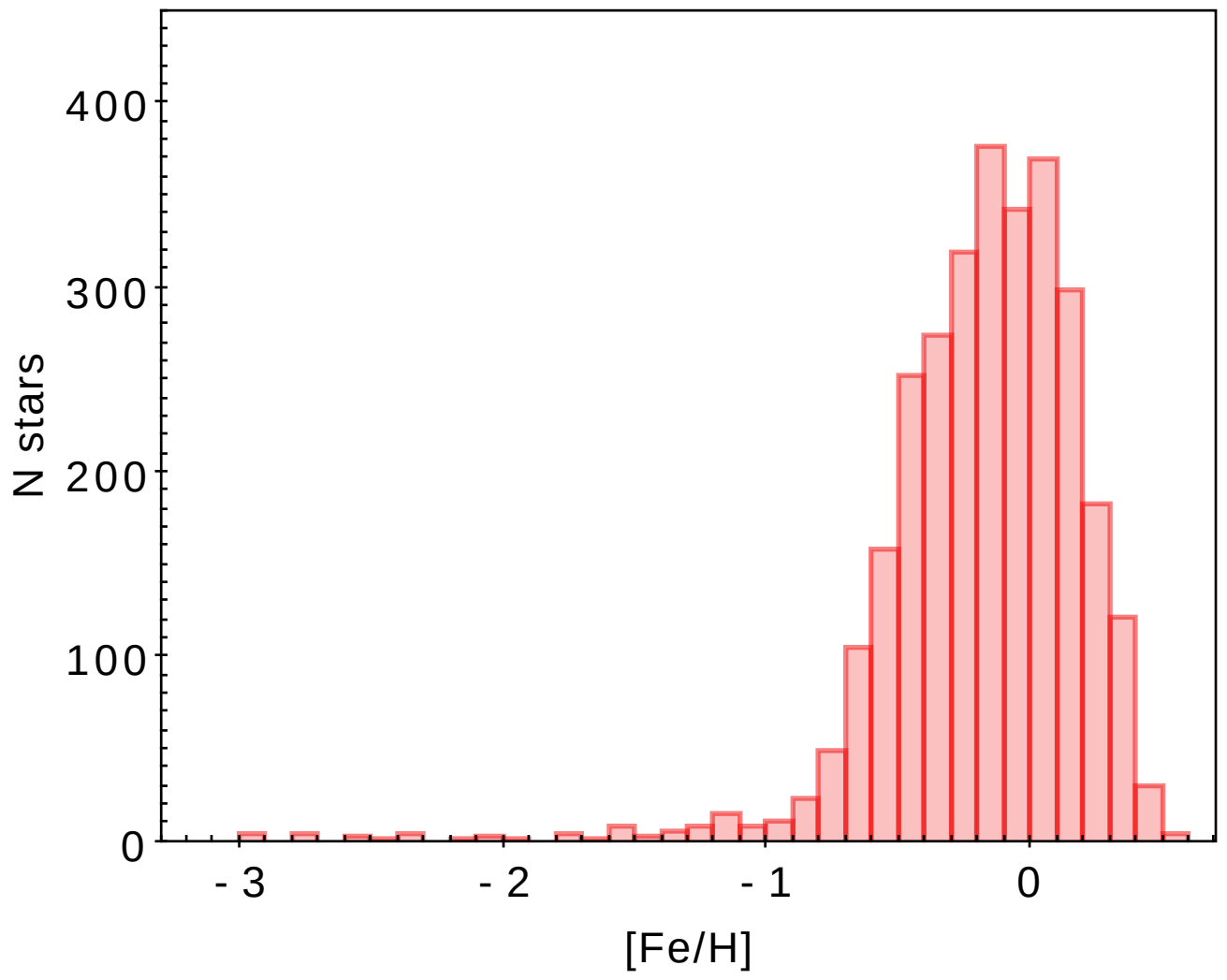

**Fig. 27.** [Fe/H] distribution of the UVES MW sample.

The cluster dataset represents a key resource for detailed studies of the formation and evolution of the thin disc, and for further investigations of the nucleosynthesis channels of many elements whose origins are not yet fully constrained. More in general, GES is yielding the first homogeneous set of RVs, abundances, rotational velocities, and ancillary stellar characteristics for such a large sample of clusters and cluster members, down to the faintest stars in nearby OCs and star forming regions and reaching, at the same time, some of the most massive stars in the Galaxy. This represents a standalone unique dataset that will allow a variety of important topics to be addressed. At the same time, the cluster dataset, again in combination with *Gaia*, will enable (and has already enabled) progress to be achieved on stellar physics and evolution.

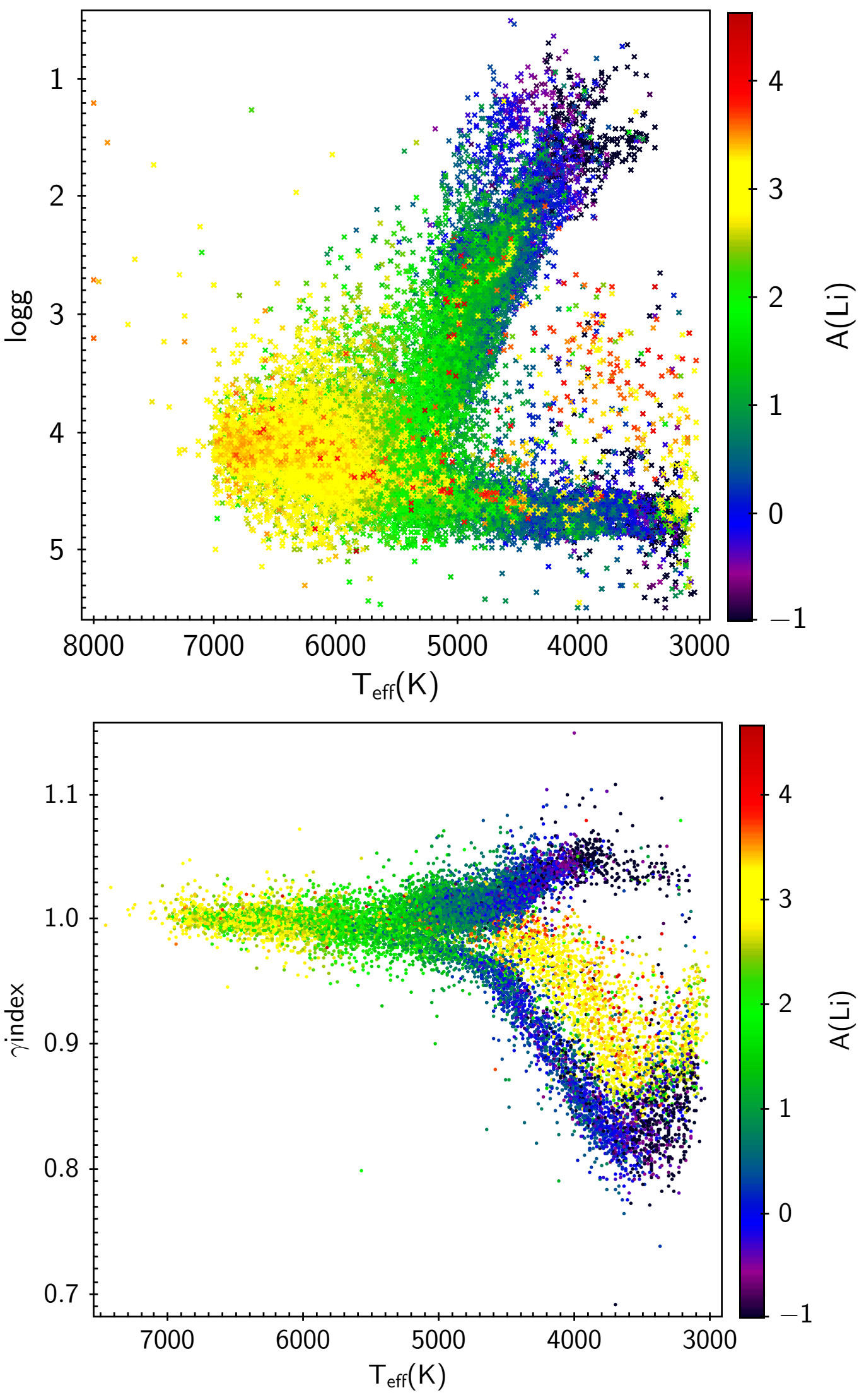

**Fig. 28.** logg vs $T_{eff}$ and $\gamma$ gravity index vs $T_{eff}$, colour-coded by lithium abundance.

One of the original goals of the cluster survey, the calibration of stellar models, including those that introduce non-standard physics, indeed maintains the highest legacy value (even more than initially foreseen); likewise, the use of GES open clusters, which sample the age range between 1 Myr and almost 10 Gyr, to calibrate age indicators and to build consistency between isochrones and asteroseismology and other age tracers (e.g. lithium, element ratios) will represent one of the long-lasting legacies of GES.

## 8. Conclusions and legacy of the Gaia-ESO Survey

After a ten-year effort by a large consortium, GES will come to an end with the delivery of the final catalogue to ESO and its publication in 2022. The catalogue will contain the advanced products for more than 100,000 stars, also covering targets retrieved from the ESO archive; for a large fraction of the sample the complete product dataset (including stellar parameters and abundances for several elements) have been derived. Membership information for the cluster targets will also be released.

Gaia-ESO, which is to date the only stellar spectroscopic survey performed on an 8m class telescope and is unique with respect to the other stellar surveys in several aspects, has encountered a number of challenges (some of them by design) and successfully overcome them. In particular, the multi-pipeline analysis strategy has implied the collaboration and interaction of several teams and researchers, as well as a well-devised parameter homogenisation strategy that has been refined at each analysis cycle. The comparison of the node and homogenised results clearly shows the success of our approach.

The science exploitation within the consortium has followed a bottom-up approach, with all survey Co-Investigators being allowed to propose science projects. This has led to many impact results and to more than 100 refereed publications covering the areas initially included in the proposal to ESO, but also presenting different topics and serendipitous discoveries. GES spectra and data have also been used by the broader community, often in combination with other datasets. The complete science potential of GES has not been fully exploited yet; the final dataset, along with *Gaia* eDR3 and future DR3 data and asteroseismic data as well, will allow the community to further address key topics in the stellar and Galactic archaeology science areas.

We conclude by highlighting that GES, in addition to its products and science, will leave a significant legacy for several years to come. Among the major, general legacy aspects we mention that GES target selection has employed data from public ESO photometric surveys such as the Vista Hemisphere Survey (VHS; McMahon et al. 2013) and the VST Photometric $H_{\alpha}$ Survey of the Southern Galactic Plane and Bulge (VPHAS+; Drew et al. 2014), hence adding value to those surveys. Moreover, GES has successfully linked spectra, *Gaia* astrometry, and asteroseismology, enhancing the science potential of each individual dataset. Whilst *Gaia* data were not needed for our target selection, processing our spectra in iDR6 considering *Gaia*'s astrometric distances and/or gravity measures as priors has improved the abundance precision, and will increase the phase-space dimensionality. GES has also exploited and extended *Gaia* benchmark stars, has produced and published a cleaned line list and lithium COGs, which will certainly be useful to the broader community. The Giraffe data reduction pipelines and processing methods have become the basis for the WEAVE and 4MOST pipelines, while UVES data reduction and identification of issues in the ESO pipeline have allowed significant collaborative enhancements to it. The calibration concept devised by GES and crucially included in our approach and strategy has now been introduced in all future stellar spectroscopic surveys.

More in general, in addition to these specific legacy aspects and the future science and papers, we believe that GES will be considered a source of inspiration (not only for the available data, but also the methods and approaches used) for those who will start complex projects.

**Note added in proof.** The ESO DR5 catatalogue has been published on May 16 2022: http://archive.eso.org/cms/eso-archive-news/Fifth-Gaia-ESO-data-release-astrophysical-parameters-of-about-115000-stars.html

*Acknowledgements.* Based on data products from observations made with ESO Telescopes at the La Silla Paranal Observatory under programmes ID 188.B-3002, 193-B-0936, and 197.B-1074. These data products have been processed by the Cambridge Astronomy Survey Unit (CASU) at the Institute of Astronomy, University of Cambridge, and by the FLAMES/UVES reduction team at INAF-Osservatorio Astrofisico di Arcetri. Public access to the data products is via the ESO Archive, and the Gaia-ESO Survey Data Archive, prepared and hosted by the Wide Field Astronomy Unit, Institute for Astronomy, University of Edinburgh, which is funded by the UK Science and Technology Facilities Council. This work was partly supported by the European Union FP7 programme through ERC grant number 320360 and by the Leverhulme Trust through grant RPG-2012-541. We acknowledge the support from INAF PRIN and Ministero dell' Universitá e della Ricerca (MUR) in the form of the grant "Premiale VLT 2012" and "Premiale Mitic". This work was partly supported by the INAF grant for mainstream projects: "Enhancing the legacy of the Gaia-ESO Survey for open cluster science". The project presented here benefited in development from discussions held during the Gaia-ESO workshops and conferences supported by the ESF (European Science Foundation) through the GREAT Research Network Programme. This research has made use of the SIMBAD database, operated at CDS, Strasbourg, France. R.Smiljanic acknowledges support from the National Science Centre, Poland (2014/15/B/ST9/03981). F.J.E. acknowledges financial support from the Spanish MINECO/FEDER through the grant AYA2017-84089 and MDM-2017-0737 at Centro de Astrobiología (CSIC-INTA), Unidad de Excelencia María de Maeztu, and from the European Union's Horizon 2020 research and innovation programme under Grant Agreement no. 824064 through the ESCAPE - The European Science Cluster of Astronomy & Particle Physics ESFRI Research Infrastructures project. T.B. was funded by the "The New Milky Way" project grant from the Knut and Alice Wallenberg Foundation. S.R.B. acknowledges support by the Spanish Government under grants AYA2015-68012-C2-2-P and PGC2018-093741-B-C21/C22 (MICIU/AEI/FEDER, UE). W. J. S. acknowledges CAPES for a PhD studentship. J.M.A. acknowledges support from the Spanish Government Ministerio de Ciencia e Innovación through grants AYA2013-40611-P, AYA2016-75931-C2-2-P, and PGC2018-095049-B-C22. T.M. and others from STAR institute, Liege, Belgium are grateful to Belgian F.R.S.-FNRS for support, and are also indebted for an ESA/PRODEX Belspo contract related to the *Gaia* Data Processing and Analysis Consortium and for support through an ARC grant for Concerted Research Actions financed by the Federation Wallonie-Brussels. This research has been partially supported by the ASI-INAF contract 2014-049-R.O: "Realizzazione attività tecniche/scientifiche presso ASDC" (PI Angelo Antonelli). V.A.acknowledges the support from Fundação para a Ciência e Tecnologia (FCT) through Investigador FCT contract nr. IF/00650/2015/CP1273/CT0001. AJK acknowledges support by the Swedish National Space Agency (SNSA). AB acknowledges support by ANID, – Millennium Science Initiative Program – NCN19_171, and FONDECYT regular 1190748. E. M. acknowledges financial support from the Spanish State Research Agency (AEI) through project MDM-2017-0737 Unidad de Excelencia "María de Maeztu" - Centro de Astrobiología (CSIC-INTA). T.Z. acknowledges financial support of the Slovenian Research Agency (research core funding No. P1-0188) and the European Space Agency (Prodex Experiment Arrangement No. C4000127986). P.J. acknowledges support FONDECYT Regular 1200703. The work of I.N. is partially supported by the Spanish Government Ministerio de Ciencia, Innovación y Universidades under grant PGC2018-093741-B-C21 (MICIU/AEI/FEDER, UE). Funding for this work has been provided by the ARC Future Fellowship FT160100402. CAP acknowledges financial support from the Spanish Government through research grants MINECO AYA 2014-56359-P, MINECO AYA2017-86389-P, and MICINN PID2020-117493GB-I00. S.F. was supported by the grants 2011-5042 and 2016- 03412 from the Swedish Research Council and the project grant "The New Milky Way" from the Knut and Alice Wallenberg Foundation. CASU is supported through STFC grants: ST/H004157/1, ST/J00541X/1, ST/M007626/1, ST/N005805/1, ST/T003081/1. Work reported here benefited from support through the GREAT-ITN FP7 project Grant agreement ID: 264895. DKF acknowledges funds from the Alexander von Humboldt Foundation in the framework of the Sofja Kovalevskaja Award endowed by the Federal Ministry of Education and Research and the grant 2016-03412 from the Swedish Research Council. A.H. acknowledges support from the Spanish Government Ministerio de Ciencia e Innovación and ERD Funds through grants PGC-2018-091 3741-B-C22 and CEX2019-000920-S. X.F. acknowledge the support of China Postdoctoral Science Foundation 2020M670023. M. L. L. Dantas acknowledges the Polish NCN grant number 2019/34/E/ST9/00133. Part of this work was funded by the Deutsche Forschungsgemeinschaft (DFG, German Research Foundation) – Project-ID 138713538 – SFB 881 ("The Milky Way System", subproject A09). MZ acknowledge support from the National Agency for Research and Development (ANID) grants: FONDECYT Regular 1191505, Millennium Institute of Astrophysics ICN12-009, BASAL Center for Astrophysics and Associated Technologies AFB-170002. R.B. acknowledges support from the project PRIN-INAF 2019 "Spectroscopically Tracing the Disk Dispersal Evolution". HMT acknowledges financial support from the Agencia Estatal de Investigación of the Ministerio de Ciencia, Innovación y Universidades through projects PID2019-109522GB-C51,54/AEI/10.13039/501100011033, and the Centre of Excellence "María de Maeztu" award to Centro de Astrobiología (MDM-2017-0737). JIGH acknowledges financial support from the Spanish Ministry of Science and Innovation (MICINN) project AYA2017-86389-P, and also from the Spanish MICINN under 2013 Ramøn y Cajal program RYC-2013-14875. V.P.D. is supported by STFC Consolidated grant ST/R000786/1. N.L. acknowledges financial support from "Programme National de Physique Stellaire" (PNPS) and the "Programme National Cosmology et Galaxies (PNCG)" of CNRS/INSU, France. A. R. C. is supported in part by the Australian Research Council through a Discovery Early Career Researcher Award (DE190100656). Parts of this research were supported by the Australian Research Council Centre of Excellence for All Sky Astrophysics in 3 Dimensions (ASTRO 3D), through project number CE170100013. PSB is Supported by the Swedish Research Council through individual project grants with contract Nos. 2016-03765 and 2020-03404. AM acknowledges funding from the European Research Council (ERC) under the European Union's Horizon 2020 research and innovation programme (grant agreement No. 772293 - project ASTEROCHRONOMETRY). JP was supported by the project RVO: 67985815. E.D.M. acknowledges the support from FCT through the research grants UIDB/04434/2020 & UIDP/04434/2020 and through Investigador FCT contract IF/00849/2015/CP1273/CT0003. This work was (partially) supported by the Spanish Ministry of Science, Innovation and University (MICIU/FEDER, UE) through grant RTI2018-095076-B-C21, and the Institute of Cosmos Sciences University of Barcelona (ICCUB, Unidad de Excelencia 'María de Maeztu') through grant CEX2019-000918-M. SLM acknowledges the support of the UNSW Scientia Fellowship program and the Australian Research Council through Discovery Project grant DP180101791. GT acknowledges financial support of the Slovenian Research Agency (research core funding No. P1-0188) and the European Space Agency (Prodex Experiment Arrangement No. C4000127986). S.G.S acknowledges the support from FCT through Investigador FCT contract nr. CEECIND/00826/2018 and POPH/FSE (EC). H.G.L. acknowledges financial support by the Deutsche Forschungsgemeinschaft (DFG, German Research Foundation) – Project-ID 138713538 – SFB 881 ("The Milky Way System", subproject A04). This work was (partially) supported by the Spanish Ministry of Science, Innovation and University (MICIU/FEDER, UE) through grant RTI2018-095076-B-C21, and the Institute of Cosmos Sciences University of Barcelona (ICCUB, Unidad de Excelencia 'María de Maeztu') through grant CEX2019-000918-M. T.K. is supported by STFC Consolidated grant ST/R000786/1. MV acknowledges the support of the Deutsche Forschungsgemeinschaft (DFG, project number: 428473034). T.M. is supported by a grant from the Fondation ULB. We acknowledge financial support from the Universidad Complutense de Madrid (UCM) and by the Spanish Ministerio de Ciencia, Innovación y Universidades, Ministerio de Economía y Competitividad, from project AYA2016-79425-C3-1-P and PID2019-109522GB-C5[4]/AEI/10.13039/501100011033. U.H. acknowledges support from the Swedish National Space Agency (SNSA/Rymdstyrelsen). D.G. gratefully acknowledges support from the Chilean Centro de Excelencia en Astrofísica y Tecnologías Afines (CATA) BASAL grant AFB-170002. D.G. also acknowledges financial support from the Dirección de Investigación y Desarrollo de la Universidad de La Serena through the Programa de Incentivo a la Investigación de Académicos (PIA-DIDULS). A. Lobel acknowledges support in part by the Belgian Federal Science Policy Office under contract No. BR/143/A2/BRASS. We acknowledge financial support from the Universidad Complutense de Madrid (UCM) and by the Spanish Ministerio de Ciencia, Innovación y Universidades, Ministerio de Economía y Competitividad, from project AYA2016-79425-C3-1-P and PID2019-109522GB-C5[4]/AEI/10.13039/501100011033. AM acknowledges the support from the Portuguese Fundação para a Ciência e a Tecnologia (FCT) through the Portuguese Strategic Programme UID/FIS/00099/2019 for CENTRA. TM acknowledges financial support from the Spanish Ministry of Science and Innovation (MICINN) through the Spanish State Research Agency, under the Severo Ochoa Program 2020-2023 (CEX2019-000920-S). EJA acknowledges funding from the State Agency for Research of the Spanish MCIU through the "Center of Excellence Severo Ochoa" award to the Instituto de Astrofisica de Andalucia (SEV-2017-0709).

[1] INAF - Osservatorio Astrofisico di Arcetri, Largo E. Fermi, 5, 50125, Firenze, Italy
e-mail: sofia.randich@inaf.it
[2] Institute of Astronomy, University of Cambridge, Madingley Road, Cambridge CB3 0HA, United Kingdom
[3] Astrophysics Group, Keele University, Keele, Staffordshire ST5 5BG, United Kingdom
[4] Institute of Theoretical Physics and Astronomy, Vilnius University, Sauletekio av. 3, LT-10257 Vilnius, Lithuania
[5] Instituto de Astrofísica de Andalucía, CSIC, Glorieta de la Astronomía SNR, Granada 18008, Spain
[6] Instituto de Astrofísica de Canarias, Vía Láctea SNR, E-38205 La Laguna, Tenerife, Spain
[7] Departamento de Astrofísica, Universidad de La Laguna, E-38205 La Laguna, Tenerife, Spain
[8] Lund Observatory, Department of Astronomy and Theoretical Physics, Box 43, SE-22100 Lund, Sweden
[9] ROB - Royal Observatory of Belgium, Ringlaan 3, B-1180 Brussels, Belgium

[10] INAF - Osservatorio di Astrofisica e Scienza dello Spazio, via P. Gobetti 93/3, 40129 Bologna, Italy
[11] INAF - Osservatorio Asronomico di Palermo, Piazza del Parlamento, 1 90134 Palermo, Italy
[12] GEPI, Observatoire de Paris, PSL Research University, CNRS, Univ. Paris Diderot, Sorbonne Paris Cité, 61 avenue de l'Observatoire, 75014, Paris, France
[13] Institute for Astronomy, Royal Observatory, University of Edinburgh, Blackford Hill, Edinburgh EH9 3HJ, UK
[14] Observational Astrophysics, Division of Astronomy and Space Physics, Department of Physics and Astronomy, Uppsala University, Box 516, 75120 Uppsala, Sweden
[15] Dipartimento di Fisica e Astronomia, Sezione Astrofisica, Universitá di Catania, via S. Sofia 78, 95123, Catania, Italy
[16] Space Science Data Center - Agenzia Spaziale Italiana, via del Politecnico, s.n.c., I-00133, Roma, Italy
[17] Université Côte d'Azur, Observatoire de la Côte d'Azur, CNRS, Laboratoire Lagrange, Bd de l'Observatoire, CS 34229, 06304 Nice cedex 4, France
[18] Nicolaus Copernicus Astronomical Center, Polish Academy of Sciences, ul. Bartycka 18, 00-716, Warsaw, Poland
[19] Institut d'Astronomie et d'Astrophysique, Université Libre de Bruxelles, CP 226, Boulevard du Triomphe, B-1050 Bruxelles, Belgium
[20] Faculty of Mathematics and Physics, University of Ljubljana, Jadranska 19, 1000 Ljubljana, Slovenia
[21] Australian Academy of Science, Box 783, Canberra ACT 2601, Australia
[22] Rudolf Peierls Centre for Theoretical Physics, Clarendon Laboratory, Parks Road, Oxford OX1 3PU, United Kingdom
[23] GEPI, Observatoire de Paris, Université PSL, CNRS, 5 Place Jules Janssen, 92190 Meudon, France
[24] Department of Physics & Astronomy, University College London, Gower Street, London WC1E 6BT, United Kingdom
[25] Institute for Astronomy, University of Edinburgh, Blackford Hill, Edinburgh EH9 3HJ UK
[26] Departamento de Física Aplicada, Facultad de Ciencias, Universidad de Alicante, 03690 San Vicente del Raspeig, Alicante, Spain
[27] European Space Agency (ESA), European Space Research and Technology Centre (ESTEC), Keplerlaan 1, 2201 AZ Noordwijk, The Netherlands
[28] Max-Planck-Institut für Astronomie, Königstuhl 17, D-69117 Heidelberg, Germany
[29] INAF - Osservatorio Astronomico di Padova, Vicolo dell'Osservatorio 5, I-35122, Padova, Italy
[30] Instituto de Física y Astronomía, Facultad de Ciencias, Universidad de Valparaíso, Chile
[31] Núcleo Milenio Formación Planetaria - NPF, Universidad de Valparaíso, Chile
[32] Niels Bohr International Academy, Niels Bohr Institute, Blegdamsvej 17, DK-2100 Copenhagen Ø, Denmark
[33] INAF - Osservatorio Astronomico di Roma, Via Frascati 33, I-00040 Monte Porzio Catone (Roma), Italy
[34] Department of Physics and Astronomy, University of Padova, v. dell'Osservatorio 2, 35122, Padova, Italy
[35] School of Physics & Astronomy, Monash University, Wellington Road, Clayton 3800, Victoria, Australia
[36] INAF - Osservatorio Astrofisico di Catania, Via S. Sofia 78, 95123 Catania, Italy
[37] Núcleo de Astronomía, Facultad de Ingeniería y Ciencias, Universidad Diego Portales, Av. Ejército 441, Santiago, Chile
[38] Department of Astronomy, Stockholm University, AlbaNova University Center, SE-106 91 Stockholm, Sweden
[39] ESO - European Organisation for Astronomical Research in the Southern Hemisphere, Alonso de Córdova 3107, Vitacura, 19001 Casilla, Santiago de Chile, Chile
[40] Departamento de Ciencias Fisicas, Universidad Andres Bello, Fernandez Concha 700, Las Condes, Santiago, Chile
[41] Instituto de Astrofísica e Ciências do Espaço, Universidade do Porto, CAUP, Rua das Estrelas, 4150-762 Porto, Portugal
[42] Observatório Nacional - MCTI (ON), Rua Gal. José Cristino 77, São Cristóvão, 20921-400, Rio de Janeiro, Brazil
[43] Department of Chemistry and Physics, Saint Mary's College, Notre Dame, IN 46556, USA
[44] Leibniz-Institut für Astrophysik Potsdam (AIP), An der Sternwarte 16, 14482 Potsdam, Germany
[45] Centro de Astrobiología (CSIC-INTA), Departamento de Astrofísica, campus ESAC. Camino bajo del castillo SNR. 28 692 Villanueva de la Cañada, Madrid, Spain.
[46] Departamento de Física de la Tierra y Astrofísica & IPARCOS-UCM (Instituto de Física de Partículas y del Cosmos de la UCM), Facultad de Ciencias Físicas, Universidad Complutense de Madrid, E-28040 Madrid, Spain
[47] Space Sciences, Technologies, and Astrophysics Research (STAR) Institute, Université de Liège, Quartier Agora, Bât B5c, Allée du 6 août, 19c, 4000 Liège, Belgium
[48] Laboratoire d'astrophysique de Bordeaux, Univ. Bordeaux, CNRS, B18N, allée Geoffroy Saint-Hilaire, 33615 Pessac, France
[49] Centro de Astrobiología (CSIC-INTA), Carretera de Ajalvir km 4, E-28850 Torrejón de Ardoz, Madrid, Spain
[50] Departamento de Astronomía, Casilla 160-C, Universidad de Concepción, Concepción, Chile
[51] INAF - Osservatorio Astrofisico di Torino, via Osservatorio 20, I-10025 Pino Torinese, Italy
[52] Stellar Astrophysics Centre, Department of Physics and Astronomy, Aarhus University, Ny Munkegade 120, DK-8000 Aarhus C, Denmark
[53] University of Vienna, Dept. Astrophysics, Türkenschanzstrasse 17, 1180 Vienna, Austria
[54] Institut de Ciències del Cosmos (ICCUB), Universitat de Barcelona (IEEC-UB), Martí i Franquès 1, E-08028 Barcelona, Spain
[55] Theoretical Astrophysics, Department of Physics and Astronomy, Uppsala University, Box 516, SE-751 20 Uppsala, Sweden
[56] Centro de Astrobiología (INTA-CSIC), Camino Bajo del Castillo SNR, 28692, Villanueva de la Cañada, Madrid, Spain
[57] Massachusetts Institute of Technology, Kavli Institute for Astrophysics and Space Research, 77 Massachusetts Ave., Cambridge, MA 02139, USA
[58] SISSA, via Bonomea, 265 - 34136 Trieste, Italy
[59] Dep. of Physics, Sapienza, University of Roma, Roma, Italy
[60] Research School of Astronomy & Astrophysics, Australian National University, ACT 2611, Australia
[61] Jeremiah Horrocks Institute, University of Central Lancashire, Preston PR1 2HE, United Kingdom
[62] INAF - Osservatorio Astronomico di Trieste, Via G.B Tiepolo, 11 I-34143 Trieste, Italy
[63] Université de Strasbourg, CNRS, Observatoire Astronomique de Strasbourg, UMR 7550, F-67000 Strasbourg, France
[64] Astronomy Department, Indiana University, 727 East 3rd St, Bloomington, IN 47405, USA
[65] The Kavli Institute for Astronomy and Astrophysics at Peking University, 100871, Beijing, PR China
[66] Instituto de Investigación Multidisciplinario en Ciencia y Tecnología, Universidad de La Serena, Avenida Raúl Bitrán SNR, La Serena, Chile
[67] Departamento de Astronomía, Facultad de Ciencias, Universidad de La Serena, Av. Juan Cisternas 1200, La Serena, Chile
[68] Max-Planck-Institute for Ex. Physics, Giessenbachstr.1, 85748 Garching, Germany
[69] Astronomisches Rechen-Institut, Zentrum für Astronomie der Universität Heidelberg, Mönchhofstr. 12–14, 69120 Heidelberg, Germany
[70] Section of Astrophysics, Astronomy and Mechanics, Department of Physics, National and Kapodistrian University of Athens, GR15784, Athens, Greece
[71] IAASARS, National Observatory of Athens, GR15236, Penteli, Greece
[72] Materials Science and Applied Mathematics, Malmö University, SE-205 06 Malmö, Sweden

[73] Institut UTINAM, CNRS UMR6213, Univ. Bourgogne Franche-Comté, OSU THETA Franche-Comté-Bourgogne, Observatoire de Besançon, BP 1615, 25010 Besançon Cedex, France
[74] School of Physics, University of New South Wales, Sydney 2052, Australia
[75] Dipartimento di Fisica e Astronomia, Università degli Studi di Bologna, Via Gobetti 93/2, I-40129 Bologna, Italy
[76] CENTRA, Faculdade de Ciências, Universidade de Lisboa, Ed. C8, Campo Grande, 1749-016 Lisboa, Portugal
[77] Departamento de Física e Astronomia, Faculdade de Ciências da Universidade do Porto, Portugal
[78] Department of Astronomy, University of Geneva, 51 chemin Pegasi, 1290 Versoix, Switzerland
[79] Université de Toulouse, Observatoire Midi-Pyrénées, CNRS, IRAP, 14 av. E. Belin, F-31400 Toulouse
[80] Astronomical Institute, CAS, Boční II 1401, 141 00 Prague 4, Czech Republic
[81] Department of Theoretical Physics and Astrophysics, Faculty of Science, Masaryk University, Kotlarska 2, 611 37 Brno, Czech Republic
[82] Physics Department, Imperial College London, Prince Consort Road, London SW7 2BZ, United Kingdom
[83] Zentrum für Astronomie der Universität Heidelberg, Landessternwarte, Königstuhl 12, 69117 Heidelberg, Germany
[84] University of Surrey, Physics Department, Guildford, GU2 7XH, UK
[85] Mullard Space Science Laboratory, University College London, Holmbury St Mary, Dorking, Surrey, RH5 6NT, United Kingdom
[86] Astronomical Observatory, Institute of Theoretical Physics and Astronomy, Vilnius University, Sauletekio av. 3, 10257 Vilnius, Lithuania
[87] Armagh Observatory and Planetarium, College Hill, Armagh BT61 9DG, United Kingdom
[88] Department of Physics & Astronomy, Johns Hopkins University, Baltimore, MD 21218, USA
[89] Institute of Astrophysics, Pontificia Universidad Católica de Chile, Av. Vicuña Mackenna 4860, Macul, Santiago, Chile
[90] Sorbonne Université, CNRS, UPMC, UMR7095 Institut d'Astrophysique de Paris, 98bis Bd. Arago, F-75014 Paris, France
[91] Department of Physics and Astronomy, Macquarie University, Sydney, NSW 2109, Australia

## Appendix A: UVES data reduction and radial velocities

As discussed in detail in Sacco et al. (2014), all UVES data were reduced using the ESO pipeline (see eso.org/sci/software/pipeline/) version 5.5.2 in combination with a pipeline developed by the INAF-Arcetri group. Specifically, the ESO pipeline was used to carry out the most standard steps of the data reduction process (bias subtraction, division by a flat field, spectra extraction, and wavelength calibration), while the Arcetri pipeline performs the sky subtraction, the barycentric correction, and the co-addition of multiple spectra of the same stars. Furthermore, the Arcetri group carried out a quality control of the spectra by means of an automatic software and visual inspection (see Sacco et al. 2014 for more details).

The RVs for the UVES spectra were calculated by cross-correlating the observed spectra with a library of templates downgraded to the UVES resolution. Since this method is not efficient for measuring the RVs of early-type stars (A, B, and O types), for this subgroup we used a different approach based on the spectral fitting described in Blomme et al. (2022). The RVs measured for early-type stars are homogenised with the RVs measured from the other UVES spectra and from the GIRAFFE spectra by the working group in charge of the homogenisation, as described in Hourihane et al. (in preparation).

As discussed in Jackson et al. (2015) and Sacco et al. (2014), for most of the spectra the major source of error is the uncertainty on the zero point of the wavelength calibration. This component was reduced for the Giraffe observations by collecting arc lamp spectra simultaneously with each OB, but given the limited number of fibres available for UVES (6 to 8 depending on the set-up), we decided not to take the simultaneous arc-lamp and perform a standard wavelength calibration using the arc-lamp taken in daytime.

After iDR4, we started correcting the zero point of wavelength calibration using the emission lines from the sky spectrum. After the introduction of this correction the median error on RVs is 0.32 km/s. The final errors on the RV of single stars also depend on the projected rotational velocities, on the spectral type of the stars, and on the SNR.

## Appendix B: HR15N radial velocity precision

Estimating the RV precision from Giraffe HR15N spectra is particularly critical since these data are used for cluster internal kinematics investigations.

### Appendix B.1: Method

As described in Jackson et al. (2015), the empirical measurement precision $E_{\rm RV} = \Delta{\rm RV}\sqrt{2}$ is characterised as a Student's t-distribution scaled by an empirical uncertainty $S_{\rm RV}$, which depends on the SNR of the spectrum and projected equatorial velocity ($v \sin i$) of the star.

The scaling constant for 'short-term repeats' (spectra taken consecutively in an OB with the same instrument set-up and wavelength calibration) is given by

$$S_{\rm RV,0} = B\,\frac{(1 + [v \sin i/C)^2]^{3/4}}{\rm SNR}\,, \qquad ({\rm B.1})$$

where $B$ is an empirically determined parameter that depends on the intrinsic stellar spectrum (largely characterised by the effective temperature $T_{\rm eff}$) and $C$ is a function of the spectrograph resolving power.

For 'long-term repeats' (e.g. spectra taken in different OBs), there is an additional contribution to the measurement uncertainty, labelled $A$, which is due to variations in instrument set-up and wavelength calibration, and which adds in quadrature to the short-term uncertainty, such that the distribution of $E_{\rm RV}$ for long-term repeats is characterised by

$$S_{\rm RV} = \sqrt{A^2 + S^2_{\rm RV,0}}\,, \qquad ({\rm B.2})$$

Jackson et al. (2015) used data for nine clusters in iDR2/3 to determine empirical values for $A$, $B$, and $C$. Both $A$ and $C$ were treated as constants over the whole analysis. The empirical analysis was repeated here using data for the 68 clusters from iDR6 to determine appropriate expressions for $A$, $B$, and $C$. This required two modifications to the analysis: the use of a reduced value of $v \sin i$ to account for changes in instrument resolving power over time, and the scaling of constant $A$ recast as $A = A_0 + A_1/(SNR)$ in order to fit data from more distant clusters where targets with lower levels of $SNR$ show additional uncertainty in long-term repeats.

### Appendix B.2: Reduced projected equatorial velocity

The GES pipeline used to estimate $v \sin i$ for iDR6 data assumes a fixed spectral resolving power, $R = 17000$ for HR15N. In practice, the effective resolution of spectra observed using the HR15N grating, measured from the line width of arc-lamp spectra varied with time over the period of the GES observations, falling from $R \sim 15000$ in January 2012 to $R \sim 13000$ in February 2015, after which a new focusing procedure for the instrument produced a consistent level of $R \sim 17000$. As a result, the pipeline values of projected equatorial velocity ($V_{\rm ROT}$) are higher than the true value of $v \sin i$ for observations made before February 2015. The effect is most pronounced for the slowest rotating stars where a $V_{\rm ROT}$ of $\sim 12\,{\rm km\,s^{-1}}$ is reported. To correct for the reduction in $R$ below the expected level a reduced value of $v \sin i$ was used to determine the effect of rotational velocity on the $RV$ measurement precision,

$$v \sin i = \sqrt{V^2_{\rm ROT} - V^2_{\rm cor}} \;\; {\rm for}\; V_{\rm ROT} > V_{\rm cor}\,, \qquad ({\rm B.3})$$

where

$$V_{\rm cor} = 0.895c\,\sqrt{\frac{1}{R^2} - \frac{1}{17000^2}}\,, \qquad ({\rm B.4})$$

$c$ is the speed of light, and $R$ the resolving power when the target was observed.

### Appendix B.3: Target temperatures for RV precision.

Jackson et al. (2015) used the $T_{\rm eff}$ values reported by the GES working groups to determine the dependence of $B$ in Eq. B.1 on $T_{\rm eff}$. At the time of this analysis the recommended temperatures were not yet available; we thus derived temperatures determined from the spectral indices, as described by Damiani et al. (2014). Gravity ($\gamma$), temperature ($\tau$), and metallicity ($\zeta$) indices were measured from the normalised stacked spectra for all iDR6 targets and used to calculate metallicity-corrected temperature $T^Z_{\rm eff}$ and uncorrected temperature $T^I_{\rm eff}$ using the expressions given in Damiani et al. (2014), which are valid over the temperature range 4000 $< T_{\rm eff} <$ 7000 K.

As a check, we performed the analysis again using the WG15 recommended $T_{\rm eff}$ values, comparing the scaling constant for RV

precision for about 20,000 stars. As expected, the differences in RV precision between the methods depend on the SNR. We found when selecting spectra with SNR> 10 that the average absolute difference in precision is very small, 0.02 km/s, with 2.2% of the targets showing a difference > 0.1 km/s. If we selected stars with SNR> 20 the average absolute difference would be 0.01 km/s.

### Appendix B.4: Fitting empirical parameters

Data from iDR6 nightly spectra (A. Hourihane, private communication) included 34176 short-term repeats and 4,478 long-term repeats (with separation <3.2 days) that were analysed to determine the empirical constants $A$, $B$, and $C$. The results of the analysis are summarised in Figs. B.1 and B.2.

Figure B.1a shows the empirical variation of constant $B$ with temperature. This approximates to a tanh type curve transitioning from a low temperature level of $B$ =3.8 km s$^{-1}$ below 4000 K to a high temperature level of $B$ = 7.9 km s$^{-1}$ above 6000 K. For the purpose of calculating parameter $B$ it is necessary to estimate $T_{\rm eff}$ for targets in the range 4000 < $T_{\rm eff}$ < 6000 K. Outside this range it is sufficient to determine whether the target $T_{\rm eff}$ is in the lower zone (< 4000 K ) or the upper zone (> 6000 K ). To calculate the RV precision, $T^{Z}_{\rm eff}$ was used in preference to $T^{I}_{\rm eff}$. In two % of the cases where $\tau$ was unresolved, these were assumed to be hot stars and a value of 8000 K was assumed. The results of this analysis gives (in units of km/s)

$$
\begin{aligned}
A &= \max(0.26, 0.04 + 13.7/\mathrm{SNR})\,; \qquad \text{(B.5)}\\
B &= 5.85 + 2.07\tanh\left((T_{\rm eff} - 5000)/500\right);\\
C &= 0.895\, c/R.
\end{aligned}
$$

The extent to which the empirical model describes the uncertainties in the short-term repeats is illustrated in Figs. B.1b and B.1c. The normalised measurement uncertainty $E_{\rm RV}/S_{\rm RV}$ is best described by a $\nu = 3$ Student's t-distribution and is shown in Fig. B.1d. This representation is robustly followed by the data to at least $\pm 3S_{\rm RV,0}$, containing $\sim$ 95 % of the probability distribution), with some evidence that it works to $\pm 4S_{\rm RV,0}$ ($\sim$ 97 % of the probability distribution). Figure B.2 summarises the analysis of the long-term repeats. Figure B.2a shows how the distribution of $E_{\rm RV}$ broadens compared to the short-term repeats. It is this broadening that is accounted for by the addition of the $A$ term in Eq. B.2. Figure B.2b shows how $A$ varies as a function of time between repeat observations. Only data for separations smaller than 3.2 days are used in the final analysis in order to mitigate the effects of binary systems; Fig. B.2c indicates how $A$ increases as SNR$^{-1}$ once the SNR falls below $\sim$ 50. Figure B.2d shows the final cumulative distribution function of $E_{\rm RV}/S_{\rm RV}$. Like Fig. 1d, this is also reasonably represented by a Student's t-distribution with $\nu = 3$, though there is some evidence that it slightly underestimates the contribution of the distribution tails; however, the tails of the uncertainty distribution are clearly enhanced with respect to a Gaussian with $\sigma = S_{\rm RV}$ (e.g. a 68 confidence error bar is $\sim \pm 1.2S_{\rm RV}$, a 90 % error bar is $\pm 2.6S_{\rm RV}$, whilst a 95 % cent error bar is $\sim \pm 3.2S_{\rm RV}$).

The value of $A$ in Eq. B.6 applies to a single OB comprising of two exposures. For stacked spectra comprising of a number of repeat OBs a constant $A/\sqrt{n}$ is used, where $n$ is the number of stacked OBs.

### Appendix B.5: Table of RV precision data

Radial velocity precision was calculated for 37930 cluster targets observed using Giraffe HR15N using iDR6 stacked spectra and associated metadata (A. Hourihane, private communication). The results are listed in the table available at the CDS.

The median RV precision is 0.43 km/s, while about 34% of the stars have a precision better than 0.3 km/s.

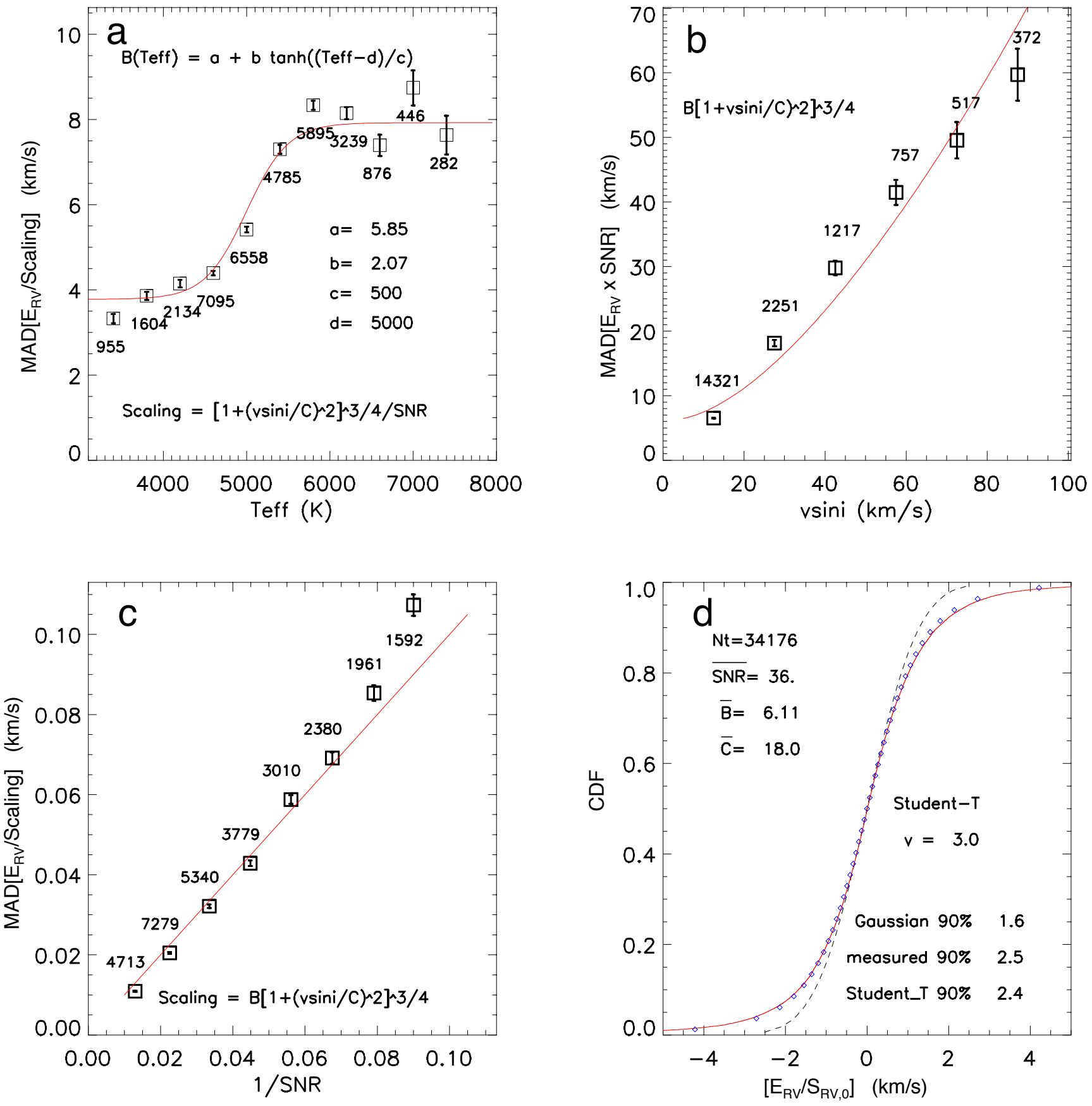

**Fig. B.1.** Fitting empirical parameters describing short-term repeats of RV to GES iDR6 data (see Eq. B.1). Plot (a) shows the variation in parameter B with target $T_{\rm eff}$. Plot (b) shows the variation in $S_{\rm RV,0}$ with $v \sin i$, and plot (c) shows the variation with SNR. Plot (d) compares the cumulative distribution of the normalised measurement uncertainty for short term repeats $E_{\rm RV}/S_{\rm RV,0}$ (red line) with a unit Gaussian distribution (dashed line) and a $\nu = 3$ Student's t-distribution (blue circles).

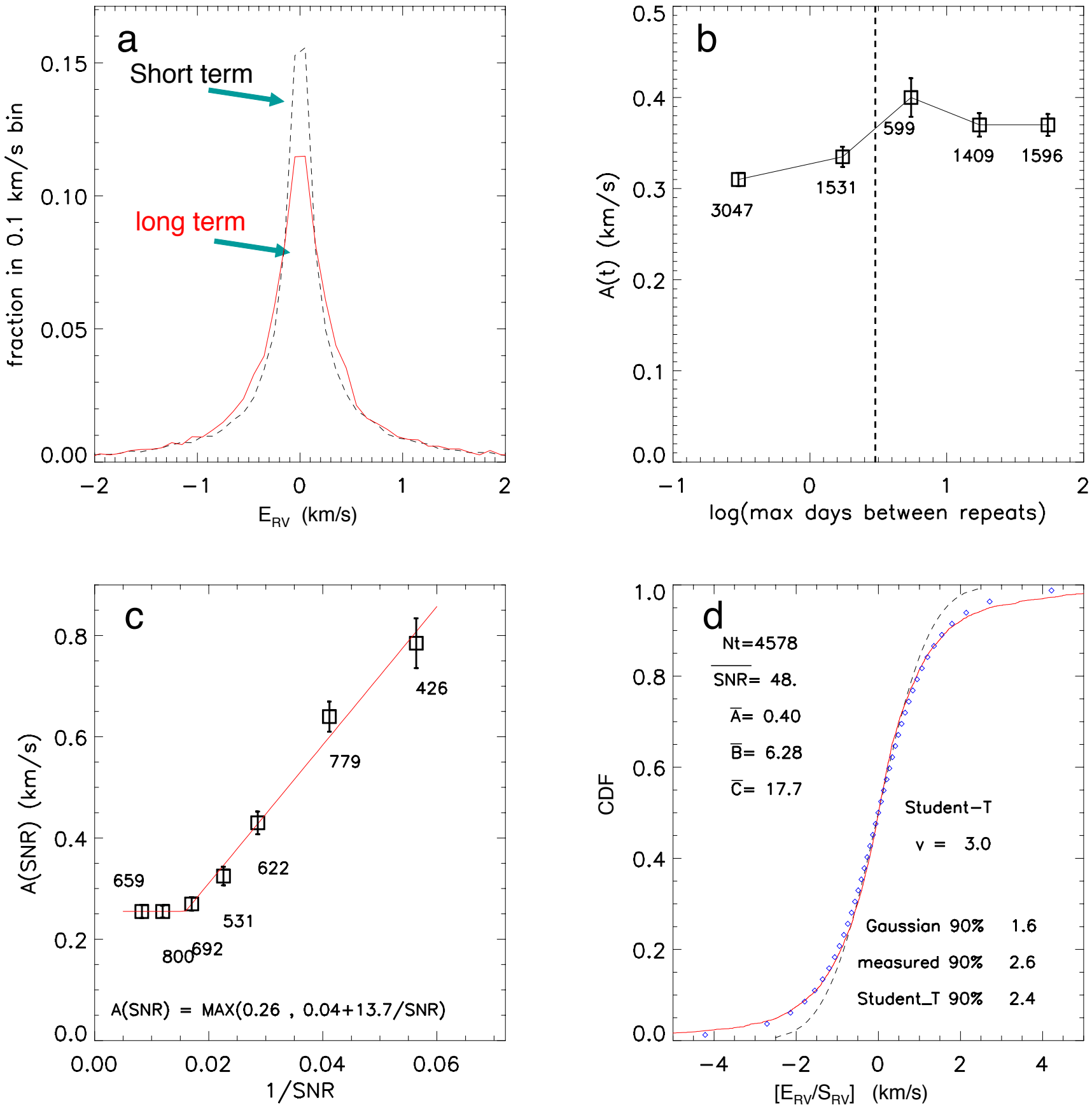

**Fig. B.2.** Fitting empirical parameters describing long-term repeats of RV to GES iDR6 data (see Eq. B.2). Plot (a) shows a histogram of $E_{\rm RV}$ for short- and long-term repeats. Plot (b) shows the variation in the mean value of $A$ as a function of time. Plot (c) shows the variation in A with SNR. Plot (d) compares the cumulative distribution of the normalised measurement uncertainty for long-term repeats $E_{\rm RV}/S_{\rm RV}$ (red line) with a unit Gaussian distribution (dashed line) and a $\nu = 3$ Student's t-distribution (blue circles).